\documentclass [11pt]{article}
 \pdfoutput=1
\usepackage{slashbox}
\usepackage{graphicx}
\usepackage{amsmath,amssymb} 
\usepackage{latexsym}
\usepackage{mathrsfs}
\usepackage{bbold}
\usepackage{color}
\usepackage{slashed}
\definecolor{red}{rgb}{1,0,0}

\makeatletter
\def\section{\@startsection {section}{1}{\z@}{-3.5ex plus -1ex minus
 -.2ex}{2.3ex plus .2ex}{\large\bf}}
\def\subsection{\@startsection{subsection}{2}{\z@}{-3.25ex plus -1ex
minus -.2ex}{1.5ex plus .2ex}{\normalsize\bf}}
\makeatother
\makeatletter

\@addtoreset{equation}{section}

\makeatother

\def\Dslash{\hspace{3pt}\raisebox{1pt}{$\slash$} \hspace{-9pt} D}

\def\bea{\begin{eqnarray}} \def\eea{\end{eqnarray}}
\def\be{\begin{equation}} \def\ee{\end{equation}} \def\nn{\nonumber}
  \def\Z{{\bf Z}}

\newcommand{\Lag}{\mathcal{L}}
\newcommand{\Tr}{\text{Tr}}

\newcommand{\promille}{%
  \relax\ifmmode\promillezeichen
        \else\leavevmode\(\mathsurround=0pt\promillezeichen\)\fi}
\newcommand{\promillezeichen}{%
  \kern-.05em%
  \raise.5ex\hbox{\the\scriptfont0 0}%
  \kern-.15em/\kern-.15em%
  \lower.25ex\hbox{\the\scriptfont0 00}}

\begin{document}

\thispagestyle{empty}

\begin{center}

\hfill SISSA-11/2012/EP \\

\begin{center}

\vspace*{0.5cm}

{\Large\bf  General Composite Higgs Models}

\end{center}

\vspace{1.4cm}

{\bf David Marzocca$^{a}$, Marco
Serone$^{a,b}$ and Jing Shu$^{a}$}\\

\vspace{1.2cm}

${}^a\!\!$
{\em SISSA and INFN, Via Bonomea 265, I-34136 Trieste, Italy} 

\vspace{.3cm}

${}^b\!\!$
{\em ICTP, Strada Costiera 11, I-34151 Trieste, Italy}

\end{center}

\vspace{0.8cm}

\centerline{\bf Abstract}
\vspace{2 mm}
\begin{quote}

We construct a general class of pseudo-Goldstone composite Higgs models, within the minimal $SO(5)/SO(4)$ coset structure, 
that are not necessarily of moose-type.
We characterize the main properties these models should have in order to give rise to a Higgs mass around 125 GeV.
We assume the existence of relatively light and weakly coupled spin 1 and 1/2 resonances.
In absence of a symmetry principle, we introduce the Minimal Higgs Potential (MHP) hypothesis:
the Higgs potential is assumed to be one-loop dominated by the SM fields and the above resonances, with a contribution 
that is made calculable by imposing suitable generalizations of the first and second Weinberg sum rules. We show that  a 125 GeV Higgs requires 
light, often sub-TeV, fermion resonances. Their presence can also be important for the models to successfully pass the electroweak precision tests.
Interestingly enough, the latter can also be passed by models with a heavy Higgs around 320 GeV.
The composite Higgs models of the moose-type considered in the literature can be seen as particular limits of our class of models.

\end{quote}

\vfill

\newpage

\tableofcontents

\section{Introduction}

A possible solution to the gauge hierarchy problem is to assume that the Higgs field is a bound state of some unspecified strongly coupled
constituents.  The mass scale where new resonances  should arise to fully unitarize the Standard Model (SM)
is parametrically higher than the one expected in Technicolor models, leading to alleviated electroweak bounds. These bounds are further mitigated if the composite Higgs
is naturally light, as it happens in models where the Higgs is a pseudo Nambu-Goldstone boson (pNGB) of an approximate spontaneously broken global symmetry of the strong sector  \cite{Kaplan}.
Moreover, the interactions of a pNGB Higgs with the SM gauge bosons are determined by chiral lagrangian techniques \cite{Coleman:1969sm} in terms of a few parameters \cite{Giudice:2007fh}.\footnote{From now on a pNGB Higgs will always be assumed.}
The price to be paid is a fine-tuning needed to keep the Higgs compositeness scale $f$
separated from the Electroweak Symmetry Breaking (EWSB) scale $v$.\footnote{This is not the case for little Higgs models where a hierarchy between $v$ and $f$ can naturally be realized, but
the explicit working implementations of this idea are a bit cumbersome. We will not consider little Higgs models in this paper.}
This price is however not very high, considering that a value of  $v/f\lesssim 1/2$ can be enough to pass the ElectroWeak Precision Tests (EWPT). 
Calculability, on the other hand,  is generically problematic. In particular, the Higgs potential, induced by the explicit breaking of the global symmetry in the composite sector, remains uncalculable.

Most of the progress in model building in composite Higgs models has been obtained in the framework of extra-dimensional  Gauge-Higgs Unification theories \cite{Hatanaka:1998yp} (most notably \cite{Agashe:2004rs}), where a composite Higgs is mapped to a Wilson line phase in the extra dimension and space-time locality automatically leads to a finite Higgs potential (to all orders in perturbation theory). 
Alternatively, as recently worked out in \cite{Panico:2011pw,DeCurtis:2011yx}, one might consider simplified deconstructed versions of 5D models where
calculability is ensured by a collective symmetry breaking mechanism \cite{nima}.  Both the 4D models and the 5D models can be schematically  interpreted as   
consisting of two sectors:
an ``elementary'' sector, including the SM gauge and fermion fields,  and
a ``composite'' strongly coupled sector, including the Higgs field and heavy resonances. The explicit breaking of the global symmetry is 
induced by gauging a part of it via the SM gauge bosons and by quadratic terms which mix the SM
fermions with fermion resonances of the  strong sector. In this particular set-up, SM vectors and fermions become partially composite and the resulting set-up
goes under the name of ``partial compositeness".\footnote{The idea of partially composite SM fermions dates back to \cite{Kaplan:1991dc}, but only extra dimensions
have allowed us to appreciate its full power \cite{Grossman:1999ra}.}  The lighter are the SM fermions, the weaker are the mixing. This simple, yet remarkable, observation allows to 
significantly alleviate most flavour bounds. 

The recent  intense Higgs searches at the LHC have ruled out a SM-like Higgs everywhere below 600 GeV, except a small window around 125 GeV, where an excess has been reported \cite{ATLAS:2012ad} and confirmed at Fermilab \cite{Fermilab}. Based on the ATLAS and CMS experimental results,  similar exclusion limits have recently been found for a composite Higgs \cite{Azatov:2012bz,Esp}.  A Higgs mass of about  320 GeV 
is allowed \cite{Azatov:2012bz}, as well as a wide open region for a fermiophobic composite Higgs in the whole region $110$ -- $500$ GeV. 
Given the reported excess, we will assume here the presence of a Higgs particle at 125 GeV, but
we will also comment on models with a heavy $320$ GeV Higgs, showing that they can also pass the EWPT.

Aim of this paper is to construct four-dimensional pNGB composite Higgs models, not directly related by deconstruction to five-dimensional models,  where the Higgs mass
can at least be assumed to be calculable, and characterize the main properties these models should have in order to give rise to a Higgs mass at around 125 GeV.
More specifically, we focus on the minimal $SO(5)/SO(4)$ coset structure and consider models with an arbitrary number of spin 1 (``vector" and ``axial") and spin 1/2
resonances. These resonances are assumed to be the only ones below the cut-off of the model at $\Lambda = 4\pi f$.
Partial compositeness is assumed. The divergencies of the Higgs potential are cancelled by imposing that certain form factors, both in the gauge and in the fermion sectors, vanish sufficiently fast 
for large euclidean values of the momentum. These conditions are straightforward generalizations of the first and second Weinberg sum rules \cite{Weinberg:1967kj}
and guarantee that the calculable part of the one-loop Higgs potential is finite. Being the Higgs potential a UV-sensitive quantity, and in absence of a symmetry mechanism protecting
it, we will simply assume that the one-loop form factors above represent the main  contributions to the potential, with higher-loop and higher-order operators
giving only a sub-leading correction.  We will denote the above assumption as the Minimal Higgs Potential (MHP) hypothesis.
This is by far the strongest assumption underlying our construction. A similar approach is known to describe quite well the pion mass difference in QCD (see \cite{Contino:2010rs} for a very nice review), in which case the knowledge of the UV theory allows to fix the asymptotic behavior of the relevant form factor for large euclidean momenta. 
There might also be other mechanisms, instead of collective symmetry breaking, protecting the Higgs potential, that effectively lead to a realization of the MHP hypothesis.
Independently of these considerations, the MHP hypothesis can be seen as an effective parametrization valid for a large class of composite Higgs models that predict a calculable Higgs potential.
As an example, we will explicitly show how the models \cite{Panico:2011pw,DeCurtis:2011yx} are particular points in the parameter space of our class of theories.\footnote{In principle, one could even impose the analogue of the first Weinberg sum rule, relaxing the second one, in which case the one-loop Higgs potential would still keep a logarithmic sensitivity to the cut-off. This allows yet more freedom, but calculability in the Higgs sector is now compromised (even within our assumption).
We will nevertheless also comment on this possibility.}

The minimal models that one can construct within our scenario are probably the simplest 4D composite Higgs models. Demanding a finite one-loop Higgs potential requires the presence of one vector and one axial spin 1 resonance and one spin 1/2 fermion resonance (for each SM fermion) mixing with the SM fermions. A non-trivial vacuum can only be obtained by tuning the gauge contribution to the potential versus the fermion one. As a result, EWSB relates the Higgs and the vector resonance masses in a linear way. The Higgs mass is also related to the
fermion resonance masses. In particular, we show that {\it a light Higgs implies light fermion resonances}. This result, already found in several 5D models, 
is proved here on general grounds  and parametrically for the simple case of one fermion resonance, and it is argued to be valid also in more complicated set-ups with more resonances, as confirmed by the study of some selected classes of models. We have just found one counter-example to the light Higgs $\rightarrow$ light fermion resonances implication, based on a chiral composite sector. In this model the right-handed (RH) top quark directly arises as a chiral massless bound state of the composite
sector and a light Higgs (actually too light) does not imply light fermion resonances.

With only one fermion and one vector/axial resonance, demanding a 125 GeV Higgs generally results to too light vector resonances and too large values of the $S$ parameter. The latter can be mildly tuned to acceptable values by considering multiple vector/axial resonances. Adding more than one fermion resonance extends the model building and allows for heavier vector resonances,
alleviating the bounds coming from $S$. Of all the models considered, we also systematically analyze the impact of the EWPT by computing, up to one-loop level,  the calculable new physics fermion contributions to the $S$ and $T$ parameters, and to $\delta g_b$, the deviation to the SM $\bar b_L Zb_L$ coupling. Given the almost unavoidable tree-level positive correction to $S$, the viable models typically require a sizable (positive) fermion contribution to the $T$ parameter. Light fermion resonances are then very welcome from EWPT
considerations as well. The direct search of $b'$-like particles from CMS \cite{Collaboration:2012ye}, which also applies to certain exotic fermions with electric charge $Q=5/3$ appearing in our models, is also included and has a significant impact in some cases.

The structure of the paper is as follows. In section 2 we describe in detail the structure of our models. In section 3 we define the MHP assumption, compute the one-loop Higgs potential and impose the generalized Weinberg sum rules to make it finite. In section 4 we give a closer look at the Higgs potential and show how it plays an important role in predicting the range of masses
for the vector and fermion resonances.
In section 5 three selected classes of models are studied in some detail and more quantitative results are reported, including the bounds coming from EWPT and direct searches.
In section 6 we make a detailed comparison of our models with previous works \cite{Panico:2011pw,DeCurtis:2011yx}.
In section 7 we conclude.
Several details of our computations, as well as the results of an analysis of other classes of models, are reported in 5 appendices. In appendix A the effect on the Higgs potential of certain mixing terms among vector resonances is discussed. In appendix B we report some details about $S$, $T$ and $\delta g_b$. In appendix C we compute the tree-level deviations of the top and bottom gauge couplings from their SM values. We show in detail in appendix D how the fermion sector of the models \cite{Panico:2011pw,DeCurtis:2011yx} 
is reproduced in our set-up. Finally, in appendix E we briefly report the basic results of a set of minimal models based on our construction.

\section{The Basic Set-Up}

We assume  the existence of an unspecified strongly interacting sector with global symmetry group $SO(5)\times U(1)_{X}$,
spontaneously broken to $SO(4)\times U(1)_{X}$.\footnote{The strongly interacting sector should also have an $SU(3)_c$ global symmetry associated to color, but
this is irrelevant for our considerations and will not be considered in what follows.} The Higgs field arises as a pNGB associated to the broken symmetries.
The global symmetry is explicitly broken by gauging a subgroup $SU(2)_L\times U(1)_Y \subset SO(4)\times U(1)_X$
and by mass mixing terms in the fermion sector.
In addition to the SM degrees of freedom, the models contain spin 1 and 1/2 resonances.
We completely neglect in the following all SM fermion fields, except the top quark (and to some extent the bottom quark, see below),
since they do not play an important role in EWSB. 

\subsection{The $\sigma$-Model}

The four pNGBs $h^{\hat{a}}$ can
be described by means of the $\sigma$-model matrix
\begin{equation}
U=\exp\left(i\frac{\sqrt{2}}{f}h^{\hat{a}}T^{\hat{a}}\right)\end{equation}
as the fluctuations along the $SO(5)/SO(4)$ broken directions.
We normalize the $SO(5)$ generators $T^A$ such that in the fundamental representation ${\rm Tr}\, T^A T^B=\delta^{AB}$, where
$A={a,\hat a}$, and $a$, $\hat a$ denote the unbroken and broken directions ($a=1,\ldots,6$, $\hat a=1,\ldots,4$) respectively.
Sometimes it is useful to consider $SU(2)_L\times SU(2)_R$ rather than $SO(4)$, in which case the index $a=(aL,aR)$,
with $aL=a, aR=a+3$ and $a=1,2,3$.
Under a transformation ${\bf g}\in SO(5)$, $U$ transforms non-linearly
as $U\rightarrow {\bf g}U{\bf h}^{\dagger}\left({\bf g}, h^{\hat{a}}(x)\right)$, where
${\bf h}\in SO(4)$. Under the unbroken $SO(4)$ subgroup, the NGBs transform
linearly as a ${\bf 4}$ of $SO(4)$. 
In the unitary gauge, where the NGBs can be taken in the form $h^{\hat{a}}=(0,0,0,h)$, 
the matrix $U$ takes the simple form
\begin{equation}
U=\left(\begin{array}{ccccc}
1 & 0 & 0 & 0 & 0\\
0 & 1 & 0 & 0 & 0\\
0 & 0 & 1 & 0 & 0\\
0 & 0 & 0 & \cos\frac{h}{f} & -\sin\frac{h}{f}\\
0 & 0 & 0 & \sin\frac{h}{f} & \cos\frac{h}{f}\end{array}\right).\label{Uunitary}
\end{equation}
At the leading order in the chiral expansion, the Lagrangian describing
the dynamics of the pNGBs is given by\begin{equation}
\mathcal{L}_\sigma=\frac{f^{2}}{4}\text{Tr}\left(\hat d_{\mu}\hat d^{\mu}\right) 
\label{eq:NGBkinetic}
\end{equation}
where $i U^{\dagger}\partial_{\mu}U=\hat d_{\mu}^{\hat{a}}T^{\hat{a}}+\hat E_{\mu}^{a}T^{a}$.\footnote{Sometimes it is also convenient to
define the linear field $\Sigma=U(h^{\hat{a}})\Sigma_{0}$, transforming as $\Sigma\rightarrow {\bf g} \Sigma$, 
where $\Sigma_{0}^{t}=(0,0,0,0,1)$, and express the $\sigma$-model in terms of this field.}
Gauging the $SU(2)_{L}\times U(1)_{Y}$ SM group simply amounts to promoting
the ordinary derivatives to covariant ones, $\partial_{\mu}\rightarrow D_\mu = \partial_{\mu}-i(g_0W_{\mu}^{a}T_{aL}+g^{\prime}_0B_{\mu}T_{3R})$ and adding
the gauge field kinetic terms (as we will see, $g_0$ and $g^\prime_0$ are only approximate SM gauge couplings, that's why the subscript $0$).
The leading order Lagrangian for the SM gauge fields and the pNGB's
reads\begin{equation}
\mathcal{L}_{\sigma_g}=-\frac{1}{4}W_{\mu\nu}^{aL}W^{aL\mu\nu}-\frac{1}{4}B_{\mu\nu}B^{\mu\nu}+\frac{f^{2}}{4}\text{Tr}\left( d_{\mu} d^{\mu}\right) \,,
\label{LSigmaModel}
\end{equation}
where $ i U^{\dagger}D_{\mu}U=d_{\mu}^{\hat{a}}T^{\hat{a}}+ E_{\mu}^{a}T^{a}$ are the gauged versions of $\hat d_\mu^{\hat a}$ and $\hat E_\mu^a$.
Their first terms in a chiral expansion for a generic $SO(4)$ gauging are \begin{equation}
\begin{cases}
d_{\mu}^{\hat{a}}= & - \frac{\sqrt{2}}{f}(D_{\mu}h)^{\hat{a}}+\ldots\\
E_{\mu}^{a}= & g_0A_{\mu}^{a}+\frac{i}{f^{2}}(h\stackrel{\leftrightarrow}{D_{\mu}}h)^{a}+\ldots
\end{cases}
\end{equation}
The hypercharge is  $Y=T_{3R}+X$ and the SM gauging corresponds to 
\begin{equation}
A_{\mu}^{aL}=W_{\mu}^{a},\qquad A_{\mu}^{3R}=\frac{g_0^\prime}{g_0}B_{\mu}\,.\end{equation}
The explicit breaking of $SO(5)$ due to the gauging and the Yukawas generates a potential for the Higgs through loop corrections and a non-vanishing  vacuum expectation value for $h$. This spontaneously breaks the EW symmetry and gives mass to the SM fermions and gauge fields. The mass of the SM $W$ bosons equals
\be
m_W = \frac{g f}{2}\sin\frac{\langle h \rangle}{f}\equiv \frac{g v}2\,,
\ee 
where $g$ is the physical gauge coupling, see eq.(\ref{gphysical}).
For simplicity of notation, we use in the following the short-hand notation 
\be
\xi \equiv s_h^2 \,, \ \ \ \ \ \ s_h = \sin\frac{\langle h \rangle}{f} \,.
\ee

\subsection{Spin-1 Resonances}

We assume that below the cut-off of the theory at $\Lambda=4\pi f$,
the theory contains spin-1 resonances parametrized by a mass $m_\rho\simeq g_\rho f$ and a coupling $1< g_\rho < 4\pi$. The coupling $g_\rho$ controls both the interactions among the resonances and the resonance-pion interactions. 

There are several ways to add vector resonances in a chiral Lagrangian. They have been shown to be all equivalent,  once field redefinitions and the addition of local counterterms is taken into account  \cite{Ecker:1989yg}. Given our assumptions, the most useful set-up is a generalization of the so-called ``hidden local symmetry" approach, where the resonances $\rho_{\mu}^L$ and $\rho_\mu^R$, in representations $({\bf 3},{\bf 1}) \oplus ({\bf 1},{\bf 3})$ of $SU(2)_L\times SU(2)_R$ respectively, transform non-linearly, while the
resonances $a_\mu$, forming $({\bf 2},{\bf 2})$ representations of $SU(2)_L\times SU(2)_R$, 
transform homogeneously. With an abuse of language, for simplicity we will denote in the following the $\rho_\mu^{L,R}$'s and the $a_\mu$ as ``vector" and ``axial" resonances, respectively, although not all $\rho_\mu^{L,R}$ and not all $a_\mu$ actually transform under parity as vector and axial gauge fields. Under a transformation ${\bf g}\in SO(5)$, we have
\begin{equation}
\quad\begin{cases}
\rho_{\mu}^L = \rho_{\mu}^{aL}T^{aL} & ,\quad\rho_{\mu}^L \rightarrow {\bf h}\rho_{\mu}^L {\bf h}^{\dagger}+\frac{i}{g_{\rho_L}} ({\bf h}\partial_{\mu}{\bf h}^{\dagger})^L, \\
\rho_{\mu}^R = \rho_{\mu}^{aR}T^{aR} & ,\quad\rho_{\mu}^R \rightarrow {\bf h}\rho_{\mu}^R {\bf h}^{\dagger}+\frac{i}{g_{\rho_R}}  ({\bf h}\partial_{\mu}{\bf h}^{\dagger})^R, \\
a_{\mu}=a_{\mu}^{\hat{a}}T^{\hat{a}} & ,\quad a_{\mu}\rightarrow {\bf h}a_{\mu}{\bf h}^{\dagger},\end{cases}
\label{eq:spin1_transf}
\end{equation}
where ${\bf h}={\bf h}({\bf g},h^{\hat a})$. At  leading order in derivatives, the most general Lagrangian allowed by eq.(\ref{eq:spin1_transf}) for $N_{\rho_L}$ multiplets in the $({\bf 3},{\bf 1})$, $N_{\rho_R}$ in the $({\bf 1},{\bf 3})$ and $N_{a}$ axial vectors in the $({\bf 2},{\bf 2})$ is 
\be
	\Lag_g =  \Lag^{v_L} + \Lag^{v_R} + \Lag^{a},
\ee
where
\begin{eqnarray}
	\Lag^{v_L} &=& \sum_{i=1}^{N_{\rho_L}} \bigg(  - \frac{1}{4} \Tr \left(\rho^{i}_{L,\mu\nu} \rho^{i, \mu\nu}_L\right) + \frac{f^2_{\rho_{L}^i}}{2} \Tr \left( g_{\rho_{L}^i} \rho_{L,\mu}^i -  E_\mu^L \right)^2+  \sum_{j<i} \frac{f^2_{\text{mix}_{ij}}}{2} \Tr \Big( g_{\rho_{L}^i} \rho_{L,\mu}^i - g_{\rho_{L}^j} \rho_{L,\mu}^j \Big)^2 \bigg), \nonumber \\
	\Lag^{v_R} &=& \Lag^{v_L}\,, \text{ with } L \rightarrow R, \nonumber \\
	\Lag^{a} &=&  \sum_{i=1}^{N_{a}} \Big(  - \frac{1}{4} \Tr \left(a^i_{\mu\nu} a^{i \mu\nu}\right) + \frac{f^2_{a_i}}{2 \Delta_i^2} \Tr \left( g_{a_i} a^{i}_\mu - \Delta_i d_\mu \right)^2   \Big).
\label{eq:LGeneralGaugeResonances}
\end{eqnarray}
In eq.(\ref{eq:LGeneralGaugeResonances}), $E_\mu^{L,R}$ are the $SU(2)_{L,R}$ components of $E_\mu$.
The field strengths and covariant derivatives are defined as
\be
 \rho_{L,\mu\nu}^i=\partial_{\mu}\rho_{L,\nu}^i-\partial_{\nu}\rho_{L,\mu}^i-i g_{\rho_L^i}[\rho_{L,\mu}^i,\rho_{L,\nu}^i], \ \ \ 
 a_{\mu\nu}=\nabla_{\mu}a_{\nu}-\nabla_{\nu}a_{\mu}, \ \  \nabla=\partial -i E.
 \ee
Note that for the axial vectors there is no need to add mass mixing terms, since one can always diagonalize the quadratic terms and bring the Lagrangian in the form above.  It is useful to define the mass parameters
\be
	m_{\rho_{L}^i}^2 = f^2_{\rho_{L}^i} g_{\rho_{L}^i}^2, \quad
	m_{\rho_{R}^i}^2 = f^2_{\rho_{R}^i} g_{\rho_{R}^i}^2, \quad
	m_{a_i}^2 = \frac{f^2_{a_i} g_{a_i}^2}{\Delta_i^2},
\ee
keeping of course in mind that the actual masses for the $\rho$'s in presence of mixing have to be obtained via a diagonalization of the quadratic terms.
The mass terms in eq.(\ref{eq:LGeneralGaugeResonances}) induce mixing terms between the vector resonances $\rho_{L,\mu}^i$ ($\rho_{R,\mu}^i$)  and the SM gauge fields $W$ ($B$),
as expected by the partial compositeness scenario \cite{Contino:2006nn},  generalized to more resonances. 
For  $N_{\rho_L}=N_{\rho_R}=1$, the actual mass eigenstates before EWSB are found by simple $SO(2)$ rotations: $W_{aL}\rightarrow W_{aL}  \cos \theta_g +\rho_{aL} \sin\theta_g $, $B\rightarrow B \cos \theta^\prime_{g^\prime}  + \rho_{3R} \sin\theta_{g^\prime}$ (and similar transformations for $\rho_{aL}$ and $\rho_{3R}$), where $\tan\theta_g = g_0/g_{\rho_L}$, $\tan\theta_{g^\prime} = g^\prime_0/g_{\rho_R}$. Alternatively, for sufficiently heavy resonances, one can keep the original $W$ and $B$ fields and integrate out the resonances.  The two descriptions are obviously equivalent, but depending on the problem at hand, one can be more convenient than the other.

 We assume that the coefficients of higher dimensional operators are dictated by Na\" ive Dimensional Analysis (NDA), where $g_\rho$
is treated as a ``weak" coupling. This should in principle be contrasted to the recent  \emph{partial UV completion} (PUVC) hypothesis, introduced in \cite{Contino:2011np}, according to which the couplings of higher dimensional operators  should not exceed the $\sigma$ model coupling, $g_* = \Lambda / f$, at the cutoff scale $\Lambda$. In particular, the NDA hypothesis
puts more severe bounds on the values of the coefficients of the higher dimensional operators.  For instance, let us consider as an illustration the ${\cal O}(p^4)$ operators $Q_1$ and $Q_2$ (in the notation of \cite{Contino:2011np}), $Q_1= {\rm Tr}\, (\rho^{\mu\nu} i [d_\mu,d_\nu])$, $Q_2 = {\rm Tr}\,(\rho^{\mu\nu}f_{\mu\nu}^+)$. The NDA and PUVC estimates of their couplings $\alpha_1$ and $\alpha_2$ are  
\be
\begin{split} \alpha_1^{(NDA)}& \simeq \; \frac{g_\rho}{16\pi^2} \,, \ \ \ \ \ \alpha_1^{(PUVC)} \leq \frac{1}{4\pi}\,,  \\
\alpha_2^{(NDA)} & \simeq \; \frac{g g_\rho}{16\pi^2} \,, \ \ \ \ \ \alpha_2^{(PUVC)} \leq 1\,.
\end{split}
\ee
We see that the two estimates are consistent with each other, but the PUVC hypothesis allows for larger coefficients.

Demanding a partial unitarization of $\mathcal{A}(h^{\hat a}h^{\hat b} \rightarrow h^{\hat c}h^{\hat d})$ by the vector resonances allows to select a definite range in the values of $f_\rho$ and $f_a$.
For example, for one vector resonance $\rho_\mu$ in the adjoint of $SO(4)$, assuming left-right (LR) $\Z_2$ symmetry, from the Lagrangian in eq.(\ref{eq:LGeneralGaugeResonances}) and eq.(\ref{eq:NGBkinetic}) one can obtain its contribution to the $hh$ scattering amplitude \cite{Contino:2011np}.
Neglecting the finite width of the resonance and for $s,t,u\gg v^2$, one has
\be \begin{split}
	\mathcal{A}(h^{\hat a}h^{\hat b} \rightarrow h^{\hat c}h^{\hat d})=&\; A(s,t,u) \delta^{\hat a\hat b} \delta^{\hat c\hat d} + A(t,s,u) \delta^{\hat a\hat c} \delta^{\hat b\hat d} + A(u,t,s) \delta^{\hat a\hat d} \delta^{\hat b\hat c}, \\
	A(s,t,u) =&\; \frac{s}{f^2} \left( 1 - \frac{3}{2} a_\rho^2 \right) -\frac{a_\rho^2}{2} \frac{m_\rho^2}{f^2} \left[ \frac{s-u}{t-m_\rho^2} + \frac{s-t}{u-m_\rho^2} \right],
\end{split}\ee
where 
\be
a_\rho \equiv \frac{f_\rho}{f}
\label{arhoDef}
\ee
 and $s,t,u$ are the usual Mandelstam variables. From this formula one can check that $\rho_\mu$ unitarizes the scattering for $a_\rho = \sqrt{2/3}$. Assuming PUVC one obtains the bounds $a_\rho \sim 1$ and $f_a / f \equiv a_a \lesssim 1$, which we will typically assume in the following.

\subsection{Spin-1/2 Resonances}

According to the formalism introduced in \cite{Coleman:1969sm}, the most general Lagrangian invariant under a non-linearly
realized group $G$, spontaneously broken to a linearly realized subgroup $H$, 
should be written using the components $d_{\mu}$ and the covariant derivative
$\nabla_{\mu}=\partial_{\mu}-iE_{\mu}$ introduced before, that act on matter fields in representations of $H$.
Therefore, we expect the Lagrangian of the spin 1/2 resonances to
be just $SO(4)\times U(1)_X$-invariant.

A source of model-dependence arises when we have to couple such composite
fermions to the elementary SM ones. We advocate here the partial compositeness scenario, 
according to which SM fermions mix with the spin 1/2 resonances \cite{Kaplan:1991dc}. Such mixing is
realized via a linear coupling $\lambda\bar{q}_{i}\mathcal{O}_{i}+h.c.$,
where $\mathcal{O}_{i}$ are composite operators made of the composite
fermions $\Psi_i$ and the pNGB's and it explicitly breaks the global symmetries of the composite sectors.

In order to simplify the possible choices of operators $\mathcal{O}_{i}$, we focus on those that transform linearly 
under the whole group $SO(5)$. Since the fermion resonances $\Psi_i$ sit in representations $r_H$ of $H$ only, this implies
that  $\mathcal{O}_{i} \sim U \Psi_i$, so that  $\mathcal{O}_{i} \rightarrow {\bf g} \mathcal{O}_{i} $. Any representation $r_H$ can be ``dressed"
with the matrices $U$ to get representations of $G$. We will not perform a systematic study of all possible $r_H$'s here, but focus on 
two representations only, the singlet and the fundamental ${\bf 4}\sim ({\bf 2}, {\bf 2})$. Let us consider  $N_S$ and $N_Q$ singlets and bi-doublets
spin 1/2 resonances $S_i$ and $Q_j$ ($i=1,\ldots, N_S$, $j=1,\ldots, N_Q$), with $U(1)_X$ charge $q_X=+2/3$.
From these fields, we can construct fermions transforming in the fundamental of  $G$
as follows:
\be
\sum_{a=1}^4 U_{A a} Q_{a,j}\,, \ \ \ U_{A 5} S_{i}\,, \ \ \ A = 1,\ldots, 5\,,
\label{PCops}
\ee
where we have explicitly reported the $SO(5)$ group indices. Each of the above two operators (\ref{PCops}) can couple to the SM fermion fields.
The latter are conveniently written in terms of spurion five-component fermions $\xi_L$ and $\xi_R$, formally transforming in the fundamental of $SO(5)$ and
with $U(1)_X$ charge $q_X=2/3$.
Keeping only the SM doublet $q_L = (t_L, b_L)^t$ and $t_R$, we can write the following two chiral spurions:
\begin{equation}
\xi_L=\frac{1}{\sqrt{2}}\left(\begin{array}{c}
b_{L}\\
-ib_{L}\\
t_{L}\\
it_{L}\\
0\end{array}\right), \ \ \ \ \  \xi_{R}=\left(\begin{array}{c}
0 \\ 0 \\ 0 \\ 0 \\  t_{R}\end{array}\right)\,.\end{equation}
The leading order Lagrangian for the SM and composite fermions  is easily constructed:
\be
\begin{split}
{\cal L}_{f,0}  & =\;   \bar q_L i \Dslash q_L +\bar t_R i \Dslash t_R +\sum_{i=1}^{N_S} \bar S_i (i \slashed\nabla - m_{iS}) S_i 
+\sum_{j=1}^{N_Q} \bar Q_j (i \slashed\nabla - m_{iQ}) Q_j + \\
&\;  \sum_{i=1}^{N_S}\Big( \frac{\epsilon^i_{tS}}{\sqrt{2}} \bar \xi_R P_L U S_i+ \epsilon^i_{qS} \bar \xi_L P_R U S_i \Big)+
  \sum_{j=1}^{N_Q}\Big( \frac{\epsilon^j_{tQ}}{\sqrt{2}} \bar \xi_R P_L U Q_i+ \epsilon^j_{qQ} \bar \xi_L P_R U Q_i \Big) + h.c. ,
  \end{split}
\label{eq:LGeneralFermions}
\ee
where a $\sqrt{2}$ factor in the definition of $\epsilon_{tS,tQ}^{i,j}$ has been introduced for later convenience and 
\be
\nabla_{\mu} = \partial_\mu - i E_\mu - i q_X g^\prime_0 B_\mu \,.
\ee
There are in general $3N_Q + 3 N_S$ complex phases appearing in eq.(\ref{eq:LGeneralFermions}), $2 N_Q + 2 N_S+1$ of which can be reabsorbed by appropriate phase redefinitions of the fermion fields, for a total of $N_Q + N_S - 1$ physical phases. Therefore, without any loss of generality, we can take the vector masses $m_{iS}$ and $m_{jQ}$ to be real and positive. 
Along the lines of \cite{Panico:2011pw}, it will be useful to rewrite the last row in \eqref{eq:LGeneralFermions} as
\be
  \sum_{i=1}^{N_S}\Big(\bar t_R E^i_{tS}P_L U S_i+  \bar q_L E^i_{qS} P_R U S_i \Big)+
  \sum_{j=1}^{N_Q}\Big( \bar t_R E^j_{tQ} P_L U Q_i+ \bar q_L E^j_{qQ} P_R U Q_i \Big) + h.c.
\ee
where the $E$'s are spurion mixing terms, transforming as follows under the enlarged group $SU(2)_L^0\times U(1)_R^0\times  U(1)_X^0\times SO(5)\times U(1)_X$, eventually broken
to $G_{SM}$ by the spurion VEV's: 
\be
E^i_{tS}, E^j_{tQ} \sim ({\bf 1},0, 2/3,  {\bf \bar 5}, -2/3), \ \ \ \ E^i_{qS}, E^j_{qQ} \sim ({\bf 2}, -1/2, 2/3, {\bf \bar 5}, -2/3)\,.
\label{Espurions}
\ee
Couplings between spin 1/2 and spin 1 resonances and additional couplings to the $\sigma$-model fields $d_\mu$ and $E_\mu$ are easily 
constructed by recalling that $g_\rho \rho_\mu-E_\mu$, $a_\mu$ and $d_\mu$, under $SO(5)$, homogeneously transform according to local $SO(4)$ transformations.
The most general leading order couplings are the following (assuming LR symmetry):
\be
\begin{split}
\mathcal{L}_{f,int}& = \!\sum_{\eta=L,R}\! \!\bigg(\! k^{V,\eta}_{ijk} \bar Q_{j} \gamma^\mu (g_{\rho^i} \rho_{\mu}^i- E_{\mu}) P_\eta Q_k+ k^{{\rm mix}}_{ijkl} \bar Q_i \gamma^\mu (g_{\rho^k} \rho^k_\mu -g_{\rho^l} \rho^l_\mu) P_{\eta} Q_j   \\
&
+  k^{A,\eta}_{ikj} \bar S_i \gamma^\mu g_{a^k} a_{\mu}^k  P_\eta Q_j + \sum_{i,j} k^{d,\eta}_{ij} \bar S_i \gamma^\mu d_\mu P_\eta Q_j +
h.c. \!\bigg),
\label{LagInt11/2}
\end{split}
\ee
where $P_\eta$ are chiral projectors.

With the above choice of fermion quantum numbers, 
$b_L$ mixes with the bi-doublet component of the fermion resonance with $T_{3R}=T_{3L}$ and potentially large contributions to $\delta g_b$ vanish \cite{Agashe:2006at}.

The total fermion Lagrangian is obtained by summing eqs.(\ref{eq:LGeneralFermions}) with (\ref{LagInt11/2}):
\be
{\cal L}_{f} = {\cal L}_{f,0} + {\cal L}_{f,int}\label{fullLag}\,.
\ee
The fermion Lagrangian (\ref{fullLag}) is easily generalized to include the couplings to other SM fermions. For instance, the bottom quark sector can be obtained by adding to eq.(\ref{fullLag}) the $b_R$ field and additional fermion singlet and bi-doublet resonances $S_i^{(d)}$ and $Q_j^{(d)}$, with $q_X=-1/3$. The latter mix to $b_R$ and $b_L$ by means of operators of the form $\bar b_{L/R} U S_{i,R/L}^{(d)}$ and $\bar b_{L/R} U Q_{j,R/L}^{(d)}$. These mixing affect the top sector, but they are safely negligible, given the smallness of the bottom mass.
They also induce a non vanishing tree-level $\delta g_b$, which is however  sub-dominant with respect to one-loop corrections coming from fermion mixing in the charge 2/3 (top) sector. 
For completeness, we report in appendix \ref{GCD} the detailed form of $\delta g_b$ at tree-level, as well as other coupling deviations, when $N_S^{(d)}$ and $N_Q^{(d)}$ singlets and bi-doublets fermion resonances with $q_X=-1/3$ are added. It is then consistent to consider the Lagrangian (\ref{fullLag}), neglecting altogether the fermion resonances $S_i^{(d)}$ and $Q_j^{(d)}$.

It is useful to discuss in some more detail the simple case $N_S=N_Q=1$.  For simplicity let us take real mixing terms $\epsilon_{t,q/Q,S}$. We see from 
eq.(\ref{eq:LGeneralFermions}) that before EWSB the LH top mixes with $Q$ through the parameter $\epsilon_{qQ}$ and the RH top mixes with $S$ through $\epsilon_{tS}$,
The degree of compositeness of the top quark can be measured by the angles $\theta_{L,R}$ \cite{Contino:2006nn} defined as:
\be
\tan \theta_L = \frac{|\epsilon_{qQ}|}{m_Q}\,, \ \ \ \tan \theta_R = \frac{|\epsilon_{tS}|}{\sqrt{2}m_S}\,.  
\label{tLtR}
\ee
The more $\tan\theta_{L/R}$ is large, the more $t_{L/R}$ is composite. For $s_h\ll 1$, the top mass is given by
\be
m_{top} \simeq \frac{\sin\theta_L \sin\theta_R}{\sqrt{2}}\Big|\frac{\epsilon_{qS}}{\epsilon_{qQ}} m_Q-\frac{\epsilon_{tQ}}{\epsilon_{tS}}m_S\Big| s_h\,.
\label{mtop}
\ee
The physical masses of the fermion resonances, before EWSB, are the following:
\be
m_{0}= \frac{m_S}{\cos\theta_R}\,, \ \ \ m_{1/6}= \frac{m_Q}{\cos\theta_L}\,, \ \ \ \ m_{7/6}= m_Q\,,
\label{mRes12}
\ee
where the subscripts  $0$, $1/6$ and $7/6$ denote the hypercharges of the singlet and of the two $SU(2)_L$ doublets forming the bi-doublet $Q$.

The case in which $t_R$ is fully composite can be studied by assuming that $t_R$ is a chiral massless fermion bound state coming from the composite sector
and directly identifying it  as the RH component of the singlet fermion resonance $S_R$ in eq.(\ref{eq:LGeneralFermions}).
In this way, $t_R$ and $S_L$, and hence the parameters $m_S$, $\epsilon_{tS}$ and $\epsilon_{tQ}$,  should be removed from eq.(\ref{eq:LGeneralFermions}). 
We will come back to this particularly simple model in section \ref{Sec:results}.
 
The total Lagrangian of the model is finally given by
\be
{\cal L}_{Tot} = {\cal L}_{\sigma_g} + {\cal L}_g + {\cal L}_f\,.
\label{LTOT}
\ee 


\section{The Minimal Higgs Potential Hypothesis and Weinberg Sum Rules}

The Higgs potential in our model is, strictly speaking, not calculable. The pNGB nature of the Higgs implies that its potential $V(h)$ depends on $s_h$ only. 
For $s_h \ll 1$, we can expand $V(h)$ up to quartic order and obtain
\be
V(h) \simeq-\gamma \, s_h^2+\beta \, s_h^4\,.\end{equation}
The coefficients $\gamma$ and $\beta$ are induced by the explicit breaking of the shift symmetries, the gauge couplings $g$ and $g^\prime$ in the gauge sector, the mixing terms
$\epsilon$ in the fermion sector and possibly other terms coming from higher dimensional operators, not appearing in the Lagrangian (\ref{LTOT}). There are generically two different contributions to $\gamma$ and $\beta$ that, with an abuse of language, we denote by IR and UV contributions.
The IR contribution is the one coming from the leading operators defining our model (\ref{LTOT}), the UV contribution is the one coming from higher dimensional operators and physics at the cut-off scale. The explicit form of $\gamma$ and $\beta$ can be deduced, in the limit of small breaking terms, by a simple spurion analysis \cite{Panico:2011pw}.  
As expected from NDA,  the IR contribution to $\gamma$ and $\beta$ shows generically quadratic and logarithmic divergencies, respectively. Instead of introducing as usual counterterms for such divergencies, leading to a loss of predictability in the Higgs sector, we can demand that the one-loop form factors defining the IR part of $\gamma$ and $\beta$, that should be integrated over all energies scales, are peaked around the resonance masses and go to zero sufficiently fast at infinity. This is done by fulfilling some generalized Weinberg sum rules.  In this way, the one-loop IR contribution to $V(h)$ can be made finite.

On the other hand, local operators, induced from states above the cut-off scale and possibly mixing with the SM fields, contribute to the UV part of $\gamma$ and $\beta$.
For small fermion mixing terms, the leading operators are\footnote{The leading fermion local operators above were not considered in \cite{Panico:2011pw}. This is probably due to the fact that the free fermion composite Lagrangian has an obvious linearly realized $SO(5)$ symmetry when $m_{iQ}=m_{iS}$. In addition, when the mixing terms are taken to be equal, $\epsilon^i_{tS}=\epsilon^i_{tQ}$ and $\epsilon^i_{qS}=\epsilon^i_{qQ}$ (as in \cite{Panico:2011pw}),
the whole Higgs field can be removed from the quadratic fermion Lagrangian by a field redefinition and hence vector mass insertions are needed to get a non-trivial one-loop potential. This is however an accident of the one-loop result and fermion operators like the ones in eq.(\ref{NDAVH}) will be anyway generated at higher loop level.}   
\be
\begin{split}
c_g f^4  \sum_{a_L=1}^3 \Sigma^t g T^{a_L} g T^{a_L} \Sigma & =   \frac 34 c_g g^2  f^4 s_h^2 \equiv \gamma_g^{(NDA)} s_h^2\,,   \\
\frac{d_g}{16\pi^2} f^4  (\sum_{a_L=1}^3 \Sigma^t g T^{a_L} g T^{a_L} \Sigma)^2&  =   \frac{9}{256\pi^2} d_g g^4  f^4 s_h^4 \equiv \beta_g^{(NDA)} s_h^4 \,, \\
c_f f^2 (E_{q S} \Sigma) (\Sigma^t E_{qS}^\dagger) & = \frac 12 |\epsilon_{qS}|^2 s_h^2 \equiv \gamma_f^{(NDA)} s_h^2 \,,  \\
\frac{d_f}{16\pi^2}  \Big((E_{q S} \Sigma) (\Sigma^t E_{qS}^\dagger)\Big)^2 & =  \frac 1{64\pi^2} |\epsilon_{qS}|^4 s_h^4 \equiv \beta_f^{(NDA)} s_h^4 \,, 
\end{split}
 \label{NDAVH}
\ee
where $c_{g,f}$ and $d_{g,f}$ are estimated by NDA to be  coefficients of ${\cal O}(1)$. Similar operators can obviously be written in terms of $g^\prime$ and the other
spurion mixing terms in eq.(\ref{Espurions}).
By comparing $\gamma_g^{(NDA)}$ and  $\beta_g^{(NDA)}$ with the typical values
one gets from the IR contribution, once made calculable (such as eqs.(\ref{gammabetaLogDiv}) and (\ref{betagammaferExp}) below), we see that $\gamma_{g,f}^{(NDA)}> \gamma_{g,f}$ and $\beta_{g,f}^{(NDA)}\simeq \beta_{g,f}$ so that calculability is still lost. In order to circumvent this problem, we  assume here that the underlying UV theory is such that $\gamma^{(NDA)}$ and
$\beta^{(NDA)}$ are sub-leading with respect to the IR part of $\gamma$ and $\beta$, so that the Higgs potential is calculable and dominated at one-loop level by the fields in our model.
The logic underlying the above assumption (that might seem too radical and strong) is that  any theory where a symmetry mechanism is at work (not only collective breaking or higher dimensions) to actually predict a calculable Higgs potential would automatically satisfy the above requirements and fall into our class of models, which can then be seen as a general parametrization of composite Higgs models. We denote the above assumption as the Minimal Higgs Potential (MHP) hypothesis.

Having explained the philosophy of our perspective, we turn to the computation of the IR contribution of the one-loop Higgs potential, from now on simply denoted by  the Higgs potential.
The latter is conveniently computed by first integrating out the heavy spin 1 and 1/2 resonances, with no need to go to a mass basis, 
and then by integrating out the remaining light degrees of freedom. This is a useful way to proceed, because the pseudo-Goldstone nature of the Higgs field and the $SO(5)\times U(1)_X$ symmetries  allow to fix in terms of a few form factors the form of the effective Lagrangian for the light states and encode there all the information of the heavy resonances.
We will be quite brief in the derivation of the Higgs potential in the following, since all the relevant steps have been repeatedly used in the literature and are by now well-known.

\subsection{Gauge Contribution}
In momentum space, the effective Lagrangian of the SM gauge fields up to quadratic order in the gauge fields and to any order in the Higgs field can be written 
in terms of 3 scalar form factors $\Pi_{W^+W^-}=\Pi_{W_3W_3}$, $\Pi_{BB}$ and $\Pi_{W_3B}$, functions of $p^2$:
\be
 \frac{P^{\mu\nu}_t}{2} \Big( 2 \Pi_{W^+W^-}W_\mu^{+}W_\nu^- + \Pi_{W_3W_3} W_\mu^{3} W_\nu^3  +\Pi_{BB}  B_\mu B_\nu + 2 \Pi_{W_3B} W_\mu^{3} B_\nu\Big),
\label{LagPi}
\ee
where $P_{t}^{\mu\nu}=\eta^{\mu\nu}-p^{\mu}p^{\nu}/p^{2}$ is the
projector on the transverse field configurations and the $\Pi$'s are form factors that also depend on the Higgs field.
The one-loop Higgs potential is easily computed from the above expression by taking the Landau gauge $\partial^{\mu}B_{\mu}=\partial^{\mu}W_{\mu}^{a}=0$.
In this gauge the longitudinal components of the gauge fields, as
well as the ghosts, decouple and can be neglected. Integrating out
the gauge fields and going to Euclidean momenta, one gets:\begin{equation}
V_g(h)=\frac{3}{2}\int\frac{d^{4}p_E}{(2\pi)^{4}}  \Big(2\:\log \Pi_{W^+W^-}(-p_E^2)+\log\left(\Pi_{BB}(-p_E^2)\Pi_{W_3W_3}(-p_E^2)-\Pi_{W_3B}^2(-p_E^2)\right)\Big)\,.
\label{Vgauge}
\end{equation}

To have an analytic understanding of the possible functional dependence on the Higgs field of the effective potential, it is useful to introduce spurionic gauge fields such that the whole $SO(5)\otimes U(1)_X$ group becomes gauged: $A_\mu = A_\mu^{\hat{a}} T^{\hat{a}} + A_\mu^{aL} T^{aL} +  A_\mu^{aR} T^{aR}$.
The most general $SO(5)\otimes U(1)_X$-invariant Lagrangian depending on the gauge fields and the NGB's, at the quadratic order in the gauge fields and in momentum space, is
\be
\begin{split}
\mathcal{L}^{eff}=\frac{P_{t}^{\mu\nu}}{2} & \left(\Pi_0^X(p^2) X_\mu X_\nu + \Pi_{0}(p^{2})\text{Tr}[A_{\mu}A_{\nu}] + \Pi_{1}(p^{2})\Sigma^{t}A_{\mu}A_{\nu}\Sigma + \right. \\
		& \left. + \Pi_{LR}(p^2) \left( \Tr[(U^\dagger A_\mu U)^L (U^\dagger A_\nu U)^L ] - \Tr[(U^\dagger A_\mu U)^R (U^\dagger A_\nu U)^R] \right) \right),
\label{eq:generalSO5}
\end{split}
\ee
where $(\ldots)^{L,R}$ implies the projection on the $({\bf 3},{\bf 1})$ and $({\bf 1},{\bf 3})$ irreducible representations inside the adjoint of $SO(4)$.\footnote{The term in the second line of \eqref{eq:generalSO5} could be generated, for example, by the operator $O_3 = \left(\Tr[E_{\mu\nu}^L E^{L\:\mu\nu}] - \Tr[E_{\mu\nu}^R E^{R\:\mu\nu}]\right)$ \cite{Contino:2011np}, or directly in a model with vector resonances $\rho_\mu^L, \rho_\mu^R$ without invariance under $L \leftrightarrow R$, see section \ref{Sec:LRasymmCase}.}
Switching off the spurionic fields, that is keeping only the components $A_{\mu}^{aL}=W_{\mu}^{a}$, $A_{\mu}^{3R}= c_X B_{\mu}$ and $X_\mu = s_X B_\mu$, where
\be
	c_X = \frac{g_X}{\sqrt{g_0^2 + g_X^2}}=\frac{g_0^\prime}{g_0}, \quad s_X = \frac{g_0}{\sqrt{g_0^2 + g_X^2}},
\ee
we obtain the most general effective Lagrangian for the gauge bosons in $SO(5)/SO(4)$ with the explicit dependence on the Higgs field:
\begin{equation}
\begin{split}
\mathcal{L}^{eff}=\frac{P_{t}^{\mu\nu}}{2} & \bigg( \Pi_{0}W_{\mu}^{a}W_{\nu}^{a} + \Pi_{1} \frac{s_h^2}{4}\left(W_{\mu}^{1}W_{\nu}^{1} + W_{\mu}^{2}W_{\nu}^{2}\right) +\\
 & + \Pi_{B} B_{\mu}B_{\nu} + \Pi_{1} \frac{s_h^2}{4}\left(\frac{g_0^\prime}{g_0} B_{\mu}-W_{\mu}^{3}\right)\left(\frac{g_0^\prime}{g_0} B_{\nu}-W_{\nu}^{3}\right)+ \\
 &  + c_h \Pi_{LR} \left( W_{\mu}^{a}W_{\nu}^{a} - \frac{g_0^{\prime 2}}{g_0^2}B_\mu B_\nu \right)  \bigg),
\end{split}\label{LgaugeLRAsym}
\end{equation}
where $\Pi_B = (s_X^2 \Pi_0^X + c_X^2 \Pi_0)$, $c_h=\cos \langle h\rangle/f$, and $g_0^\prime = g_0 c_X$.
From this Lagrangian one obtains 
\be \begin{split}
	\Pi_{W^+W^-} = \Pi_{W_3W_3} =&\; \Pi_0 +\frac{s_h^2}4 \Pi_1 + c_h \Pi_{LR}, \\
	\Pi_{BB} =&\; \Pi_B+ c_X^2\frac{s_h^2}4\Pi_1 -c_X^2 c_h \Pi_{LR}, \\
	\Pi_{W_3B} =&\; -c_X\frac{s_h^2}4\Pi_1\,.
\end{split}
\label{eq:GaugeFormFactorHiggsDep}
\ee
The form factor $\Pi_{W_3 B}$ is related to the $S$-parameter \cite{Peskin:1991sw}:
\be
\Delta S =  - \frac{16\pi}{g g^\prime} \Pi_{W_3B}^\prime(0) \simeq \frac{16\pi}{g^2}\frac{s_h^2}{4} \Pi^\prime_1(0)\,,
\label{Spara}
\ee
where $\Delta S=S-S_{SM}$ (see appendix \ref{App:EWPT}).
It is well known that  $\Delta S$ is the main phenomenological electroweak bound constraining Composite Higgs Models, that requires $s_h<1$.
As we will show below, a necessary condition to kill the quadratic divergence in the potential is to demand $\lim_{p_E\rightarrow \infty} \Pi_{LR} = 0$. In order to ensure this condition and to keep the model simple, in the following we impose a LR symmetry in the strong sector, that automatically implies $\Pi_{LR}=0$.

The explicit form of the form factors is obtained by integrating out the heavy vector resonances at tree-level and quadratic order (the one relevant at one-loop level). 
This is not straightforward to do for an arbitrary number of vector resonances, due to the last term in ${\cal L}^{v_L}$, eq.(\ref{eq:LGeneralGaugeResonances}).
Let us then set $f_{\rm mix}=0$ in the following (see appendix \ref{Mixtwovect} for the effect of this term in the two vector case).
In this simple case, we get
 \be
 \begin{split}
\Pi_{1}(p^{2})&=  g_{0}^{2}f^{2}+2 g_{0}^{2}p^{2}\left[ \sum_{i=1}^{N_a} \frac{f_{a_i}^{2}}{(p^{2}-m_{a_i}^{2})}-\sum_{j=1}^{N_\rho}\frac{f_{\rho^j}^{2}}{(p^{2}-m_{\rho^j}^{2})}\right],
 \\
\Pi_{0}(p^{2})&=  -p^{2}+g_{0}^{2}p^{2}\sum_{j=1}^{N_\rho} \frac{f_{\rho^j}^{2}}{(p^{2}-m_{\rho^j}^{2})},\ \ \ \ \ \ \Pi_0^X(p^2) = - p^2\,.
\end{split}
\label{PiFFSS}
\ee
The physical SM gauge couplings are modified by the contribution of the resonances and given by:
\be
g^{2}  =  -\frac{g_{0}^{2}}{\Pi_{0}^{\prime}(0)}= g_0^2 \Big(1+\sum_{j=1}^{N_\rho} \frac{g_0^2}{g_{\rho^j}^2} \Big)^{-1}
\,,\ \ \ \ 
g^{\prime2} =  -\frac{g_{0}^{2}}{\Pi_{B}^{\prime}(0)}=g_0^{\prime 2} \Big(1+\sum_{j=1}^{N_\rho} \frac{g_0^{\prime 2}}{g_{\rho^j}^2} \Big)^{-1}\,,
\label{gphysical}
\ee
where ${}^\prime$ stands for a derivative with respect to $p^2$. It is straightforward to get from the above relations the form of the gauge contribution to $\gamma_g$
and $\beta_g$ to the Higgs potential:\footnote{We have inserted the IR cut-off $\mu_g\simeq m_W$ to regulate a logarithmic divergence appearing in $\beta_g$. 
This is a spurious divergence arising from the expansion (the full potential is manifestly IR-finite) and does not play an important role in what follows.
We have checked that our results do not sensitively depend on the choice of $\mu_g$\label{f1}.} 
\be
\begin{split}
\gamma_g & =  -\frac{3}{8(4\pi)^{2}}\int_{0}^{\infty}dp_E^{2}\, p_E^{2}\left(\frac{3}{\Pi_{0}}+\frac{c_X^2}{\Pi_{B}}\right)\Pi_{1},\\
\beta_g & =  -\frac{3}{64(4\pi)^{2}}\int_{\mu_g^{2}}^{\infty}dp_E^{2}\, p_E^{2}\left(\frac{2}{\Pi_{0}^{2}}+\left(\frac{1}{\Pi_{0}}+\frac{c_X^2}{\Pi_{B}}\right)^{2}\right)\Pi_{1}^{2}.
\end{split}
\label{betagammaDef}
\ee
For large Euclidean momenta, the form factors $\Pi_{0}\propto \Pi_{0}^{B}\propto p_E^2$, while $\Pi_1\propto p_E^0$, indicating that 
all higher terms in the $s_h$ expansion are UV finite. On the other hand, $\gamma_g$ and $\beta_g$  are respectively quadratically and 
logarithmically divergent in the UV, in general. Their UV properties are fixed by the single form factor $\Pi_1$.
Without imposing any condition, the form factor $\Pi_{1}$ goes to
a constant at high energy and the potential diverges quadratically.
However, the form-factor $\Pi_{1}(p^{2})$ is an order parameter of
the spontaneous symmetry breaking (being proportional to the difference of the form factors of gauge fields along the unbroken and broken generators 
\cite{Agashe:2004rs}), so for energies much higher than
the symmetry breaking scale $f$, it should go to zero, assuring that
the potential will diverge only logarithmically. Imposing this condition, we
obtain the first Weinberg sum rule \cite{Weinberg:1967kj}:\begin{equation}
\lim_{p_E^{2}\rightarrow\infty} g_0^{-2} \Pi_{1}(-p_E^{2})=f^{2}+2 \sum_{i=1}^{N_a} f_{a_i}^{2} -2 \sum_{j=1}^{N_\rho} f_{\rho^j}^{2} \equiv0\,. \ \ \ ({\rm I})
\label{SRI} 
\end{equation}
Demanding that $\Pi_{1}$ goes to zero faster than $p_E^{2}$
(finite potential) for large Euclidean momenta
gives the second Weinberg sum rule:\footnote{The sum rules (\ref{SRI}) and (\ref{SRII}) are also valid for the general case $f_\text{mix} \neq 0$ when $N_\rho =2$.}
\begin{equation}
\lim_{p_E^{2}\rightarrow\infty}g_0^{-2} p_E^{2}\Pi_{1}(-p_E^{2}) = 2 \sum_{i=1}^{N_a} f_{a_i}^{2} m_{a_i}^2 - 2 \sum_{j=1}^{N_\rho} f_{\rho^j}^{2} m_{\rho^j}^2
\equiv0. \ \ \ ({\rm II})
\label{SRII}
\end{equation}
Notice that the first sum rule requires the presence of at least one vector resonance $\rho_\mu$, while the second sum rule requires at least one axial resonance $a_\mu$.
There is a qualitative difference between the Weinberg sum rules (I) and (II). While the former must be unavoidably imposed (at high energies the global symmetry
is by assumption restored), the latter can be relaxed, leaving a mild logarithmic UV-sensitivity of the Higgs potential.\footnote{The second sum rule was originally derived by assuming that the broken and unbroken currents behave as free fields in the UV \cite{Weinberg:1967kj}. This assumption holds for asymptotically-free gauge theories but can break down if, say, the UV theory is a strongly interacting CFT. In particular, it has been pointed out in \cite{Orgogozo:2011kq},  where an approach similar to ours has been advocated in Higgsless models, that the second Weinberg sum rule does not hold in Conformal Technicolor.}
From eqs.\eqref{Spara} and \eqref{PiFFSS}, we get the tree-level contribution to the $S$-parameter:
\be
\Delta S \simeq 8 \pi s_h^2 \left( \sum_{j=1}^{N_\rho} \frac{f_{\rho^j}^{2}}{m_{\rho^j}^2}- \sum_{i=1}^{N_a}\frac{ f_{a_i}^{2}}{ m_{a_i}^2} \right) \,,
\label{SparaGauge}
\ee
where we have approximated $g_0\simeq g$ for simplicity.

The explicit form of $\gamma_g$ and $\beta_g$ is readily computed for $N_\rho=N_a=1$. Setting for simplicity $g^\prime = 0$, $a_\rho=1$ and expanding at leading order in $(g/g_\rho)^2$ (and in $\mu_g(=m_W)/m_\rho$ in $\beta_g$), 
we get
\be
\gamma_g   \simeq   - \frac{9 f^2 g^2 m_\rho^2  \log 2}{64\pi^2}  \,,  \ \ \ \ \ 
\beta_g  \simeq    \frac{9f^4g^4}{1024\pi^2}\Big(5+\log \frac{ m_W^2}{32m_{\rho}^2}\Big) \,.
\label{gammabetaLogDiv}
\ee
For $N_\rho=N_a=1$, when both eqs.(\ref{SRI}) and (\ref{SRII}) are imposed, $\Delta S$ can be rewritten as
 \be
 \Delta S =8 \pi s_h^2\frac{f^2}{m_\rho^2} \left(1 - \frac{f^2}{4 f_\rho^2} \right) 
  \label{SIplusII} 
 \ee
and, as eq.\eqref{SRI} imposes $f_\rho > f/\sqrt{2}$, it is manifestly positive definite. 
As expected, for $s_h=1$, eq.(\ref{SIplusII}) agrees with the vector dominance estimate in technicolor theories derived in \cite{Peskin:1991sw}.  In holographic 5D models, $\Delta S$ is positive as well. For $N_\rho$ or $N_a> 1$, on the other hand, $\Delta S$ can in principle have any sign.
Since as far as we know there is no general proof about the positivity of $\Delta S$ (neither in Higgsless Technicolor theories nor in Composite Higgs Models)
we will also consider in the following one model (with $N_\rho=1$, $N_a=2$) in the  ``exotic" region where $\Delta S$ can be negative.  
The UV uncalculable contribution to $\Delta S$ is easily estimated by using NDA:
\be
\Delta S^{(NDA)} \sim \frac{1}{\pi}s_h^2\,.
\label{Snda}
\ee
As expected, this is the value one gets from eq.(\ref{SparaGauge}) (modulo accidental cancellations or enhancements), when the vector and axial couplings approach $4\pi$.

A possible constrain on the form factor $\Pi_1$ comes from the results of \cite{Witten:1983ut}.  A straightforward generalization of the proof given there implies  
that any composite Higgs model, UV-completed by vector-like gauge theories, cannot give rise to EWSB  without additional contributions to the Higgs potential 
(such as those given by fermion resonances).
In other words, for $s_h<1$, $\gamma_g$ in eq.(\ref{betagammaDef})  should be negative definite. This condition (always satisfied in 5D models) 
is automatically satisfied when both (I) and (II) hold for $N_\rho=N_a=1$ (see eq.(\ref{gammabetaLogDiv})).\footnote{On the contrary, if one imposes only the sum rule (I), even for $N_\rho=N_a=1$, $\gamma_g$ (and $\Delta S$) can have any sign.}
On the other hand, when $N_\rho$ or $N_a> 1$, $\gamma_g$ can be positive and induce EWSB by itself (although these regions are never found in our numerical scans). 

\subsubsection{Left-Right Asymmetric Case}
\label{Sec:LRasymmCase}

Let us study in this section what are the consequences of having a LR asymmetric model. We consider the simplest example, with $N_{\rho_L}=N_{\rho_R}=1$, which already shows all the important  aspects. 
From eq.\eqref{eq:LGeneralGaugeResonances} and eq.\eqref{eq:GaugeFormFactorHiggsDep} we get:
\be
\begin{split}
\Pi_0(p^2) &= -p^2 + \frac{g_0^2 f_{\rho_L}^2 p^2}{2(p^2 - m_{\rho_L^2})} + \frac{g_0^2 f_{\rho_R}^2 p^2}{2(p^2 - m_{\rho_R^2})},  \\
\Pi_B(p^2) &= -p^2 + \frac{g_0^{\prime 2} f_{\rho_L}^2 p^2}{2(p^2 - m_{\rho_L^2})} + \frac{g_0^{\prime 2} f_{\rho_R}^2 p^2}{2(p^2 - m_{\rho_R^2})},  \\
\Pi_1(p^2) &= g_0^2 \left(f^2  - \frac{ f_{\rho_L}^2 p^2}{p^2 - m_{\rho_L^2}}  - \frac{ f_{\rho_R}^2 p^2}{p^2 - m_{\rho_R^2}} \right), \\
\Pi_{LR}(p^2) &= \frac{g_0^2 f_{\rho_L}^2 p^2}{2(p^2 - m_{\rho_L^2})}  - \frac{g_0^2 f_{\rho_R}^2 p^2}{2(p^2 - m_{\rho_R^2})}.
\end{split}
\label{eq:GeneralGaugeFormFactors}
\ee
The form factor $\Pi_{LR}$ goes to a constant for large Euclidean momenta, and it induces a quadratic divergence in the Higgs potential. Since the functional dependence related to this form factor is $c_h$, see eq.(\ref{LgaugeLRAsym}), this divergence is present at any order in the expansion for small $s_h^2$. Similarly to $\Pi_1$, $\Pi_{LR}$ is an order parameter for the symmetry breaking and should hence go to zero at high energies. From the expression above we get
\be
	\lim_{p_E \rightarrow \infty} \Pi_{LR}(-p_E^2) = \frac{g_o^2}{2} \left( f_{\rho_L}^2 -  f_{\rho_R}^2\right) - \frac{g_0^2}{2 p_E^2}  \left( f_{\rho_L}^2  m_{\rho_L}^2-  f_{\rho_R}^2  m_{\rho_R}^2\right) + \mathcal{O}(p_E^{-4})\,.
\ee
Canceling the quadratic and logarithmic divergence  requires  $f_{\rho_L} = f_{\rho_R}$ and $m_{\rho_L} = m_{\rho_R}$, respectively, 
which is equivalent in this case to impose a complete LR symmetry, for which $\Pi_{LR}=0$ identically. Note that by adding more copies of vector resonances, however, one might be able to have a finite potential even without imposing a LR symmetry.

\subsection{Fermion Contribution}

The top quark effective Lagrangian up to quadratic order in the fermions and to any order in the Higgs field can be written, in momentum space, as
\begin{equation}
\bar{t}_{L}{\slashed p} \, \Pi_{t_{L}}t_{L}+\bar{t}_{R}{\slashed p}\, \Pi_{t_{R}}t_{R}-(\bar{t}_{L}\Pi_{t_{L}t_{R}} t_{R}+h.c.)\,,\end{equation}
resulting in the following contribution to the Higgs potential:
\begin{equation}
V_f(h)=-2N_{c}\int\frac{d^{4}p_{E}}{(2\pi)^{4}}\log\left(p_E^2\Pi_{t_{L}}(-p_{E}^{2})\Pi_{t_{R}}(-p_{E}^{2})+\left|\Pi_{t_{L}t_{R}}(-p_{E}^{2})\right|^{2}\right).
\label{eq:ColWeinFermPot}
\end{equation}
Integrating out the fermion resonances $S_i$ and $Q_j$, we get the following expression for the form factors:
\be
\Pi_{t_L} = \Pi_Q + s_h^2 \Pi_{1Q}\,, \ \ \ \ \ \Pi_{t_R} = \Pi_S + s_h^2 \Pi_{1S}\,, \ \ \ \ \ \Pi_{t_Lt_R} = i s_h c_h \Pi_{QS}\,,
\label{pipipi}
\ee
where
\bea
\Pi_{Q}(p^2)& = &  1-\sum_{j=1}^{N_Q} \frac{ |\epsilon_{qQ}^j|^2}{p^2- m_{jQ}^2}\,, \ \ \ \ 
\Pi_{1Q}(p^2)  =  -\frac{1}{2}  \bigg( \sum_{i=1}^{N_S}   \frac{ |\epsilon_{qS}^i|^2}{p^2- m_{iS}^2}- \sum_{j=1}^{N_Q}   \frac{ |\epsilon_{qQ}^j|^2}{p^2- m_{jQ}^2} \bigg) \,, \nn \\
\Pi_{S}(p^2)& = & 1-\sum_{i=1}^{N_S} \frac{ |\epsilon_{tS}^i|^2}{2(p^2- m_{iS}^2)}\,,  \ \ \ \ 
\Pi_{1S}(p^2)  =  \frac{1}2   \bigg( \sum_{i=1}^{N_S}   \frac{ |\epsilon_{tS}^i|^2}{p^2- m_{iS}^2}- \sum_{j=1}^{N_Q}   \frac{ |\epsilon_{tQ}^j|^2}{p^2- m_{jQ}^2} \bigg)
\,, \nn \\
\Pi_{QS}(p^2) & = &  \frac{1}{2} \bigg( \sum_{i=1}^{N_S}  \epsilon_{tS}^{i *}  \epsilon_{qS}^i   \frac{m_{iS}}{p^2- m_{iS}^2}- \sum_{j=1}^{N_Q}  \epsilon_{tQ}^{j *}  \epsilon_{qQ}^j   \frac{m_{jQ}}{p^2- m_{jQ}^2} \bigg) \,.
\eea
Similarly to the gauge case, for $s_h \ll 1$, we can expand $V_f$ up to quartic order:
\be
V_f(h) \simeq-\gamma_fs_h^2+\beta_f s_h^4,\end{equation}
with
\be
\begin{split}
\gamma_f & = \frac{2N_c}{(4\pi)^{2}}\int_{0}^{\infty}dp_E^{2}\, p_E^2  \left( \frac{\Pi_{1Q}}{\Pi_Q}+ \frac{\Pi_{1S}}{\Pi_S}+
 \frac{\Pi_{QS}^2}{p_E^2\Pi_Q \Pi_S} \right)\,,  \\
\beta_f & =  \frac{N_c}{(4\pi)^{2}}\int_{\mu_f^2}^{\infty}dp_E^{2}\, p_E^2\Bigg(  \left(\frac{\Pi_{QS}^2}{p_E^2\Pi_Q\Pi_S}+\frac{\Pi_{1Q}}{\Pi_Q}+\frac{\Pi_{1S}}{\Pi_S}\right)^2-\frac{2(p_E^2 \Pi_{1Q}\Pi_{1S}-\Pi_{QS}^2)}{p_E^2\Pi_Q\Pi_S} \Bigg)\,.
\end{split}
\label{gammafbetaf}
\ee
For large Euclidean momenta  $\Pi_{Q,S}\propto p_E^0$, $\Pi_{1Q,1S}\propto p_E^{-2}$, $\Pi_{QS}\propto p_E^{-2}$. It then follows that
the terms involving $\Pi_{QS}$ in eq.(\ref{gammafbetaf}) are all finite.  The factor $\mu_f$ is an IR-cutoff curing a spurious logarithmic divergence arising from the expansion of the potential.
We fix it to be around the top mass (see footnote \ref{f1}). 
All higher terms in the $s_h$ expansion are UV finite.  We can impose the fermion analogue of the Weinberg sum rules, demanding that 
the divergencies in $\gamma_f$ and $\beta_f$ above cancel. The cancellation of the logarithmic divergence in $\beta_f$ requires 
\be \begin{split}
\lim_{p_E^{2}\rightarrow\infty} (-2) p_E^2 \frac{\Pi_{1S}}{\Pi_S}&  =\;  \sum_{i=1}^{N_S}  |\epsilon_{tS}^{i}|^2-   \sum_{j=1}^{N_Q}  |\epsilon_{tQ}^{j}|^2= 0  \,, \\
\lim_{p_E^{2}\rightarrow\infty} 2p_E^2 \frac{\Pi_{1Q}}{\Pi_Q} & = \; \sum_{i=1}^{N_S}  |\epsilon_{qS}^{i}|^2-   \sum_{j=1}^{N_Q}  |\epsilon_{qQ}^{j}|^2= 0\,. 
\end{split} \qquad ({\rm III}) \label{SRIII} 
\ee
When eq.(\ref{SRIII}) is satisfied, the quadratic divergence in $\gamma_f$ is automatically cancelled. Imposing the cancellation of the logarithmic divergence in $\gamma_f$
requires the second condition
\be
\lim_{p_E^{2}\rightarrow\infty} 2 p_E^4 \Big(\frac{\Pi_{1S}}{\Pi_S}+\frac{\Pi_{1Q}}{\Pi_Q}\Big)  =  \sum_{i=1}^{N_S} m_{iS}^2 \Big( |\epsilon_{tS}^{i}|^2- |\epsilon_{qS}^{i}|^2\Big)
-  \sum_{j=1}^{N_Q} m_{jQ}^2 \Big( |\epsilon_{tQ}^{j}|^2-  |\epsilon_{qQ}^{j}|^2\Big)
 = 0  \ \ \ ({\rm IV}) \,. \label{SRIV}
\ee
It is useful to consider in some detail the case $N_Q=N_S=1$, taking all the mixing parameters to be real, for simplicity.
Assuming $m_S\neq m_Q$, a solution to eqs.(\ref{SRIII}) and (\ref{SRIV}) is
\be
\epsilon_{tS} = \epsilon_{tQ} = \epsilon_{qS} =- \epsilon_{qQ} \equiv \epsilon \,. 
\label{Sol1}
\ee
Other solutions with different sign choices can also be considered. We take $\epsilon_{qQ}$ of opposite sign with
respect to the other $\epsilon$'s so that  the top mass is maximized, see eq.(\ref{mtop}).\footnote{Otherwise, in order to get the correct top mass,
a larger degree of top compositeness is needed and the Higgs mass turns out to be always heavier than 125 GeV.} 
The coefficients $\gamma_f$ and $\beta_f$ are now easily computed in analytic form, but the resulting expressions are too lengthy  to be reported.
For illustration, we just show here their approximate form in the limit of small mixing, $\theta_{L,R}\ll 1$. At leading order we get\footnote{Contrary to the expansion in $g/g_\rho$ in the gauge contribution (\ref{gammabetaLogDiv}), that is always a sufficiently accurate approximation, the explicit forms (\ref{betagammaferExp}) are not always useful. When $t_L$ and/or $t_R$ significantly mix with the composite sector, different limits should be considered.} 
\be
\begin{split}
\gamma_f & =  \frac{N_c \epsilon^4}{32\pi^2}\frac{1-x^2+\Big(x^2+2x+2\Big)\log x^2}{x^2-1}\,, \ \ \ \ \ \ x = \frac{m_Q}{m_S}\,,
  \label{betagammaferExp} \\
\beta_f & =   \frac{N_c \epsilon^4}{32\pi^2}\frac{(1+x)\log x^2}{x-1}\,. 
\end{split}
\ee
Notice that the $\epsilon^4$ behaviour of $\gamma_f$ is an accident of the $N_Q=N_S=1$ case, the typical scaling being $\propto \epsilon^2$.

The generalized Weinberg sum rules (I-IV) must be satisfied by any composite Higgs model where a symmetry mechanism is at work to realize the MHP hypothesis.
They are clearly also satisfied in the notable case of five-dimensional theories, where locality in the extra dimension forbids any local Higgs potential to all orders in perturbation theory
 (thus implementing in full the MHP hypothesis). However, when one has to sum over an {\it infinite} set of fields, with increasing mass,  
such as in the 5D models,  the sum rules written as in (I-IV) are not very useful. It is more convenient to first sum over the infinite set of fields and then take
the limit of large euclidean momenta.\footnote{The higher-dimensional symmetries demand that one has to sum over the whole infinite tower of states, despite the 
limited regime of validity of the 5D effective theory.}  In doing that, one finds that the form factors such as $\Pi_1$, $\Pi_{1S}$, $\Pi_{1Q}$ and $\Pi_{QS}$ introduced before, all go to zero exponentially
for $p_E\rightarrow \infty$. For instance, in the simplest set-up of a 5D theory on a flat interval of length $L$, one gets $\Pi_1(p_E) \propto p_E /\sinh (2 L p_E)$ (see e.g.\cite{Serone:2009kf} for an introduction and further examples).


\section{Analysis of the Higgs Potential}
\label{Sec:analysis of potential}

The total Higgs potential up to ${\cal O}(s_h^4)$ is given by 
\be
V(h) = V_g(h)+V_f(h) = - \gamma s_h^2 + \beta s_h^4\,,
\label{Vhhh}
\ee
where we have denoted $\gamma = \gamma_g+\gamma_f$ and $\beta=\beta_g+\beta_f$.
The potential has three extrema: $s_h=0$ (no EWSB), $s_h=1$ (maximal EWSB) and 
\be
s_h^2 = \xi = \frac{\gamma}{2\beta}\,.
\label{min}
\ee
The one at $\xi=1$ should be discarded because it is outside the regime of validity of eq.(\ref{Vhhh}) (and leads anyway
to massless  SM fermions, $\Pi_{t_Lt_R}=0$ in eq.(\ref{pipipi})).
The extremum (\ref{min}) is a local minimum of the potential when $\gamma > 0$ and,  at the same time, $\gamma< 2 \beta$. Demanding a sufficiently small value of $\xi$, as suggested by the EWPT, requires to tune $\gamma < \beta$.
The Higgs mass at the non trivial minimum (\ref{min}) equals
\begin{equation}
m_{h}^{2}=\frac{8\beta}{f^{2}}\xi\left(1-\xi\right).
\label{eq:HiggsMass}
\ee
It is very useful to parametrically understand what are (if any) the generic relations among the Higgs mass and the masses of the vector and fermion resonances.
For simplicity, we first consider the set-up where $N_\rho=N_a=N_S=N_Q=1$ and set $g^\prime=0$.
Assuming PUVC,  we take $a_\rho$, as defined in eq.(\ref{arhoDef}), equal to one. 
When the Weinberg sum rules (I-II) in the gauge sector are imposed, the axial mass and decay constant are completely determined in terms of the vector mass $m_\rho$, which is the
only mass scale in the spin 1 sector. We choose to solve the sum rules (III-IV) as in eq.(\ref{Sol1}), so that the fermion sector is characterized
by three mass scales: the mixing parameter $\epsilon$ and the vector masses $m_S$ and $m_Q$.

From eq.(\ref{gammabetaLogDiv}),  we see that the following parametric expressions for $\gamma_g$ and $\beta_g$ approximately hold: 
\be
\gamma_g \sim -\frac{g^2 f^2 m_\rho^2}{16\pi^2}\,, \ \ \ \  \beta_g \sim \frac{g^4 f^4}{16\pi^2}\sim |\gamma_g| \Big(\frac{g}{g_\rho}\Big)^2\ll |\gamma_g| \,.
\label{gammabetag}
\ee
For $\xi\ll 1$, using eqs.(\ref{min}) and (\ref{eq:HiggsMass}) we have
\be
\frac{m_\rho^2}{m_H^2} \simeq \frac{4\pi^2}{g^2} \frac{|\gamma_g|}{\gamma}\,.
\label{mrhomh}
\ee
Given the bounds coming from the $S$ parameter, we parametrically require $\gamma\ll | \gamma_g|$, as well as $\gamma\ll \beta$.
This implies a fine-tuning at work, so that
$\gamma$ is small because the fermion and the gauge contribution compensate with each other, $\gamma_f \simeq - \gamma_g$.
As we will shortly see, $|\gamma_f | \simeq |\beta_f|$, while $\beta_g \sim  \gamma_g (g/g_\rho)^2$, implying that generally $\beta_g\ll \beta_f$  and can be neglected.
The fermion sector, with three different mass scales, is more involved. It is useful to parametrize it in terms of $\omega_L\equiv \tan\theta_L$ and $\omega_R\equiv \tan \theta_R$, introduced in eq.(\ref{tLtR}), and one mass scale. We can split the fermion parameter space in $3\times 3 = 9$ regions, $\omega_L\ll 1$ (elementary $t_L$), $\omega_L\simeq 1$ (semi-composite $t_L$), and $\omega_L\gg 1$ (fully composite $t_L$) and similarly for $\omega_R$. We always take $\omega_L$ and $\omega_R$ to scale in a similar fashion, so 
that $\omega_L \simeq \omega_R$ for $(\omega_L\ll1,\omega_R\ll1)$ and  $(\omega_L\gg 1,\omega_R\gg 1)$, and $\omega_L \omega_R\simeq 1$  for $(\omega_L\ll1,\omega_R\gg1)$ and  $(\omega_L\gg 1,\omega_R\ll 1)$.\footnote{It is important to keep in mind that physically there is actually no way to take the formal parametric limit $\omega_{L,R}\rightarrow 0$ or $\omega_{L,R}\rightarrow \infty$, 
because, at fixed top mass, some fermion resonance mass becomes infinitely massive.
The maximal value of a fermion mass  in the effective theory should be less than $\Lambda=4\pi f$, above which we should integrate out the heavy field. 
In light of that,  the actual allowed range for $\omega_{L,R}$ is
\be
\frac{1}{4\pi} \lesssim \omega_{L,R} \lesssim 4\pi \,.
\ee
\label{thetabounds}}
In each region we choose as mass scale the physical mass of the Lightest Fermion Resonance (LFR), denoted by $m_{L}$, as given by eq.(\ref{mRes12}). 
This is always either $m_0$ or $m_{7/6}$. We then define the  parameters
\be
m_{top}^2 \equiv k_t(\omega_L,\omega_R) m_L^2 \xi, \ \ \ \ \gamma_f \equiv \frac{N_c m_L^4}{16\pi^2} k_\gamma(\omega_L,\omega_R)\,, \ \ \ \ 
\beta_f \equiv \frac{N_c m_L^4}{16\pi^2} k_\beta(\omega_L,\omega_R)\,.
\ee
We report in table 1 the parametric dependence of $k_t$, $k_\gamma$ and $k_\beta$ on $\omega_L$ and $\omega_R$, as well as $m_L$, in each region.
Notice that the table is not symmetric under the exchange $\omega_L \leftrightarrow \omega_R$ and $m_Q \leftrightarrow m_S$, because of the presence of the bi-doublet
with $Y=7/6$, whose mass is $m_Q$, independently of $\omega_L$ and $\omega_R$. 
\begin{table}
\centering
\begin{tabular}[t]{|c|c|c|c|}\hline
\backslashbox{\vspace{0.1cm}$\omega_L$}{\vspace{-0.1cm}$\omega_R$} & $\ll 1$ & $\simeq 1$ & $\gg 1$\\[0.05in]\hline 
 $\ll 1$ &   $(m,\omega^4, \omega^4,\omega^4)$ &   $(m_0,1,1,1)$ &  $(m_0,1,1,1)$   \\[0.05in] \hline
 $\simeq 1$ &   $(m_{7/6},1,1,1)$   &   $(m,1,1,1)$  &   $(m,1,1,1)$   \\[0.05in] \hline
 $\gg 1$ & $(m_{7/6}, \omega_L^2, \omega_L^4,\omega_L^4)$  &  $(m_{7/6},\omega_L^2, \omega_L^4,\omega_L^4)$ &  $(m_{7/6},1, \omega_L^2,\omega_L^2)$  \\[0.05in]
\hline
\end{tabular}
\caption{Values of $m_L$, $k_t$, $k_\gamma$ and $k_\beta$ (in order) for the parametric limits of elementary, semi-composite and fully composite $t_L$, $t_R$. 
For simplicity, we have omitted the subscripts $0,7/6$ on $m$, and $L,R$ on $\omega$, when not necessary.}
\label{table:groups}
\end{table}
Given the mixing parameters and $\xi$, everything else is parametrically determined, namely $m_\rho$, $m_L$ and $m_H$. In particular, we have
\be
 m_\rho^2\simeq \frac{N_c m_{top}^2}{4m_W^2}\frac{k_\gamma}{k_t^2 \xi}m_{top}^2 \,, \ \ \ \ \
  m_H^2 \simeq \frac{g^2 N_c m_{top}^2}{8\pi^2 m_W^2}\frac{k_\beta}{k_t^2} m_{top}^2\,, \ \ \ \ m_L^2 = \frac{m_{top}^2}{k_t \xi}\,,
  \label{MhMrhoMfRelations1}
\ee
\be
 \frac{m_\rho^2}{m_H^2}\simeq \frac{2\pi^2}{g^2\xi}\frac{k_\gamma}{k_\beta}\,, \ \ \ \ \
  \frac{m_\rho^2}{m_L^2}\simeq \frac{N_c m_{top}^2}{4m_W^2}\frac{k_\gamma}{k_t}\,, \ \ \ \ 
    \frac{m_H^2}{m_L^2}\simeq \frac{g^2 N_c m_{top}^2}{8\pi^2 m_W^2} \frac{k_\beta}{k_t}\xi\,.
  \label{MhMrhoMfRelations2}
  \ee
In all regions, except ($\omega_L\ll1,\omega_R\ll 1)$ and  ($\omega_L\gg1,\omega_R\gg 1)$, $k_\beta/k_t^2\sim 1$ and the Higgs
is parametrically determined in terms of $m_{top}$ to be quite light (below the LEP bound, taking eq.(\ref{MhMrhoMfRelations1}) literally).\footnote{Needless to say,  the considerations above are quite schematic and are only valid parametrically. They are not  accurate enough for a more quantitative description.}
In all these regions, for reasonably natural values of $\xi$ (say, $\xi\simeq 1/10$), the LFR (singlet $S$ or exotic doublet $Q_7$, depending on the region) is always 
light, of order $1/\sqrt{\xi}$ times the top mass, or even too light, of order  $1/(\omega_L \sqrt{\xi})$, with $\omega_L\gg 1$.
For  ($\omega_L\gg 1,\omega_R\gg 1)$ the Higgs is heavier and yet the fermion resonance $Q_7$ is light. 
Finally, when  ($\omega_L\ll 1,\omega_R\ll 1)$, both the Higgs and the resonance masses (vector and fermion)  increase as $1/\omega^2$.
In all regions, $k_\beta = k_\gamma$, implying that $m_\rho/m_H$ is independent of the fermion sector and determined, at fixed $\xi$. 
Finally, since $k_{\beta,\gamma}\geq k_t$ in all regions,  we can conclude that {\it a light Higgs implies light fermion and vector resonances}. The latter are always heavier than the former, as can be seen from eq.(\ref{MhMrhoMfRelations2}) that, taken literally, predict vector masses roughly twice heavier than fermion masses.
The converse is not always true. In particular, for a strongly composite top, we can have light fermion resonances and an heavy Higgs.\footnote{
The direct link between $m_H$ and $m_\rho$ can be problematic for these minimal models with just one resonance.
In fact, a more detailed analysis reveals that $m_\rho$ is always below 2 TeV for a 125 GeV Higgs mass (see eqs.(\ref{mrhomh1model})-(\ref{mrhoBound}) and fig. \ref{fig:SpectrumMinModel} (b,d) in appendix E), leading generally to a too large $S$ parameter.}

Let us now consider the generalizations to models with multi vector and fermion resonances. 
When more spin 1 resonances are considered, a too large $S$ parameter can be circumvented by either 
some tuning between the axial and vector resonances or by an increase in the vector resonance mass.
For illustration, let us consider how the latter situation can be realized with 2 vectors and 1 axial resonance (see section \ref{Sec:results} and appendix E for a discussion of a model based on this gauge sector).
For simplicity, we take $f_{\rho^1}=f_{\rho^2}=f$ and $f_{mix}=0$. Imposing the sum rules (I) and (II) allows to determine $m_a$ and $f_a$ as a function of $f$ and of the two vector 
masses $m_{\rho^1}$ and $m_{\rho^2}$.
A simple calculation gives as leading expression in an expansion in $(g/g_\rho)^2$
\be
\gamma_g = -\frac{9f^2 g^2 m_{\rho}^2}{64\pi^2} \Big( (1+x^2) \log\Big(\frac 23 (1+x^2)\Big) -2 x^2 \log x \Big)\,,
\label{gammag2}
\ee
where $m_\rho=m_{\rho^1}$ and $x=m_{\rho^2}/m_{\rho^1}$. For an appropriate range in $x$, the coefficient multiplying $f^2 g^2 m_\rho^2$ in eq.(\ref{gammag2})
can be significantly smaller than the one in eq.(\ref{gammabetaLogDiv}). At fixed $\gamma_f$, this implies the possibility of increasing $m_\rho$ and hence decreasing the value of $\Delta S$ within the allowed range.
One can also check that in the case of 2 axials and 1 vector resonance,  $\Delta S$ can be made small
when one of the two axial resonances is quite light  (see eq.(\ref{Stwoaxials})). 

When more fermion resonances are involved, $N_Q$ and/or $N_S$ greater than one, the analysis is greatly complicated by the large number of parameters involved.
The main qualitative feature, as already mentioned, comes from $\gamma_f$ that for small mixing terms scales as $\epsilon^2$. 
This implies that parametrically $\gamma_f\gg \beta_f$, in tension with eq.(\ref{min}), that would favour regions where $\gamma\ll \beta$. 
On the other hand, a larger $\gamma_f$ is welcome, because it implies a larger $\gamma_g$ (in order to tune $\gamma_f+\gamma_g$ to be small) 
and hence spin one resonance masses heavy enough to keep $\Delta S$ under control, although at the expense of a higher fine-tuning.
We still expect the Higgs to be light when the LH and RH top are substantially composite $(\epsilon_i \gtrsim m_i)$ and at least 
one fermion resonance, barring accidental cancellations, to be light and parametrically related to the top mass by $m_L^2 \sim m_{top}^2/\xi$.
On the other hand, when we approach the region of an elementary top, both the Higgs mass and the fermion resonances related to the top become heavy.
We then expect that the implication light Higgs $\rightarrow$ light fermion resonances continue to apply. 
We will provide more accurate estimates of the relation among Higgs and fermion resonance masses in the next section, where we consider in more detail some specific classes of models.

\begin{figure}[!t]
\begin{center}
\hspace*{-0.65cm} 
\begin{minipage}{0.5\linewidth}
\begin{center}
	\hspace*{0cm} 
	\fbox{\footnotesize $N_Q=N_S=2$, $N_\rho = N_a = 1$ - $\xi = 0.1$} \\[0.05cm]
	\includegraphics[width=70mm]{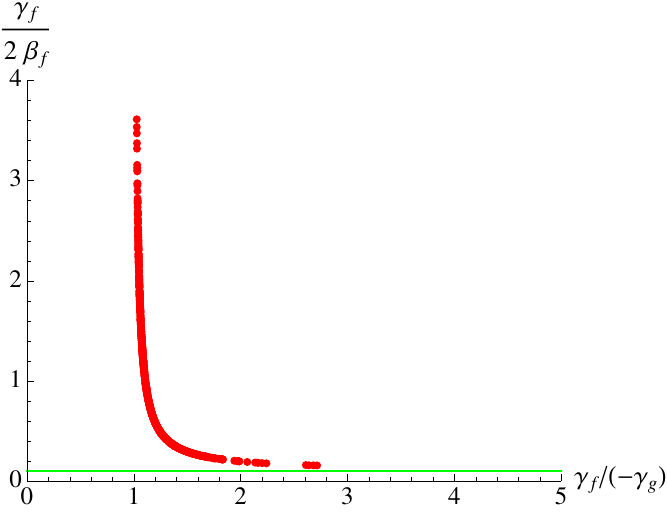}\\
	\mbox{\footnotesize (a)} \\
\end{center}
\end{minipage}
\hspace{0.25cm}
\begin{minipage}{0.5\linewidth}
\begin{center}
	\hspace*{0cm} 
	\fbox{\footnotesize 3-sites \cite{Panico:2011pw}:  $N_Q=N_S=N_\rho = 2$, $N_a = 1$ - $\xi \simeq 0.1$} \\[0.05cm]
	\includegraphics[width=70mm]{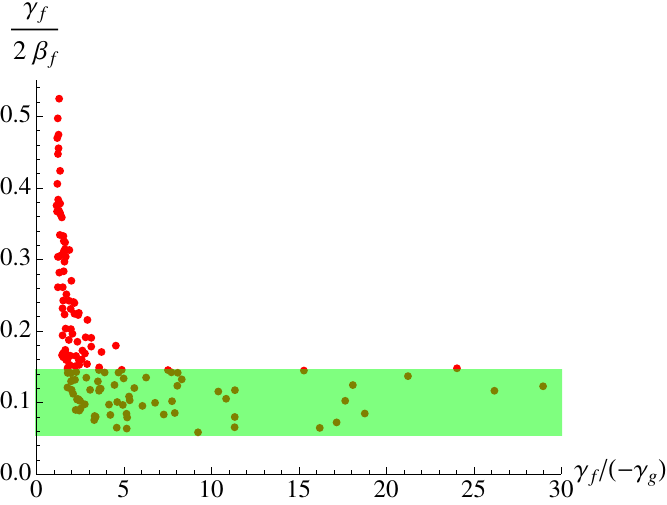}\\
	\mbox{\footnotesize (b)} \\
\end{center}
\end{minipage}
\end{center}
\vspace*{-0.2cm}
\caption{\label{fig:FermVSGauge} 
\small
Values of $\gamma_f/(-\gamma_g)$ versus $\gamma_f/(2\beta_f)$, that is the value of $\xi$ one would get by neglecting the gauge contribution to the Higgs potential.
 The points are obtained by a numerical scan, requiring $m_H \in [100,150]$ GeV.
(a) The range of the parameters is taken as follows: $m_{iQ},m_{iS} \in [0,8f]$, $\theta_{qQ},\theta_{tQ},\theta_{qS},\theta_{tS} \inÊ[0, 2\pi]$, $a_\rho \in [1/\sqrt{2}, 2]$. $\epsilon_t$, as defined in eq.(\ref{eq:WeinSolNs2Nq2}),  has been obtained by fixing $m_{top}$ while $m_\rho$ by fixing $\xi$. The green line represents $\xi=0.1$. In most of the points $\gamma_g\simeq -\gamma_f$ and  it is never possible to go in the region where $\gamma_f \gg - \gamma_g$.
(b) The range of the parameters is taken, in the notation of \cite{Panico:2011pw}, as follows: $g_*, \tilde g_* \in [0,8]$, $M_Q, M_S, m, \Delta \in [0,8f]$, $y_R/(\sqrt{2}y_L) \in [0.3, 0.6]$ and $y_L$ has been obtained fixing $m_{top}$, cutting for $\xi \in [0.05, 0.15]$.  The green band represents the actual values of $\xi\in [0.05, 0.15]$. In most of the points still $\gamma_g\simeq -\gamma_f$, but now there is a region where the gauge contribution is negligible.
}
\end{figure}
Non-minimal models with more vectors and fermions allow the possibility to tune $\xi \ll 1$ in a different way. Since with more vectors, as we have just seen, the estimate
(\ref{gammabetag}) does not necessarily hold, there is the possibility to have $\gamma_f \gg |\gamma_g|$ (and yet heavy enough vector resonances), so that 
the whole gauge contribution to the Higgs potential is sub-leading with respect to the fermion one. All the tuning is at work in the fermion sector to get $\gamma_f \simeq 2 \xi \beta_f \ll \beta_f$. This is possible, in the region of small mixing, if {\it both} the coefficients of the leading quadratic and next-to-leading quartic terms in the mixing in $\gamma_f$ are tuned to be small,  so that $\gamma_f \ll \beta_f$. In such regions a double tuning is at work, needed to get a small hierarchy between $v$ and $f$.
See fig.\ref{fig:FermVSGauge} for a comparison between the multi-fermion and multi-gauge model (e.g. the 3-sites theory of \cite{Panico:2011pw}), where this kind of tuning can occur, and the multi fermion (but minimal-gauge) model.

\section{Three Examples of Selected Models}
\label{Sec:results}

The framework introduced in the previous sections opens up a huge set of possibilities for model building. In fact, not only the number of spin 1 and spin 1/2 resonances to be introduced below the cutoff is free, but also the Weinberg sum rules have often physically different possible solutions. Studying in detail each of these models is well beyond the scope of this work and, as the simplest cases are already able to produce working models which pass the EWPT and display all the interesting aspects, we focus in the following on the case where $N_Q, N_S, N_\rho, N_a \leq 2$. A schematic presentation of the results for all the different cases will be presented in the appendix \ref{App:results}.
The simplest realization of our framework, that is the model with $N_\rho = N_a = N_Q = N_S = 1$ described in section \ref{Sec:analysis of potential} and in appendix \ref{App:results}, does not grossly pass the EWPT for $m_H \in [100, 150]$ GeV, because of a too large tree-level $S$ parameter, induced by (relatively) too light vector resonances, $m_\rho \lesssim 2$ TeV, as can be seen in fig.\ref{fig:SpectrumMinModel} $(b,d)$. This is a direct consequence of 
the first relation in eq.\eqref{MhMrhoMfRelations2} and of the fact that $k_\gamma \simeq k_\beta$ in this model. On the other hand, this model passes EWPT and the direct bound (\ref{DSbound}) for $m_H\simeq  320$ GeV (see appendix \ref{App:results}).

A straightforward way to circumvent this problem is to add more freedom either in the gauge sector or in the fermionic sector.
In the rest of this section we  consider three models. The first two, in our opinion, offer the best compromise between simplicity and viability, that is $N_\rho=2$, $N_a = N_Q = N_S = 1$ and $N_S=2$, $N_Q = N_\rho = N_a = 1$. The third one is actually the simplest possible model, with $N_\rho = N_Q  = 1$ and $N_a = N_S=0$.
Here the composite sector is assumed to contain a massless chiral bound state, identified with the RH top quark.
As we will see, this model is not realistic because it predicts a too light Higgs, but it is a counterexample to the statement that a light Higgs predicts light fermion resonances.

For the first two models presented here and those in the appendix \ref{App:results} we have performed a scan of the parameters imposing the generalized Weinberg sum rules, setting the ratio $v^2/f^2 = \xi = 0.1,0.2$ and requiring a light Higgs boson, $m_H \in [100, 150]$ GeV, or $m_H \simeq 320$ GeV, still allowed for a composite Higgs \cite{Azatov:2012bz}. 
In all our scans we set the top mass (roughly at the TeV scale) to be $m_{top}({\rm TeV}) \simeq 150$ GeV.
For all the points which satisfy the constraints above we have computed the new physics 1-loop fermion contribution to $\Delta S$ and $\Delta T$ and the deviation to $\Delta \delta g_{Z}(b_L)$ (for more details on the EWPT see appendix \ref{App:EWPT}). We then performed a combined $\chi^2$ analysis using the same fit already used in  \cite{safari}, based on the $\epsilon_i$ parameters \cite{Altarelli:1990zd}.\footnote{We have checked that our fit, restricted to the $S$ and $T$ parameters, reproduces with good accuracy the fit provided by the Particle Data Group for different values of the Higgs mass, figure 10.4 of \cite{Nakamura:2010zzi}.}

Direct search bounds on the fermion resonance masses should also be taken into account. 
The bi-doublets and singlets contain  the $t'$, $b'$ and the exotic fermion $\chi$ (the upper component of the bi-doublet with $Y=7/6$, with electric charge $Q=5/3$). The latter has no mixing with the SM fields and would be always lighter than any other fermion from the same bi-doublet. Therefore, the LFR would be either the lightest $\chi$ or $t'$ from the singlet.  The exotic $\chi$ has a 100\% decay branching into $t W^+$ which implies a stringent constrain from the same-sign dilepton and trilepton events with $b$ tags.
At present, the best bound for $\chi$ is the one from CMS $b'$ search $m_{b'} >  611$ GeV \cite{Collaboration:2012ye} which also applies to the $\chi$ search \cite{Wulzer}. 
The lightest $t'$ coming from an $SO(4)$  singlet, however, has three different decay channels: $t' \rightarrow b W^+$, $t' \rightarrow t Z$ and $t' \rightarrow t h$, where only the first one has a significant bound. The constraints on $t'$ largely depend on its decay branching ratio and are weak. For instance, if Br$(t' \rightarrow b W^+ < 35 \%$), we find that the CMS bound \cite{CMStpbWll} would imply $m_{t'} < 350$ GeV which is outside the range of the CMS search. Therefore, throughout the paper, we only include the direct search bound for $\chi$, imposing 
\be
m_{7/6} >  611\;\; {\rm  GeV}. 
\label{DSbound}
\ee
 In the appendix \ref{App:results} we also comment on the models where the generalized second Weinberg sum rules are relaxed and the Higgs potential is
logarithmically divergent.

\subsection{Two-vector model}

The models with $N_\rho=2$, $N_a = N_Q = N_S = 1$, are the simplest models passing the EWPT 
within our set-up. A similar model with $N_\rho=N_Q=N_S=1$ and $N_a=2$, considered in the appendix \ref{App:results}, also pass the EWPT,  but it is theoretically less motivated than the $N_\rho=2$, $N_a=1$ case. Indeed, while the gauge sector of the latter can be realized, for instance, in a deconstructed model (such as the 3-sites model of \cite{Panico:2011pw}), the former appears to be more exotic and unconventional. For this reason, we have decided to focus on the $N_\rho=2$, $N_a=1$ model in the following.
We assume invariance under $LR$ symmetry, so that  $\Pi_{LR} $ in the last row of eq.(\ref{LgaugeLRAsym}) vanishes.
In the fermion sector  we take eq.\eqref{Sol1} to satisfy the two sum rules (\ref{SRIII}) and (\ref{SRIV}),  and keep $m_S \neq m_Q$. This solution allows us to explore both the regions of parameter space where the LFR is a $t^\prime$ or $\chi$.

As explained in section \ref{Sec:analysis of potential}, adding a second vector resonance allows for a higher overall mass scale for the vectors, keeping $\gamma_g$ fixed, and alleviate the constraints coming from the $S$ parameter. This can be explicitly seen in the approximation $f_\text{mix} = 0$ and $f_{\rho_1} = f_{\rho_2} = f$, where we obtain the expression \eqref{gammag2} for $\gamma_g$, which is negative in the range $0.4\lesssim x \equiv m_{\rho^2} / m_{\rho^1} \lesssim 2.5$ and positive otherwise. It is therefore possible to tune $x\simeq 2.5$ (or $x\simeq 0.4$) and at the same time increase $m_{\rho^1}$ to keep $\gamma_g$ fixed. A posteriori, the numerical scan shows that $a_\text{mix} \equiv f_\text{mix}/f \lesssim 0.3$, so that the approximation used above is valid.

The fermion sector of this model is simple enough that it is not hard to write simple analytic formulas for the top and Higgs mass, that go beyond the parametric estimate given in section 4.
In particular, this allows us to explicitly check that a light Higgs requires light fermion resonances.  Let us first consider the elementary $t_{L,R}$ region, with $\omega_{L,R}<1$.
In this region, at leading order in $\omega_L\sim \omega_R$, we have
\be
m_{top}^2 \simeq \frac 12  m_S^2 \omega_R^2 (\omega_L+\sqrt{2}\omega_R)^2\xi=\frac 14 \epsilon^4 \frac{(m_Q+m_S)^2}{m_Q^2 m_S^2}\xi\,.
\label{mtopExp}
\ee
Using eq.(\ref{eq:HiggsMass}) for $\xi\ll1$ and expanding $\beta_f$ at leading order in $\omega_{L,R}$, we immediately get\footnote{A similar formula was already obtained 
in \cite{Wulzer}.}
\be
m_H^2 \simeq \frac{N_c}{\pi^2 f^2}\frac{m_S^4 \omega_R^4 (\omega_L+\sqrt{2}\omega_R)^2 \xi}{2\omega_R^2-\omega_L^2}\log\Big(\frac{2\omega_R^2}{\omega_L^2}\Big) = \frac{N_c}{\pi^2 f^2} \frac{m_Q^2 m_S^2}{m_Q^2-m_S^2}  \log\Big(\frac{m_Q^2}{m_S^2}\Big) m_{top}^2\,,
\label{mhExp1}
\ee
where in the last relation we have used eqs.(\ref{tLtR}) and (\ref{mtopExp}). It is straightforward to derive from eq.(\ref{mhExp1}) an upper bound for the LFR mass $m_L$:
\be
m_L \leq \frac{\pi f}{\sqrt{N_c}} \frac{m_H}{m_{top}}\,.
\label{mfbound1}
\ee
Let us now consider the region $\omega_L<1$, $\omega_R\simeq 1$ (elementary $t_L$, semi-composite $t_R$, often found in the numerical scan).
In this region the LFR is necessarily $t^\prime$, with $m_L=m_0\simeq \sqrt{2} m_S$. Expanding in $\omega_L<1$, we have
\begin{figure}[!t]
\begin{center}
\hspace*{-0.65cm} 
\begin{minipage}{0.5\linewidth}
\begin{center}
	\hspace*{0cm} 
	\fbox{\footnotesize $N_\rho=2$, $N_a=N_Q=N_S=1$ - $\xi = 0.1$} \\[0.05cm]
	\includegraphics[width=70mm]{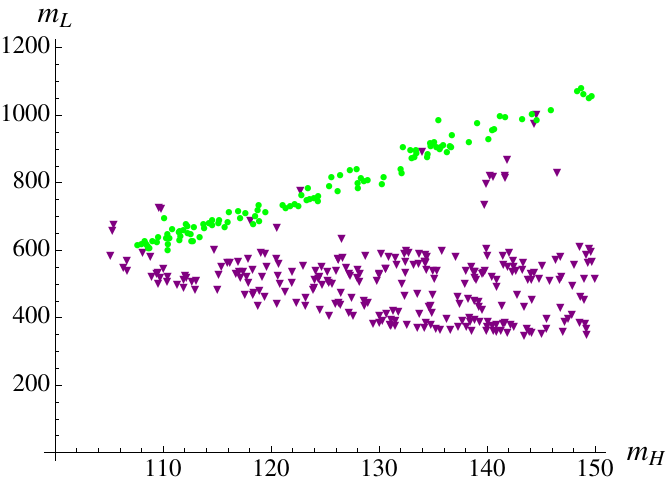}\\
	\mbox{\footnotesize (a)} \\
\end{center}
\end{minipage}
\hspace{0.25cm}
\begin{minipage}{0.5\linewidth}
\begin{center}
	\hspace*{0cm} 
	\fbox{\footnotesize  $N_\rho=2$, $N_a=N_Q=N_S=1$ - $\xi = 0.2$} \\[0.05cm]
	\includegraphics[width=70mm]{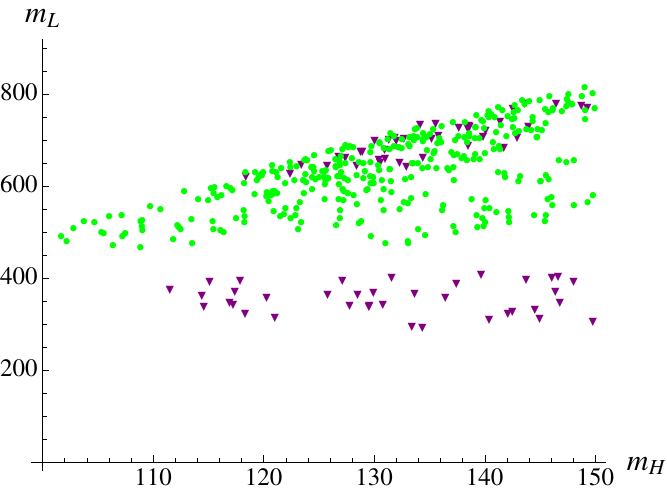}\\
	\mbox{\footnotesize (b)} \\
\end{center}
\end{minipage}
\end{center}
\vspace*{-0.2cm}
\caption{\label{fig:mLNr2Na1Ns1Nq1} 
\small
Mass of the LFR (in GeV), before EWSB, as a function of the Higgs mass  (in GeV). The green circles represent the singlet while the purple triangles represent the exotic doublet with $Y=7/6$. 
The masses $m_Q, m_{\rho_1}$ and $m_{\rho_2}$  are taken in the range $[0, 8 f]$, $a_{\rho_1}, a_{\rho_2} \in [1/2,2]$ and $a_\text{mix} \in [0,5]$; $\epsilon$ and $m_S$ have been obtained by fixing $m_{top}$ and $\xi$. EWPT and  the bound (\ref{DSbound}) have not been imposed.}
\end{figure}
\be
\begin{split}
m_{top}^2 & \simeq   \frac{m_L^2}{4}\xi \,, \\
m_H^2  & \simeq  \frac{N_c}{8\pi^2 f^2} m_L^2\Big(\log\xi^{-1}+8\log\Big(\frac{m_Q^2}{m_S^2}\Big) +\log 4-1\Big) m_{top}^2 \,,
\end{split}
\label{mtopExp2}
\ee
and gives the upper bound
\be
m_L \leq  \frac{2\sqrt{2}\pi f}{\sqrt{N_c}\sqrt{\log\xi^{-1}}} \frac{m_H}{m_{top}}\,.
\label{mfbound2}
\ee
\begin{figure}[!t]
\begin{center}
\hspace*{-1cm} 
\begin{minipage}{0.5\linewidth}
\begin{center}
	\hspace*{1cm} 
	\fbox{\footnotesize $N_\rho = 2$,  $N_a = N_Q = N_S = 1$ - $\xi = 0.1$} \\[0.05cm]
	\includegraphics[width=70mm]{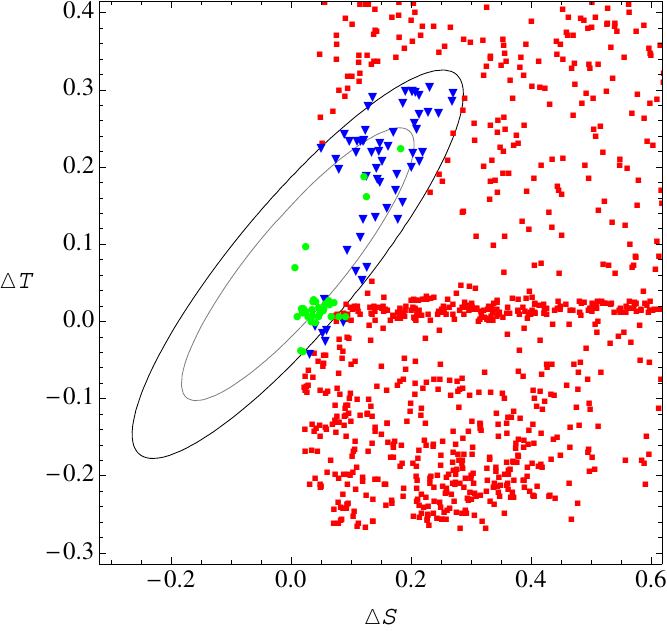}\\
	\mbox{\footnotesize (a)} \\
\end{center}
\end{minipage}
\hspace{0.25cm}
\begin{minipage}{0.5\linewidth}
\begin{center}
	\hspace*{1cm} 
	\fbox{\footnotesize $N_\rho = 2$,  $N_a = N_Q = N_S = 1$ - $\xi = 0.2$} \\[0.05cm]
	\includegraphics[width=70mm]{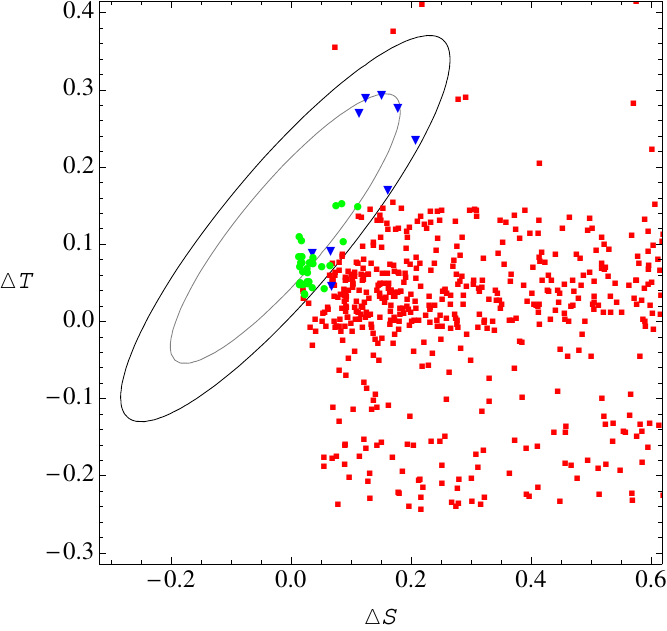}\\
	\mbox{\footnotesize (b)} \\
\end{center}
\end{minipage}
\end{center}
\vspace*{-0.2cm}
\caption{\label{fig:STNr2Na1Ns1Nq1} 
\small
$S$ and $T$ parameters for the points of the numerical scan with a light Higgs: $m_H \in [ 100, 150]$ GeV. The ellipses are the 99\% and 90\% C.L., for a mean value of $m_H = 125$ GeV. The green circles are the points which pass both EWPT and the bound (\ref{DSbound}), the blue triangles pass EWPT but are ruled out by the bound (\ref{DSbound}) and the red squares  don't pass EWPT. The range of the input parameters is as indicated in fig.\ref{fig:mLNr2Na1Ns1Nq1}.
}
\end{figure}
We performed the parameter scan for a light Higgs, both for $\xi=0.1$ and $\xi=0.2$. We show in fig.\ref{fig:mLNr2Na1Ns1Nq1} the relation between the LFR mass, $m_L$, and the Higgs mass, $m_H$, in the light Higgs region, obtained by a numerical scan over the parameter space.
In the case of $\xi=0.1$, approximately 4\% of the points produced by the scan are able to pass the EWPT, as indicated in fig.\ref{fig:STNr2Na1Ns1Nq1}, where we show the reduction of our fit to $(\Delta S, \Delta T)$ by marginalizing with respect to $\delta g_b$ (as done in figs.\ref{fig:STNr2Na1Ns1Nq1_HH}, \ref{fig:STNr1Na1Ns2Nq1} and \ref{fig:STNr1Na1Ns2Nq1_HH}).\footnote{Notice that the corrections to $\Delta S$ and $\Delta T$ due to the compositeness of the Higgs have been absorbed in the definition of the fit.  
In other words, $\Delta S_{plot}= \Delta S_{fit}-\Delta S_H(m_H)$ and $\Delta T_{plot}=\Delta T_{fit}-\Delta T_H(m_H)$, where $\Delta S_{fit}$ and $\Delta T_{fit}$ are defined in eqs.(\ref{DeltaSfit}) and (\ref{DeltaTfit}).}
The $\chi$ and $t^\prime$ fields are respectively  the lightest states (with $m_{7/6} \simeq 500$ GeV and $m_0 \simeq 800$ GeV) in the region of positive and sizable $\Delta T$ and of small $\Delta T$. The points which pass the EWPT are evenly distributed in these two regions but the bound $m_{7/6}>611$ GeV rules out most of the region
with a light $\chi$. As explained above, the vector masses can be arbitrarily heavy, so passing the constraints on the $S$ parameter is not an issue for this model. The points which pass the EWPT show a lightest vector resonance always above $\sim 1.5$ TeV. Also in this case, the tuning to get a successful EWSB is between the gauge and the fermion contribution to the Higgs potential, $\gamma_g$ and $\gamma_f$.  As expected, a smaller portion of points pass the EWPT for $\xi=0.2$ ($\sim 1\%$). The scan shows that in this case the EWPT prefer the points where $t^\prime$ is the LFR, with a mass $m_0 \sim 600$ GeV.
\begin{figure}[!t]
\begin{center}
\begin{minipage}{0.5\linewidth}
\begin{center}
	\hspace*{1cm} 
	\fbox{\footnotesize $N_\rho = 2$,  $N_a = N_Q = N_S = 1$ - Heavy Higgs - $\xi = 0.15$} \\[0.05cm]
	\includegraphics[width=70mm]{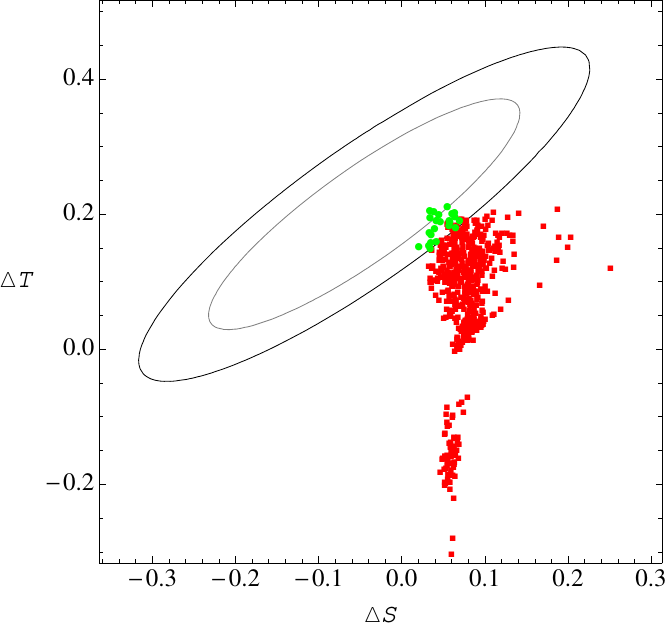}
\end{center}
\end{minipage}
\end{center}
\vspace*{-0.2cm}
\caption{\label{fig:STNr2Na1Ns1Nq1_HH} 
\small
$S$ and $T$ parameters  for the points of the numerical scan for a heavy Higgs ($m_H \in [300, 350]$ GeV) and $\xi = 0.15$. The ellipses are the 99\% and 90\% C.L., for a mean value of $m_H = 325$ GeV. The green circles are the points which pass the EWPT (and  the bound (\ref{DSbound})) while the red squares are the points which don't pass EWPT. 
The ranges of the input parameters is as indicated in fig.\ref{fig:mLNr2Na1Ns1Nq1}.
}
\end{figure}
The analysis of the bounds on the Higgs mass performed in \cite{Azatov:2012bz} shows that in the Composite Higgs Models we are considering (within the so called MCHM5 class) there is still an allowed region for $m_H \sim 320$ GeV, if $\xi \gtrsim 0.1$.
We then also performed a scan for this case, fixing $\xi = 0.15$ and cutting for $m_H \in [300,350]$ GeV. We find that the LFR can be as heavy as $2.2$ TeV and it can be both $t^\prime$  or $\chi$.  Interestingly enough, despite the heavy Higgs mass, the model passes the EWPT. Approximately $\sim 4 \%$ of the points pass the EWPT (see fig.\ref{fig:STNr2Na1Ns1Nq1_HH}), which prefer $t^\prime$ as LFR with $m_0 \sim 1.7$ TeV and the spin 1 resonances with masses above 3 TeV.

\subsection{Two-singlet model}

Adding a second composite fermion, singlet of $SO(4)$, is the minimal choice to go beyond the simplest setup in the fermionic sector. This is already enough to increase $\gamma_f$ and therefore to obtain heavier vector resonances and smaller tree-level contribution to the $S$~parameter.

The fermionic Lagrangian we start with is the one of eq.\eqref{eq:LGeneralFermions} with $N_Q=1$, $N_S = 2$. The most general solution to the first fermionic sum rule, eq.\eqref{SRIII}, is given in terms of two angles and two mixings:
\be \begin{split}
	 \epsilon_{qQ} = \epsilon_q\,, & \qquad	\vec \epsilon_{qS} = (\epsilon_q \cos \theta_{q},\; \epsilon_q\sin \theta_{q}), \\
	 \epsilon_{tQ} = \epsilon_t \,, & \qquad		\vec \epsilon_{tS} = (\epsilon_t \cos \theta_{t},\; \epsilon_t \sin \theta_{t}). \\
\end{split}\ee
There are various ways to satisfy the second fermionic sum rule eq.\eqref{SRIV}. One possibility is to solve for one of the remaining parameters, say $\epsilon_q$, in terms of the remaining ones. In this way, we get 
\be
\epsilon_q = \epsilon_t \sqrt{\frac{m_Q^2 - m_{1S}^2 \cos^2 \theta_t - m_{2S}^2 \sin^2 \theta_t  }{m_Q^2 - m_{1S}^2 \cos^2 \theta_q - m_{2S}^2 \sin^2 \theta_q } }.
\ee
Once we impose this relation, for small mixing $\epsilon_t$ we have $\gamma_f \propto \epsilon_t^2$ and $\beta_f \propto \epsilon_t^4$, in contrast to the 1-singlet case where $\gamma_f, \beta_f \propto \epsilon^4$. In particular, we get
\be
	\gamma_f \propto \epsilon_t^2 (m_Q^2 - m_{1S}^2) (m_Q^2 - m_{2S}^2) (m_{2S}^2 - m_{1S}^2) (\cos 2\theta_q - \cos 2\theta_t).
\ee
\begin{figure}[!t]
\begin{center}
\hspace*{-0.65cm} 
\begin{minipage}{0.5\linewidth}
\begin{center}
	\hspace*{0.0cm} 
	\fbox{\footnotesize $N_\rho = N_a = N_Q = 1$, $N_S = 2$ - $\xi = 0.1$} \\[0.05cm]
	\includegraphics[width=70mm]{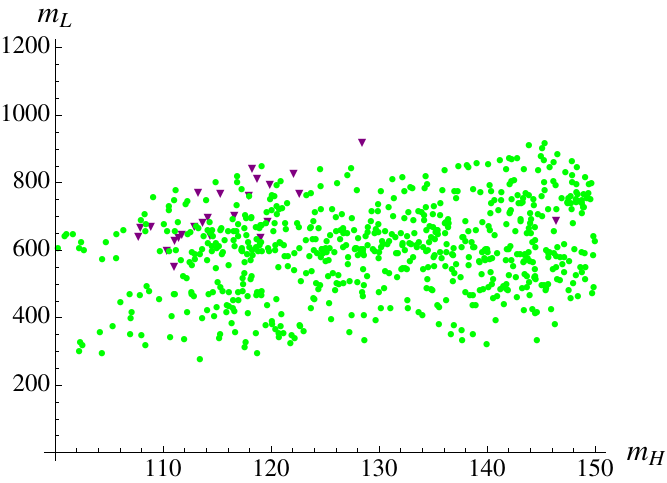}\\
	\mbox{\footnotesize (a)} \\
\end{center}
\end{minipage}
\hspace{0.25cm}
\begin{minipage}{0.5\linewidth}
\begin{center}
	\hspace*{0cm} 
	\fbox{\footnotesize $N_\rho = N_a = N_Q = 1$, $N_S = 2$ - $\xi = 0.2$} \\[0.05cm]
	\includegraphics[width=70mm]{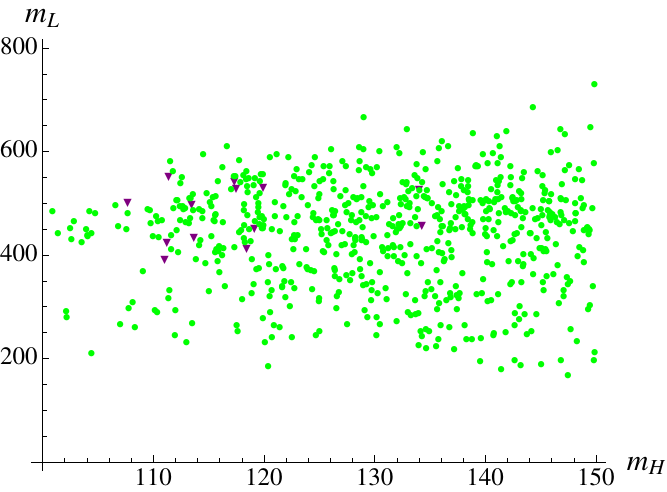}\\
	\mbox{\footnotesize (b)} \\
\end{center}
\end{minipage}
\end{center}
\vspace*{-0.2cm}
\caption{\label{fig:mLNr1Na1Ns2Nq1} 
\small
Mass of the LFR (in GeV), before EWSB, as a function of the Higgs mass  (in GeV). The green circles represent the singlet while the purple triangles represent the exotic doublet with $Y=7/6$. 
All the fermion masses are taken in the range $[0, 6 f]$, the angles $\theta_q, \theta_t \in [0,2 \pi]$ and $a_{\rho} \in [1/\sqrt{2},2]$. The mixing $\epsilon_t$ and the mass $m_\rho$ have been obtained by fixing $m_{top}$ and $\xi$ respectively. EWPT and  the bound (\ref{DSbound})  have not been imposed.}
\end{figure}
This implies that $\gamma_f$ can be enhanced with respect to the estimate in eq.\eqref{betagammaferExp}. But $\gamma_g$ cannot increase too much, leading otherwise to too heavy vector resonances, and hence the enhancement of $\gamma_f$ should be kept small. This is confirmed by the numerical scan where we get small deviations from the exact cancellation. In this simple, yet fundamental, observation lies the reason why this model, like all the ones with more fermionic resonances, is able to pass the EWPT.
\begin{figure}[!t]
\begin{center}
\hspace*{-0.65cm} 
\begin{minipage}{0.5\linewidth}
\begin{center}
	\hspace*{1cm} 
	\fbox{\footnotesize $N_\rho = N_a = N_Q = 1$, $N_S = 2$ - $\xi = 0.1$} \\[0.05cm]
	\includegraphics[width=70mm]{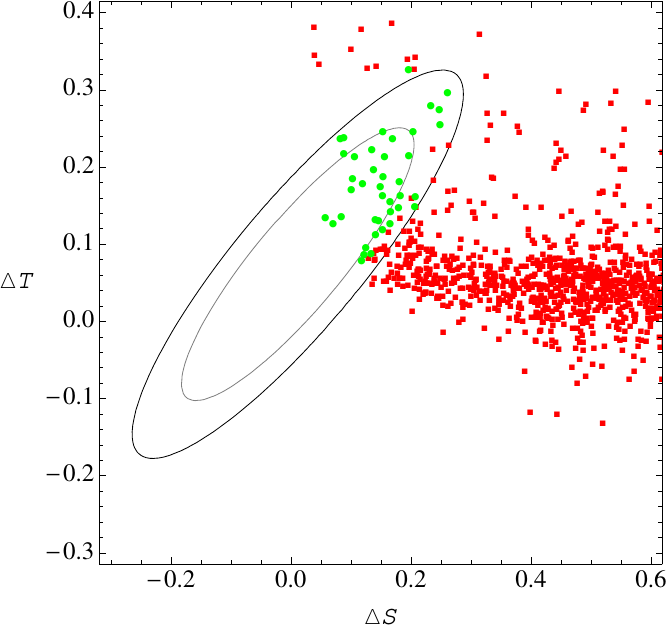}\\
	\mbox{\footnotesize (a)} \\
\end{center}
\end{minipage}
\hspace{0.25cm}
\begin{minipage}{0.5\linewidth}
\begin{center}
	\hspace*{1cm}
	\fbox{\footnotesize $N_\rho = N_a = N_Q = 1$, $N_S = 2$ - $\xi = 0.2$} \\[0.05cm]
	\includegraphics[width=70mm]{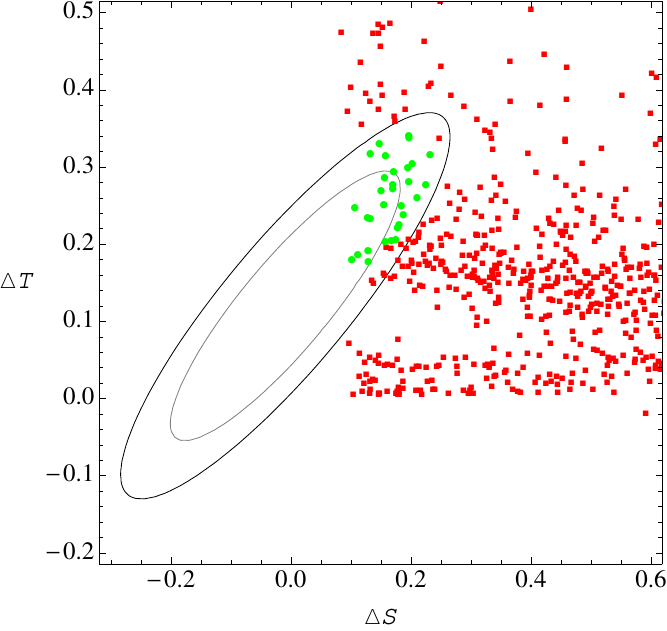}\\
	\mbox{\footnotesize (b)} \\
\end{center}
\end{minipage}
\end{center}
\vspace*{-0.2cm}
\caption{\label{fig:STNr1Na1Ns2Nq1} 
\small
$S$ and $T$ parameters for the points of the numerical scan with a light Higgs: $m_H \in [100, 150]$ GeV. The ellipses are the 99\% and 90\% C.L., for a mean value of $m_H = 125$ GeV. 
 The green circles are the points which pass EWPT (and  the bound (\ref{DSbound})) and the red squares  don't. The ranges of the input parameters is as indicated in fig.\ref{fig:mLNr1Na1Ns2Nq1}.
}
\end{figure}

Let us consider a specific region in parameter space selected by EWSB,  where $\epsilon_q \sim \epsilon_t \sim\epsilon$, $m_Q \sim m_{2S} \sim M$, $\theta_q \sim \pi$ and $\theta_t \sim 0$, with both $m_{1S}$ and $\epsilon$ much smaller than $M$.  In this region the coefficient of the $\epsilon_t^2$ term in $\gamma_f$ is suppressed.
We get
\be \begin{split}
	\gamma_f &\simeq\; \frac{N_c}{32 \pi^2} \epsilon^4 \left( \log \frac{M^2}{m_L^2} - 1 \right) =  \frac{N_c}{32 \pi^2} \epsilon^4 b_\gamma,\\
	\beta_f &\simeq\; \frac{N_c}{32 \pi^2} \epsilon^4  \left( \log \frac{M^2}{m_L^2} + \frac{\epsilon^4}{8 m_L^4}  \left( \log \frac{m_L^2}{\mu_f^2} - 1 \right) \right)= \frac{N_c}{32 \pi^2} \epsilon^4 b_\beta, \\
	m_{top}^2 &\simeq\; \frac{\xi}{4} \frac{ \epsilon^4}{m_L^2}\,,
\end{split}\ee
where $\mu_f$ is the IR regulator of the spurious IR divergence arising from $\beta_f$ (see eq.(\ref{gammafbetaf}) and footnote \ref{f1}) and $m_L$ denotes the mass of the LFR, that is clearly the singlet $S_1$ in this region: $m_L \simeq \sqrt{m_{S_1}^2 + \epsilon^2/2}$.
From these relations we obtain the estimate
\be
	m_L \simeq f \sqrt{\frac{\pi}{b_\beta}} \frac{m_H}{m_{top}}\,.
\ee
Since $b_\beta >  \log \frac{M^2}{m_L^2} \gtrsim 2$ for at least $M > 3 m_L$, the singlet has an upper bound of $m_L\lesssim 800$ GeV for $\xi = 0.1$. We therefore obtain that also in this case \emph{a light Higgs boson implies light fermionic resonances}.
For both $\xi = 0.1$ and 0.2 we find that the singlet is the LFR, with a mass in the range $\sim 300-800$ GeV, see fig.\ref{fig:mLNr1Na1Ns2Nq1}. Even though the bulk of the points show a vector mass in the same range as in the minimal model, there are nevertheless points with bigger values of $m_\rho$ so that the model can pass the EWPT, see fig.\ref{fig:STNr1Na1Ns2Nq1}. For $\xi = 0.1$, approximately $2.75 \%$ of the points pass the EWPT, with $m_\rho \gtrsim 2.5$ TeV and $m_L \in [400, 700]$ GeV. For $\xi = 0.2$, less than $1 \%$ of the points pass the EWPT, the typical value of $m_\rho$ and $m_L$ being analogous to the $\xi = 0.1$ case.
 
In the heavy Higgs case, that is for $m_H \in[300, 350]$ GeV and for $\xi = 0.15$, the LFR is still the first singlet, but with a mass range $400\, \text{GeV } \lesssim m_L \lesssim 2$ TeV. In this model $\sim 7 \%$ of the points pass the EWPT, see fig.\ref{fig:STNr1Na1Ns2Nq1_HH}, the preferred region being for $1 \,\text{TeV } \lesssim m_L \lesssim 1.5$ TeV and $m_\rho \gtrsim 3$ TeV. The fraction of points which pass is surprisingly high also because, for bigger Higgs masses, most of the points are naturally in a region of heavy $\rho$ (that is, small $\Delta S$), as can be understood from the estimate in \eqref{MhMrhoMfRelations2}.

\begin{figure}[t]
\begin{center}
\begin{minipage}{0.5\linewidth}
\begin{center}
	\hspace*{0.5cm}
	\fbox{\footnotesize $N_\rho = N_a = N_Q = 1$, $N_S = 2$ - Heavy Higgs - $\xi = 0.15$} \\[0.05cm]
	\includegraphics[width=70mm]{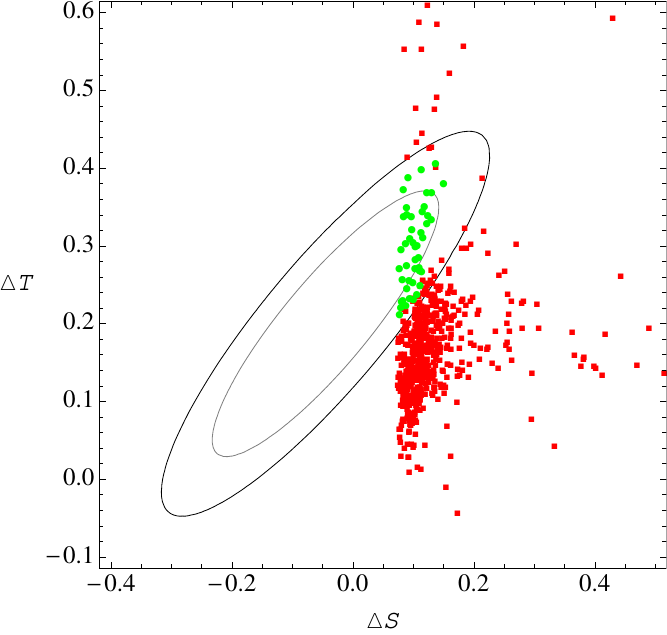}
\end{center}
\end{minipage}
\end{center}
\vspace*{-0.2cm}
\caption{\label{fig:STNr1Na1Ns2Nq1_HH} 
\small
$S$ and $T$ parameter for the points of the numerical scan, for a heavy Higgs ($m_H \in [300, 350]$ GeV) and $\xi = 0.15$. The ellipses are the 99\% and 90\% C.L., for a mean value of $m_H = 325$ GeV. The green circles are the points which pass the EWPT (and  the bound (\ref{DSbound})) while the red squares are the points which don't. The ranges of the input parameters is as indicated in fig.\ref{fig:mLNr1Na1Ns2Nq1}.
}
\end{figure}

\subsection{A Counter-Example: a Light Higgs and Heavy Resonances}

We consider in the following a model where the RH top quark is fully composite.
As already mentioned at the end of section 2.3, this model is built assuming that the $t_R$ is a chiral composite state in the singlet representation of $SO(4)$ and adding one composite fermion in the bidoublet representation, $Q$. No singlet fields $S$ are present, $N_S=0$. The leading fermion Lagrangian is\footnote{One might think that the Lagrangian (\ref{LfulltR}) can be obtained from eq.(\ref{eq:LGeneralFermions}) with $N_Q=N_S=1$, in the limit $\epsilon_{tS,tQ}\rightarrow \infty$, in which case the singlet becomes ultra-heavy and can be integrated out. This is however not the case, because the Weinberg sum rule (III) would imply $\epsilon_{qS,qQ}\rightarrow \infty$, and hence a ultra-heavy doublet as well.} 
 \be
{\cal L}_{f,0}  =  \bar q_L i \Dslash q_L + \bar t_R i \slashed \nabla t_R +\bar Q (i \slashed \nabla - m_{Q}) Q + \epsilon_{qS} \bar \xi_L U  P_S t_R + \epsilon_{qQ} \bar \xi_L U P_Q Q_R + h.c. \,.
\label{LfulltR}
\ee
\begin{figure}[!t]
\begin{center}
\begin{minipage}{0.5\linewidth}
\begin{center}
	\hspace*{0cm} 
	\fbox{\footnotesize $N_a = N_S = 0$, $N_\rho = N_Q = 1$ - $\xi = 0.1$} \\[0.05cm]
	\includegraphics[width=80mm]{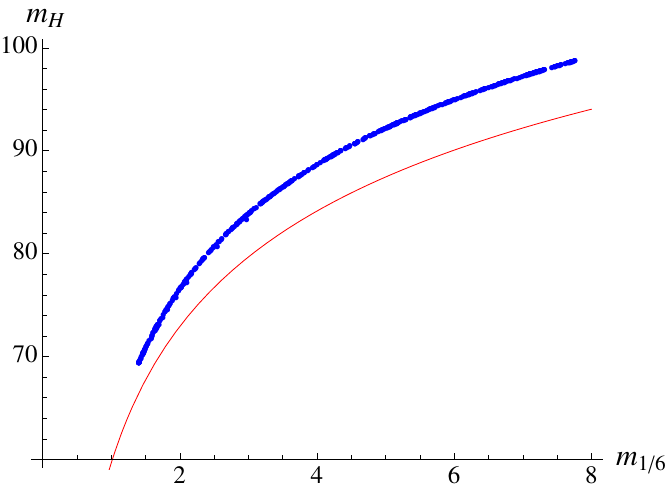}
\end{center}
\end{minipage}
\end{center}
\vspace*{-0.2cm}
\caption{\label{fig:m16comptR} 
\small
Higgs mass (in GeV) as a function of the $Y=1/6$ doublet mass (in TeV) in the composite $t_R$ model, for $\xi = 0.1$. The blue points are obtained by a numerical scan, while the thin red line represents the analytic estimate eq.\eqref{eq:mhm16comptR}. The two results are compatible, up to a $\sim 5\%$ error, due to the expansion for small $\xi$ in eq.\eqref{eq:mhm16comptR}. In the numerical scan, the mass $m_Q$ has been taken in the range $[0, 10f]$, while the mixing parameter $\epsilon$ has been obtained by fixing $m_{top}$.}

\end{figure}
The Weinberg sum rules (III) and (IV) obtained in section 3 do not apply in this case with $N_S=0$, but the expressions for the form factors and the 1-loop Higgs potential are particularly simple. Demanding the cancellation of the quadratic divergence in the fermion sector requires $|\epsilon_{qQ}| = |\epsilon_{qS}| \equiv \epsilon$. 
Demanding also the cancellation of the logarithmic divergence in $\gamma_f$ would require $\epsilon = 0$, which is not a viable possibility. We are therefore forced to keep the logarithmic divergence, which, as we explained in the appendix E, means that $\gamma$, and thus $\xi$, is not calculable.
We then proceed assuming a given value for $\xi$ and computing only $\beta$. Since $\beta_g \ll \beta_f$, we can completely neglect the gauge sector.\footnote{Since $\xi$ is not calculable, we can also relax the Weinberg sum rule (II) in the gauge sector, in which case we can assume that no axial resonance is present at all.\label{foottR}}
In this approximation, and at first order in $\xi$, we obtain the expression for the Higgs and top masses:
\be
	m_H^2 \simeq \frac{N_c}{8 \pi^2} \frac{\epsilon^4 m_Q^4}{f^2 m_{1/6}^4} \xi \left( \log \frac{m_{1/6}^2}{\mu_f^2} - 1\right), \qquad 
	m_{top}^2 \simeq \frac{\epsilon^2 m_Q^2}{2 m_{1/6}^2} \xi\,,
\ee
where $m_{1/6}^2 = m_Q^2 + \epsilon^2$ is the physical mass of the composite $Y=1/6$ doublet before EWSB.
From these expressions we get the estimate for the Higgs mass as
\be
	m_H \simeq \sqrt{ \frac{N_c}{2 \pi^2} } \frac{m_{top}^2}{v} \sqrt{ \log \frac{m_{1/6}^2}{\mu_f^2} - 1} \simeq 36 \sqrt{ \log \frac{m_{1/6}^2}{\mu_f^2} - 1} \text{  GeV}.
	\label{eq:mhm16comptR}
\ee
As can be noticed immediately, the Higgs is always too light ($m_H \simeq 90$ GeV for $m_{1/6} \simeq 6$ TeV). This conclusion has also been checked by a numerical scan of the model, which gives results in agreement with the estimate above, see fig.\ref{fig:m16comptR}.
In this model the LFR is $\chi$, with a mass (before EWSB) $m_{7/6} = m_Q$. It is interesting to notice that a light Higgs \emph{doesn't} imply a light fermionic resonance, at least for models with a chiral composite sector.

\section{Comparison with Previous Works}

In the previous sections we constructed a general framework for composite Higgs models, based only on the assumptions of $SO(5)/SO(4)$ symmetry breaking pattern and the MHP hypothesis. The aim of this section is to explicitly show how this general setup is able to reproduce the physics of two deconstructed composite Higgs models.
\subsection{Discrete Composite Higgs Model}
\label{Sec:3sitesPW}

Let us start with the two and three sites deconstructed models described in \cite{Panico:2011pw}. The two sites model is based on the coset $SO(5)_L \otimes SO(5)_R / SO(5)_V$, where the SM group is embedded in $SO(5)_L$. From this coset one has 10 Goldstones $\pi^A$, transforming in the adjoint of $SO(5)_V$. The $SO(4)$ subgroup of $SO(5)_R$ is gauged by introducing six gauge fields $\tilde \rho_\mu^a$, which become massive by eating the six Goldstone bosons $\pi^a$. The Lagrangian of this model is (in the notation of \cite{Panico:2011pw})
\be
	\Lag^{g,2-sites}_{PW} = \frac{\tilde{f}^2}{4} \Tr\left[ (D_\mu U)^t D^\mu U \right] - \frac{1}{4} \Tr \left[ \tilde{\rho}_{\mu\nu} \tilde{\rho}^{\mu\nu} \right] + \Lag^{gauge}_{SM},
\ee
where the Goldstone matrix is $U = \exp \left[ i \sqrt{2} \pi^A T^A / f_\pi \right]$, the covariant derivative is $D_\mu U = \partial_\mu U - i (g_0 W_\mu^\alpha T^\alpha_L + g_0^\prime B_\mu T^3_R)U + i \tilde{g}_* U \tilde{\rho}_\mu^a T^a$ and $ \Lag^{gauge}_{SM}$ is the usual gauge Lagrangian for the SM EW gauge bosons.
Going to the ``holographic"  gauge, where $\pi^{a}=0$, this model is described by the Lagrangian of eq.\eqref{eq:LGeneralGaugeResonances}, with one vector multiplet in the adjoint of $SO(4)$, no axial resonances, and fixing the parameters as (imposing invariance under $LR$ symmetry):
\be
	\text{\emph{2-sites:}}\qquad f = f_\pi = \tilde{f}, \qquad g_\rho = \tilde g_*, \qquad m_\rho^2 = \frac{1}{2} \tilde{g}_*^2 \tilde{f}^2, \qquad f_\rho^2 = \frac{\tilde{f}^2}{2}.
	\label{eq:PW2sGaugeDictionary}
\ee
One can check that only the first Weinberg sum rule is satisfied and the gauge contribution to the Higgs potential remains logarithmically divergent.

In order to get a finite potential, the authors of \cite{Panico:2011pw} add to the model another site, doubling the coset to $(SO(5)_L^1 \otimes SO(5)_R^1) / SO(5)_V^1 \times (SO(5)_L^2 \otimes SO(5)_R^2) / SO(5)_V^2$. From this symmetry breaking pattern 20 Goldstone bosons arise and can be parametrized by two $SO(5)$ matrices $U_1=U(\pi_1^A)$ and $U_2=U(\pi_2^A)$.
Sixteen NGB's are eaten by the gauging of $SO(4) \subset SO(5)_R^2$ by $\tilde{\rho}_\mu^a$ and of the diagonal combination of $SO(5)_R^1 \otimes SO(5)_L^2$ by the gauge field $\rho_\mu^A$:
\be \begin{split}
	D_\mu U_1 &= \partial_\mu U_1 - i (g_0 W_\mu^\alpha T^\alpha_L + g_0^\prime B_\mu T^3_R) U_1 + i g_* U_1 \rho_\mu^A T^A, \\
	D_\mu U_2 &= \partial_\mu U_2 - i g_* \rho_\mu^A T^A U_2 + i \tilde{g}_* U_2 \tilde{\rho}_\mu^a T^a.
\end{split} \ee
The Lagrangian of this model is
\be
	\Lag^{g,3-sites}_{PW} = \frac{\tilde{f}^2}{4} \Tr\left[ (D_\mu U_1)^t D^\mu U_1 \right] +  \frac{\tilde{f}^2}{4} \Tr\left[ (D_\mu U_2)^t D^\mu U_2 \right] - \frac{1}{4} \Tr \left[ \tilde{\rho}_{\mu\nu} \tilde{\rho}^{\mu\nu} \right] - \frac{1}{4} \Tr \left[ \rho_{\mu\nu} \rho^{\mu\nu} \right] + \Lag^{gauge}_{SM}.
\ee
In the holographic gauge where $\pi_2^A = \pi_1^a = 0$, one obtains the Lagrangian of \eqref{eq:LGeneralGaugeResonances} for two vectors and one axial resonances, with LR symmetry and the following parameters:
\be \;\; \text{\emph{3-sites:}}  \left\{  \quad \begin{split}
	 &f = f_\pi = \frac{\tilde{f}}{\sqrt{2}}, \qquad f_a = \frac{\tilde{f}}{2}, \qquad f_{\rho_1} = \frac{\tilde{f}}{\sqrt{2}}, \qquad f_{\rho_2} = 0, \\
	 & f_{mix} = \frac{\tilde{f}}{\sqrt{2}}, \qquad g_a = g_{\rho_1} = g_*, \qquad g_{\rho_2} = \tilde{g}_*, \qquad \Delta = - \frac{1}{2}.
\end{split} \right. \label{eq:PW3sGaugeDictionary}
\ee
Both Weinberg sum rules \eqref{SRI} and \eqref{SRII} are now satisfied. Notice that the term proportional to $f_{\rho_2}$ is absent in the deconstructed model because it would correspond to a non-local interaction in field space.
For completeness, we report in appendix \ref{Comparison} the detailed map for the fermion sector.

\subsection{Minimal 4D Composite Higgs}

Let us now write a similar dictionary for the deconstructed model described in \cite{DeCurtis:2011yx}. This model is based on a two-coset Lagragian: $SO(5)_L \otimes SO(5)_R / SO(5)_D$, described by the NGB matrix $\Omega_1 = \exp( i \sqrt{2} \tilde \pi^A T^a / f_1 )$, and another coset $SO(5)/SO(4)$, described by the matrix $\Omega_2 = \exp( i \sqrt{2} \bar \pi^{\hat{a}} T^{\hat{a}} / f_2 )$. The SM gauging is embedded in $SO(5)_L$ and to absorb the 10 exceeding NGB's, the diagonal subgroup of $SO(5)_R \otimes SO(5)$ is gauged by the field $\rho^A_\mu$. In the notation of \cite{DeCurtis:2011yx}, the Lagrangian is
\be
	\Lag = \frac{f_1^2}{4} \Tr \left| D_\mu \Omega_1\right|^2 + \frac{f_2^2}{2} (D_\mu \Phi_2)^t D^\mu \Phi_2  - \frac{1}{4 g_\rho^2} \rho_{\mu\nu}^A \rho^{A\mu\nu},
\ee
where $\Phi_2 = \Omega_2 \phi_0$ ($\phi_0 = (0,0,0,0,1)^t $) and 
\be
	D_\mu \Omega_1 = \partial_\mu \Omega_1 - i A_\mu \Omega_1 + i \Omega_1 \rho_\mu, \qquad
	D_\mu \Omega_2 = \partial_\mu \Omega_2 - i \rho_\mu \Omega_1.
\ee
Going again to the holographic gauge, where $\Omega_2 = 1$ and $\Omega_1 \equiv U = \exp( i \sqrt{2}  \tilde \pi^{\hat{a}} T^{\hat{a}} / f_1 )$, and redefining the NGB fields as $\tilde \pi^{\hat{a}} = f_1 / f \pi^{\hat{a}}$, one can write the Lagrangian as in eq.\eqref{eq:LGeneralGaugeResonances}:
\be
	\Lag = \frac{f_1^2 f_2^2}{4(f_1^2 + f_2^2)} \Tr \left[ d_\mu d^\mu \right] + \frac{f_1^2}{4} \Tr \left[ (g_\rho \rho_\mu - E_\mu)^2 \right] + \frac{f_1^1 + f_2^2}{4}\Tr \left[ (g_\rho a_\mu - \frac{f_1^2}{f_1^2+f_2^2} d_\mu)^2 \right] -\frac{1}{4} \rho_{\mu\nu}^2 - \frac{1}{4} a_{\mu\nu}^2,
\ee
from which we obtain the dictionary for $N_\rho = N_a = 1$:
\be \left\{  \quad \begin{split}
	 &f^2 = \frac{f_1^2 f_2^2}{f_1^2 + f_2^2}, \qquad f_\rho^2 = \frac{f_1^2}{2}, \qquad g_a = g_\rho, \\
	 & \Delta =\frac{f_1^2}{f_1^2+f_2^2},\qquad \frac{f_a^2}{\Delta^2} = \frac{f_1^2+f_2^2}{2}.
\end{split} \right.
\label{eq:RediGaugeDictionary}
\ee
It is straightforward to check that both Weinberg sum rules are satisfied with these parameters.
The map for the fermion sector is reported in appendix \ref{Comparison}.

\section{Discussion and Conclusions}

We have constructed a general class of composite Higgs models,  in the context of the minimal $SO(5)/SO(4)$ coset structure,
and introduced the MHP hypothesis that allows to predict the Higgs potential in terms of the parameters defining the model.
We have argued that any composite Higgs model based on the partial compositeness paradigm and leading to a calculable Higgs potential should satisfy the generalized
Weinberg sum rules (I-IV) and should be described by our Lagrangian (\ref{LTOT}), or straightforward generalizations thereof.
We emphasize that our approach allows to considerably enlarge the possibilities for model building and the parameter space for each model.
For instance, models where the fermion resonance representations do not form complete $SO(5)$ multiplets, also with $N_S\neq N_Q$, obviously allowed from effective field theory considerations, are easily constructed in our framework, while they are not easily obtained in deconstructed models.

We have explicitly shown the main properties of the simplest models one can construct within our framework. We argued that for non-chiral composite fermion sectors,  a light Higgs, around 125 GeV, implies the presence of at least one light, often sub-TeV, fermion resonance of charge 5/3 or 2/3, independently of EWPT considerations.  When the latter are taken into account, on the other hand, one realizes that these fermion resonances can play an important role in determining the viability of the model, given mainly by their sizable contribution  to the $T$ parameter. Models where the LFR has charge 5/3 are significantly constrained by the direct search bound
(\ref{DSbound}) , see e.g. fig.\ref{fig:STNr2Na1Ns1Nq1}(a). 
We have also shown that  models with a 320 GeV composite Higgs, yet not excluded by the current ATLAS and CMS bounds \cite{Azatov:2012bz}, can pass both EWPT and direct search bounds.  A heavy Higgs is actually welcome to increase the vector resonance mass and hence to decrease $\Delta S$.

There are various obvious ways in which our paper can be generalized. From a bottom-up perspective, fermion resonances in representations that are not only $SO(4)$ singlets or fundamentals can be considered, as well as less minimal cosets, such as $SO(6)/SO(5)$ \cite{Gripaios:2009pe} or others. From a top-down perspective, it would be very interesting to find new symmetry principles, other than collective breaking or 5D locality, that lead to a (at least partial) UV-completion of some of our models, realizing the MHP hypothesis.

We have decided to omit a phenomenological study of our models, because a careful analysis would require a paper on its own, given also the various different possibilities at hand. We just mention here that the current searches at LHC for heavy fermions start to  put significant bounds on models. In particular, the bound for the exotic $Q=5/3$ state $\chi$, $m_{7/6}>611$ GeV, already excludes sizable regions in the parameter space in some of our models. It is definitely important to study in more detail the actual bounds on $t^\prime$ coming from the $SO(4)$ singlet, less constrained by the current analyses, given that such fermion is often the lightest  composite particle.

The Higgs hunt at the LHC is probably coming to an end, with some evidence around 125 GeV, that hopefully will be confirmed or ruled out soon. 
If confirmed, the new era of understanding the properties of this particle will start.
We expect that future improvements in the heavy vector quark searches would help us to discriminate whether the Higgs is an elementary or a composite particle.

\vskip 10pt

{\bf Note added}: 

While this work was at the final stages of its completion, two papers appeared, refs.\cite{MPW} and \cite{RediTesi}, that have some overlap with our work. In particular, about the correlation light Higgs $\rightarrow$ light fermion resonances in Composite Higgs Models. 

\section*{Acknowledgments}

M.S. thanks Stefano Bertolini, Giuliano Panico, Enrico Trincherini and Andrea Wulzer, and J.S. thanks Tao Liu and Carlos Wagner for useful discussions.
We also thank the authors of \cite{MPW} for sharing with us the results of their work before the submission to the archive.  The work of J.S. is supported by the ERC Advanced Grant no. 267985 ``DaMESyFla''.


\appendix

\section{ Effect of $f_{\rm mix}$ for two vectors}
\label{Mixtwovect}

Let us consider a model with two copies of vector fields in the adjoint of $SO(4)$, $\rho^1_\mu$ and $\rho_\mu^2$, and $N_a$ axial resonances, assuming LR symmetry so that $\Pi_{LR} = 0$. This is a subcase of the generic Lagrangian of eq.\eqref{eq:LGeneralGaugeResonances} where the left and right parameters are identified.
Before integrating out the heavy vectors, we have to diagonalize the $\rho$ mass matrix:
\be
	m_\pm^2 = a + b \pm \sqrt{(a-b)^2 + 4 c^2}, \quad \tan \theta = \frac{b - a - \sqrt{(b-a)^2 + 4 c^2}}{2c},
\ee
where
\be
	a = g_{\rho^1}^2 \frac{f_{\rho^1}^2 + f^2_\text{mix}}{2}, \quad
	b = g_{\rho^2}^2 \frac{f_{\rho^2}^2 + f^2_\text{mix}}{2}, \quad
	c = g_{\rho^1} g_{\rho^2} \frac{f^2_\text{mix}}{2}.
\ee
The mass eigenstates, before EWSB, are given by the linear combinations
\be
\left\{ \begin{split}
	\rho^-_\mu =&\; c_\theta \rho^1_\mu + s_\theta \rho^2_\mu, \\
	\rho^+_\mu =&\; -s_\theta \rho^1_\mu + c_\theta \rho^2_\mu.
\end{split} \right.
\ee
In terms of these fields, the $\rho$ Lagrangian in momentum space and to quadratic order in the fields is
\be \begin{split}
	\Lag	\supset &\; - p^2 \frac{P_t^{\mu\nu}}{2} (\rho_\mu^{+a} \rho_\nu^{+a} + \rho_\mu^{-a} \rho_\nu^{-a}) + \frac{m_+^2}{2} \rho_\mu^{+a} \rho^{\mu +a} + \frac{m_-^2}{2} \rho_\mu^{-a} \rho^{\mu-a} \\
	&- f^2_{\rho^+} g_{\rho^+} \rho_\mu^{+a} E^{\mu a} - f^2_{\rho^-} g_{\rho^-} \rho_\mu^{-a} E^{\mu a} + \frac{f_{\rho^1}^2 + f_{\rho^2}^2}{2} E_\mu^a E^{\mu a},
\end{split} \ee
where
\be 
	f^2_{\rho^-} g_{\rho^-} \equiv c_\theta f_{\rho^1}^2 g_{\rho^1} + s_\theta f_{\rho^2}^2 g_{\rho^2} \quad \text{and} \quad
	f^2_{\rho^+} g_{\rho^+} \equiv -s_\theta f_{\rho^1}^2 g_{\rho^1} + c_\theta f_{\rho^2}^2 g_{\rho^2}.
\ee
Now we can integrate out these vectors and the two axial vectors, going to Euclidean momenta we obtain
\bea
\Pi_0(p) &=& -p^2 + g_0^2 f_{\rho^1}^2 +  g_0^2 f_{\rho^2}^2  + \frac{g_0^2 f_{\rho^-}^4 g_{\rho^-}^2 }{p^2 - m_{\rho^-}^2} + \frac{g_0^2 f_{\rho^+}^4 g_{\rho^+}^2 }{p^2 - m_{\rho^+}^2}, \nonumber \\
\Pi_B(p) &=& -p^2 + g_0^{\prime 2} f_{\rho^1}^2 +  g_0^{\prime 2} f_{\rho^2}^2  + \frac{g_0^{\prime 2} f_{\rho^-}^4 g_{\rho^-}^2 }{p^2 - m_{\rho^-}^2} +\frac{g_0^{\prime 2} f_{\rho^+}^4 g_{\rho^+}^2 }{p^2 - m_{\rho^+}^2},  \\
\Pi_1(p) &=& g_0^2 \left(f^2 - 2 f_{\rho^1}^2 - 2 f_{\rho^2}^2  - \frac{2 f_{\rho^+}^4 g_{\rho^+}^2 }{p^2 - m_{\rho^+}^2} 
- \frac{2 f_{\rho^-}^4 g_{\rho^-}^2 }{p^2 - m_{\rho^-}^2} 
+ \sum_{i=1}^{N_a} \frac{2 f_{a_i}^2 p^2 }{p^2 - m_{a_i}^2}
\right).\nonumber
\label{eq:GeneralGaugeFormFactorsmix}
\eea
One can check explicitly that in this case the Weinberg sum rules are the same as in eq.\eqref{SRI},\eqref{SRII}.
The SM gauge couplings are modified by the contribution of the resonances and given by:
\be \begin{split}
g^{2}  =&- \frac{g_{0}^{2}}{\Pi_{WW}^{\prime}(0)}=\frac{g_{0}^{2}}{1 + \frac{g_{0}^{2} f_{\rho^-}^4 g_{\rho^-}^2}{m_{\rho^-}^4} + \frac{g_{0}^{2} f_{\rho^+}^4 g_{\rho^+}^2}{m_{\rho^+}^4}  },\\ 
g^{\prime2} =& -  \frac{g_{0}^{2}}{\Pi_{BB}^{\prime}(0)}=\frac{g_{0}^{\prime 2}}{1 + \frac{g_{0}^{\prime 2} f_{\rho^-}^4 g_{\rho^-}^2}{m_{\rho^-}^4} + \frac{g_{0}^{2\prime } f_{\rho^+}^4 g_{\rho^+}^2}{m_{\rho^+}^4}  }.
\end{split} \ee

\section{Comments on the ElectroWeak Precision Tests}
\label{App:EWPT}

ElectroWeak Precision Tests put strong indirect constrains on new physics beyond the SM.  The most relevant parameters are $S$ and $T$. We have neglected the constraints coming from the $W$ and $Y$ parameters \cite{Shat}, since in our model they are parametrically suppressed with respect to $S$ by a factor $(g/g_\rho)^2$. 
A non-universal important bound comes from $\delta g_b$, the deviation of the $ \bar b_L Z b_L$ 
coupling from its SM value. 
Imposing a custodial symmetry and a proper mixing of $b_L$ with the fermion resonances allow to suppress the tree-level values of $T$ and $\delta g_b$.
More precisely, in the (oblique) basis where the contributions to $\delta g_b$  coming from vector resonance mixing (universal for any SM fermion) vanish, $T$ exactly vanishes.
The explicit expression of the tree-level contribution to $\delta g_b$ coming from fermion resonance mixing is reported in appendix \ref{GCD}. It  does not vanish, but 
it is always sufficiently suppressed to be safely neglected. At tree-level, then, the only dominant parameter is $S$, as given by eq.(\ref{SparaGauge}).
Since the custodial symmetries protecting $T$ and $\delta g_b$ at tree-level are explicitly broken in the full Lagrangian, mainly by Yukawa couplings in the
top sector, one-loop corrections to $T$ and $\delta g_b$ cannot be neglected \cite{Carena:2006bn,Barbieri:2007bh}. 

We define by $\Delta T$, $\Delta S$ and $\Delta \delta g_{b}$ the contribution given by new physics only, with the SM contribution subtracted:
\be
\Delta T= T-T_{SM}\,, \ \ \ \  \Delta S= S-S_{SM}, \ \ \ \ \ \Delta \delta g_{b}= \delta g_b-\delta g_{b,SM}\,,
\label{STgbDef}
\ee
where
\bea
g_{b,SM} & = & -\frac 12 +\frac 13 \sin^2\theta_W^2 \,, \ \ \ \  T_{SM}   \simeq   \frac{N_c r}{16\pi \sin^2\theta_W} \,, \ \ \ \ \  
S_{SM} =  \frac{N_c}{18\pi} \left(3+\log \left(\frac{m_{bottom}^2}{M_{top}^2}\right)\right)\,, \nn \\
\delta g_{b,SM} & = & \frac{\alpha_{em}}{16\pi \sin^2\theta_W} \frac{r(r^2-7r+6+(2+3r)\log r)}{(r-1)^2}\,, \ \ \ r\equiv \frac{M_{top}^2}{m_W^2}\,,
\eea
where $M_{top}$ is the pole top mass, $M_{top}= 173.1$ GeV \cite{:2009ec}, not to be confused with $m_{top}$ at the high scale, always taken around 150 GeV in our paper.

Due to the non-renormalizable nature of our theory, strictly speaking $\Delta S$, $\Delta T$ and $\Delta \delta g_b$ are not calculable. 
It is possible to disentangle an IR, calculable part, from the uncalculable part and use NDA and a spurionic analysis to estimate the size of the latter. We will not report some details of our estimate, that can be found, e.g., in \cite{Panico:2011pw}.

Let us start by estimating $\Delta T$.  The hypercharge coupling $g^\prime$ is the only custodial breaking parameter in the gauge sector. The NDA estimate for the uncalculable contribution
to $\Delta T$ coming exclusively from the gauge sector is 
\be
\Delta T_g^{(NDA)} \sim \frac{s_h^2}{8\pi \cos\theta_W^2}  \,.
\label{DeltaTg}
\ee
Eq.(\ref{DeltaTg}) also coincides with the NDA estimate for the contribution of the vector and axial resonances, because their couplings $g_\rho, g_a<4\pi$ and their masses $m_\rho \simeq g_\rho f$, $m_a \simeq g_a f$, precisely compensate in the contribution to $\Delta T$ to reproduce eq.(\ref{DeltaTg}).
Another contribution arises from the modified couplings of the Higgs with the SM gauge bosons \cite{Barbieri:2007bh}. This can be computed by introducing running $\Delta S$ and $\Delta T$ parameters and demanding  that they vanish at the scale $\Lambda$. In this way one gets
\be
\Delta T_H(\mu) = -\frac{3s_h^2 }{8\pi\cos^2\theta_W} \log \frac{\Lambda}{\mu}\,, \ \ \ \ 
\Delta S_H(\mu) = \frac{s_h^2 }{6\pi} \log \frac{\Lambda}{\mu}\,.
\label{STHiggsMod}
\ee
For $\Lambda\gg \mu\simeq m_H$, eq.(\ref{STHiggsMod}) captures the calculable  ``leading log" deviations to $\Delta S$ and $\Delta T$ due to a composite Higgs.\footnote{We have explicitly checked the reliability of eq.(\ref{STHiggsMod}) by computing the whole deviations to $\Delta S$ and $\Delta T$ due to the modified Higgs couplings, obtained from the full  SM one-loop Higgs+gauge boson contributions to the gauge vacuum polarizations amplitudes computed in dimensional regularization (see e.g. \cite{Aoki:1982ed}) and replacing the $1/\epsilon$ pole with $2\log \Lambda$. 
We have found that the ${\cal O}(1)$ deviations to eq.(\ref{STHiggsMod}) are always of the same order
of the uncalculable contributions (\ref{Snda}) and (\ref{DeltaTg}) and can thus be reabsorbed in a change of UV boundary conditions for the values of $\Delta S$ and $\Delta T$ at the cut-off scale.}
Finally we have the fermion contribution. The uncalculable fermion contribution is easily shown to be sub-leading, in the limit of small mixing $\epsilon^i$, and can be neglected. 
The calculable contribution to $T$ due to the fermion resonances is given by (see e.g. the appendix of \cite{safari} for some explicit expressions of fermion contributions to $T$)
\be
\Delta T_f  \sim \frac{N_c }{2\pi \sin^2 \theta_W g^2}\frac{\lambda^4 f^2 s_h^2}{m_L^2} 
\label{DeltaTf}
\,,
\ee
where $\lambda$ is the Yukawa coupling between the top and a fermion resonance, $\lambda \sim \epsilon/f$, and $m_f$ is its vector-like mass.
We  get
\be
\frac{\Delta T_f  }{\Delta T_g^{(NDA)} } \sim \frac{4N_c\cos^2\theta_W}{g^2 \sin^2\theta_W} \frac{\epsilon^4}{f^2 m_L^2} \sim  \frac{4N_c\cos^2\theta_W}{g^2 \sin^2\theta_W} \lambda_{top}^2 \,,
\ee
where in the last equality we have used eq.(\ref{mtop}). The calculable fermion contribution is hence the dominant contribution to $\Delta T$.
We have then included in our fit 
\be
\Delta T_{fit} = \Delta T_H(m_H)+\Delta T_f \,.
\label{DeltaTfit}
\ee

A similar analysis applies to $\Delta S$. The uncalculable gauge contribution, as well as the vector and axial one,  is given in eq.(\ref{Snda}), while the fermion one is negligible. The calculable fermion resonance contribution can be estimated as
\be
\Delta S_f \simeq \frac{N_c}{4\pi} \frac{\lambda^2 v^2}{m_f^2} \simeq \frac{N_c}{4\pi} s_h^2 \frac{\epsilon^2}{m_f^2}\,.
\label{DeltaSf}
\ee
For $\theta_{L,R}\sim {\cal O}(1)$ this is roughly of the same size of eq.(\ref{Snda}), but we have kept it in our fit, because it is calculable and in some region in parameter space the actual value of $\Delta S_f$ can be significantly larger than the estimate (\ref{DeltaSf}).
We have then included in our fit 
\be
\Delta S_{fit} = \Delta S_0+ \Delta S_H(m_H)+\Delta S_f \,.
\label{DeltaSfit}
\ee
where $\Delta S_0$ is the tree-level value (\ref{SparaGauge}).

Let us now consider $\Delta \delta g_b$. First of all, we have to distinguish between universal and non-universal gauge coupling deviations.
The universal calculable and uncalculable deviations can be rotated in $\Delta S$ and $\Delta T$ and can be shown to be of the same order as $\Delta T_g^{(NDA)}$ and $\Delta S^{(NDA)}$. The calculable contribution we have computed arises from loops where a SM $W$ is exchanged, $\delta g_b^W$, that 
can be estimated as
\be
\delta g_b^W \simeq  \frac{|\lambda|^4 v^2}{16\pi^2 m_L^2}\simeq \frac{|\epsilon|^4 s_h^2}{16\pi^2 f^2 m_L^2} \,.
\label{deltagbW}
\ee
In addition to that,  we also have a calculable contribution where a vector resonance is exchanged in the loop, and the usual uncalculable contribution. 
The latter is estimated by NDA. It arises when the spurions (\ref{Espurions}) are inserted in the fermion bilinears. 
There are several local operators one can construct. For example, one contributing to $\delta g_b$ is the following:
\be
\frac{c_g}{f^4(16\pi^2)^2}\Big(\bar q_L E_{qQ}\gamma^\mu E_{qQ}^\dagger  q_L\Big)\sum_{\alpha=1}^3  (\Sigma^t E_{qQ}^{\dagger,\alpha}) (E_{qQ}^{\alpha} D_\mu \Sigma) =-
\frac{c_g|\epsilon_{qQ}|^4s_h^2}{4(16\pi^2)^2 f^4}\frac{g}{\cos\theta_W}\bar q_L\slashed Z q_L
\ee
with $c_g$ an ${\cal O}(1)$ coefficient and $\alpha$ the $SU(2)_L^0$ index (see \cite{Panico:2011pw} for details), leading to
\be
\delta g_b^{(NDA)} \sim \frac{ |\epsilon_{qQ}|^4}{(16\pi^2)^2 f^4} s_h^2\,,
\label{deltagbNDA}
\ee
which is sub-leading with respect to eq.(\ref{deltagbW}). The one-loop deviations where a vector and a fermion resonance are exchanged in the loop are induced by the couplings
(\ref{LagInt11/2}). They are estimated to be
\be
\delta g_b^\rho \simeq  \frac{k^2 g_\rho^2|\epsilon|^2 |\lambda|^2 v^2}{16\pi^2 m_L^2 m_\rho^2}\simeq \frac{k^2 |\epsilon|^4 s_h^2}{16\pi^2 f^2 m_L^2} \,,
\label{deltagbrho}
\ee
where $k$ generically represents the   ${\cal O}(1)$ $k$ coefficients in eq.(\ref{LagInt11/2}). In general $\delta g_b^\rho\sim \delta g_b^W$ and both should be taken into account.
However, $\delta g_b^\rho$ depends on the couplings (\ref{LagInt11/2}) that are otherwise irrelevant in our analysis. For simplicity, we then assume that $k\lesssim 1$ so that $\delta g_b^W$ marginally dominates over $\delta g_b^\rho$. Under this assumption, we have included in our fit 
\be
(\Delta \delta g_b)_{fit} = \delta g_b^W \,,
\ee
neglecting the vector resonance contribution. 
We have not inserted in our EWPT fit the tree-level correction (\ref{deltagL}) to $\delta g_b$, because it depends on several other parameters (the down-type mixing $\epsilon^{(d)}$ and  the couplings $\tilde k_{ij}$) that do not play any other role in our analysis.
This omission is justified by noticing that $\delta g_b$ in eq.(\ref{deltagL}), in addition to the $s_h^2$ factor, is suppressed by the small mixing of the $d$-type, $\epsilon^{(d)}$. Neglecting the second term coming from the axial resonance (depending also on $\tilde k_{ij}$), we have checked for $N_S^{(u,d)}=N_Q^{(u,d)}=1$ that the tree-level correction (\ref{deltagL}) is typically 2-3 times smaller than $\delta g_b^W$. 

As we see, the calculability of $ \delta g_b$ is not on the same footing as that of $S$ and $T$. 
Nevertheless the effect of $\Delta \delta g_b$ in our fit is sub-dominant, the main effects coming from $\Delta S$ and $\Delta T$.

\section{Gauge Coupling Deviations}
\label{GCD}

In this appendix we report the tree-level deviations from their SM values of the top and bottom trilinear couplings to the SM gauge fields.
They are computed in the basis where we keep as light fields directly the fields $q$, $W$ and $B$ in the Lagrangian (``oblique" or ``holographic basis").  In this basis, the deviations are all proportional to the fermion mixing parameters $\epsilon$ introduced in eq.(\ref{eq:LGeneralFermions}), the universal effect induced by vector resonances 
being ``shifted" in $S$ and $T$.
We compute the deviation due to dimension six operators involving the Higgs field, ${\cal O}(h^2/f^2)$, neglecting higher derivative dimension 6 operators with no Higgs. 
The latter give a sub-dominant effect, suppressed by a factor $(g/g_\rho)^2$ with respect to the former. We can effectively set all ordinary derivatives for the gauge resonances to zero. 
Vector and axial resonances can contribute to the deviations by means of non-universal contributions induced by the couplings (\ref{LagInt11/2}).
For simplicity, we compute in the following the axial contribution only, the vector one being in general complicated by the mixing $f_{{\rm mix},ij}$.

As discussed in the main text, including the bottom sector requires the addition of other
fermion resonances and mixing. The total fermion Lagrangian can still be written in the form (\ref{fullLag}), provided we add an extra index $a$, taking values $a=u,d$, that distinguish the 
top and bottom sector. In other words, $Q_j\rightarrow Q_j^{(a)}$, $m_{jQ}\rightarrow m_{jQ}^{(a)}$,  $S_i\rightarrow S_i^{(a)}$, $m_{iS}\rightarrow m_{iS}^{(a)}$, 
 $\epsilon_{tS,tQ}^i \rightarrow \epsilon_{tS,tQ}^{i(a)}$,  $\epsilon_{qS,qQ}^j \rightarrow \epsilon_{qS,qQ}^{j(a)}$, and similarly for the couplings in eq.(\ref{LagInt11/2}).

Integrating out the axial resonances gives 
$a_\mu^i = \Delta_i /g_{a_i} d_\mu+\ldots $. Plugging back this relation in the fermion Lagrangian generates the operators
\be
\tilde k_{ij}^{\eta(a)} \bar S_i^{(a)} \gamma^\mu d_\mu P_\eta Q_j^{(a)} + h.c.
\label{ExtraTerm}
\ee
that contribute to $\delta g$. In eq.(\ref{ExtraTerm})
\be
\tilde k_{ij}^{\eta (a)} =\sum_{k=1}^{N_a}  k_{ikj}^{A(a),\eta} \Delta_k + k_{ij}^{d(a),\eta}
\ee
are the effective coupling constants of the above operators, functions of the couplings appearing in eq.(\ref{LagInt11/2}).
Integrating out the fermion resonances gives, upon rescaling the SM fields to get canonical kinetic terms, the following results, at leading order for $s_h\ll 1$:
\bea
\delta g_Z(b_L) & = & \frac{s_h^2}{8Z_{q_L}} \bigg(\sum_{i=1}^{N_S^{(d)}} \frac{|\epsilon_{qS}^{i(d)}|^2}{m_{iS}^{(d)2}}+\sum_{j=1}^{N_Q^{(d)}} \frac{|\epsilon_{qQ}^{j(d)}|^2}{m_{jQ}^{(d)2}} 
-2\sum_{i=1}^{N_S^{(d)}}\sum_{j=1}^{N_Q^{(d)}} \frac{\epsilon_{qS}^{i(d)}\epsilon_{qQ}^{j(d)*}}{m_{iS}^{(d)}m_{jQ}^{(d)}} \tilde k_{ij}^{L(d)}+h.c.
 \bigg)\,, \nn  \\
\delta g_Z(t_L) & = & -\frac{s_h^2}{8Z_{q_L}} \bigg(\sum_{i=1}^{N_S^{(u)}} \frac{|\epsilon_{qS}^{i(u)}|^2}{m_{iS}^{(u)2}}+\sum_{j=1}^{N_Q^{(u)}} \frac{|\epsilon_{qQ}^{j(u)}|^2}{m_{jQ}^{(u)2}} 
-2\sum_{i=1}^{N_S^{(u)}}\sum_{j=1}^{N_Q^{(u)}} \frac{\epsilon_{qS}^{i(u)}\epsilon_{qQ}^{j(u)*}}{m_{iS}^{(u)}m_{jQ}^{(u)}} \tilde k_{ij}^{L(u)}+h.c.
 \bigg)\,,  \label{deltagL}  \\
\delta g_W(t_L b_L) & =& - \frac{s_h^2}{4Z_{q_L}}\sum_{a=u,d} \bigg(\sum_{i=1}^{N_S^{(a)}} \frac{|\epsilon_{qS}^{i(a)}|^2}{m_{iS}^{(a)2}}+\sum_{j=1}^{N_Q^{(a)}} \frac{|\epsilon_{qQ}^{j(a)}|^2}{m_{jQ}^{(a)2}} -2\sum_{i=1}^{N_S^{(a)}}\sum_{j=1}^{N_Q^{(a)}} \frac{\epsilon_{qS}^{i(a)}\epsilon_{qQ}^{j(a)*}}{m_{iS}^{(a)}m_{jQ}^{(a)}} \tilde k_{ij}^{L(a)}\bigg) \,, \nn
\eea
 where 
 \be
Z_{q_L} =  {1+ \sum_{a=u,d} \sum_{j=1}^{N_Q^{(a)}}\frac{|\epsilon_{qQ}^{j(a)}|^2}{m_{jQ}^{(a)2}}}\,.
\ee
In eq.(\ref{deltagL}) $\delta g \equiv g - g_{SM}$ and
\be
g_{Z,SM}(q_L) =   T_{3L} - \sin^2\theta_W Q\,, \ \ \ g_{W,SM}(t_L b_L) = 1\,.
\ee
As expected by gauge invariance, no correction proportional to $\sin^2\theta_W$ arises.
It is straightforward to show that no deviations occur to the RH fields at tree-level, so that
\be
\delta g_Z(b_R) = \delta g_Z(t_R)  =g_W(t_R b_R) = 0\,.
\ee

\section{The Fermion Sector of the Deconstructed Models}
\label{Comparison}

\subsection{Discrete Composite Higgs Model} 
The fermionic sector of \cite{Panico:2011pw} can be studied directly in the holographic gauge. As we are interested only in the leading contribution to the 1-loop Higgs potential, we neglect in the following interactions between fermions and spin-1 fields (gauge bosons, vector and axial resonances) as well as composite fermions necessary to give mass to SM fermions other than the top.
In the two sites model the authors introduce a complete multiplet in the fundamental of $SO(5)_R$, $\tilde \psi = \tilde Q + \tilde S$, with a mass term that is only $SO(4)_R$ invariant:
\be
	\Lag^{f,2-sites}_{PW} = \Lag^{elem} + \Lag^{comp} + \Lag^{mix},
\ee
where $\Lag^{elem}$ is the kinetic term for the SM fermions,
\be
	\Lag^{comp} = i \bar{\tilde Q} \slashed{D} \tilde{Q}  + \tilde{m}_Q \bar{\tilde Q} \tilde{Q} + i \bar{\tilde S} \slashed{D} \tilde S + \tilde{m}_T \bar{\tilde S} \tilde{S},
\ee
\be
	\Lag^{mix} = y_L \tilde{f} \bar{\xi_L} U \left(  \tilde Q + \tilde S \right) + y_R \tilde{f} \bar{\xi_R} U \left( \tilde Q + \tilde S \right) + h.c.
\ee
Comparing this Lagrangian to the general one of eq.\eqref{eq:LGeneralFermions}, it is immediate to recognize that the models are the same once we fix $N_Q = N_S = 1$ and
\be
	\text{\emph{ 2 sites:}} \qquad \epsilon_{qQ} = \epsilon_{qS} = y_L f, \qquad  \epsilon_{tQ} = \epsilon_{tS} = \sqrt{2} y_R f, \qquad m_Q = - \tilde m_Q, \qquad m_S = - \tilde m_T.
	\label{eq:PW2sFermDictionary}
\ee
One can check that the sum rules of \eqref{SRIII} are satisfied while the one in eq.\eqref{SRIV} is generically not, so that the potential is logarithmically divergent. One could however impose the finiteness of the one loop potential setting $y_L = \sqrt{2} y_R$.

In the three sites model there are two composite fermionic multiplets, one in the fundamental of $SO(5)_R^1$, $\psi = Q + S$, and another one in the fundamental of $SO(5)_R^2$, $\tilde \psi = \tilde Q + \tilde S$. In the holographic gauge, the Lagrangian is
\be \begin{split}
	\Lag^{comp} =&\;  i \bar{\tilde Q} \slashed{D} \tilde{Q} + i \bar{\tilde S} \slashed{D} \tilde S  +  i \bar{ Q} \slashed{D} Q  + i \bar{ S} \slashed{D} S + \\
		& \tilde{m}_Q \bar{\tilde Q} \tilde{Q} + \tilde{m}_T \bar{\tilde S} \tilde{S} + m (\bar{Q} Q  + \bar{S} S) + \Delta (\bar{Q} \tilde{Q} + \bar{S} \tilde{S}) + h.c. \; ,\\
	\Lag^{mix} =&\;  y_L \tilde{f} \bar{\xi}_L U \left(  Q +  S \right) + y_R \tilde{f} \bar{\xi}_R U \left(  Q +  S \right) + h.c.
\end{split} \ee
Note that $\Delta$, as well as the gauging by $\rho_\mu^A$, explicitly breaks $SO(5)_R^1 \otimes SO(5)_L^2$ to the diagonal subgroup $SO(5)_D$. As the composite mass terms are not diagonal, one needs to diagonalize them before comparing this model with our setup:
\be
\left\{ \begin{split}	
	Q_1 =&\; c_{\theta_Q} \tilde Q + s_{\theta_Q} Q, \\
	Q_2 =&\; -s_{\theta_Q} \tilde Q + c_{\theta_Q} Q,
	\end{split}
\right. , \qquad 
\left\{ \begin{split}	
	S_1 =&\; c_{\theta_S} \tilde S + s_{\theta_S} S, \\
	S_2 =&\; -s_{\theta_S} \tilde S + c_{\theta_S} S.
	\end{split}
\right. 
\ee
After doing that, we obtain that this three sites model can be described by the Lagrangian \eqref{eq:LGeneralFermions} for $N_Q = N_S = 2$ and
\be \;\;\; \text{\emph{3-sites:}} \left\{ \qquad \begin{split}
	m_{1,2\: Q} =&\; \frac{1}{2} \left( m + \tilde{m}_Q \mp \sqrt{(m - \tilde{m}_Q)^2 + 4 \Delta^2} \right), \\
	m_{1,2\: S} =&\; \frac{1}{2} \left( m + \tilde{m}_S \mp \sqrt{(m - \tilde{m}_S)^2 + 4 \Delta^2} \right), \\
	\tan \theta_Q =&\; \frac{\Delta}{\sqrt{\Delta^2 + (m - \tilde{m}_Q) \left( m - \tilde{m}_Q + \sqrt{(m - \tilde{m}_Q)^2 + 4 \Delta^2}\right) }}, \\
	\tan \theta_S =&\; \frac{\Delta}{\sqrt{\Delta^2 + (m - \tilde{m}_S) \left( m - \tilde{m}_S + \sqrt{(m - \tilde{m}_S)^2 + 4 \Delta^2}\right) }}, \\
	\epsilon_{qQ}^1 =&\; y_L \tilde{f} s_{\theta_Q}, \qquad \epsilon_{qQ}^2 = y_L \tilde{f} c_{\theta_Q}, \\
	\epsilon_{qS}^1 =&\; y_L \tilde{f} s_{\theta_S}, \qquad \epsilon_{qS}^2 = y_L \tilde{f} c_{\theta_S}, \\
	\epsilon_{tQ}^1 =&\; \sqrt{2} y_R \tilde{f} s_{\theta_Q}, \qquad \epsilon_{tQ}^2 =  \sqrt{2} y_R \tilde{f} c_{\theta_Q}, \\
	\epsilon_{tS}^1 =&\; \sqrt{2} y_R \tilde{f} s_{\theta_S}, \qquad \epsilon_{tS}^2 =  \sqrt{2} y_R \tilde{f} c_{\theta_S}.
\end{split} \right. \label{eq:PW3sFermDictionary}
\ee
One can check that the sum rules \eqref{SRIII} and \eqref{SRIV} are satisfied.
One can also check that the fermion contribution to the potential has a leading mass term proportional to the square of the mixings, which can be tuned away for $y_L \simeq \sqrt{2} y_R$, allowing for a successful EWSB, confirming what stated in \cite{Panico:2011pw}. 

\subsection{Minimal 4D Composite Higgs}

The fermion sector of \cite{DeCurtis:2011yx}, as far as the top is concerned, consists of the elementary SM fields and two complete multiplets in the fundamental of $SO(5)$: $\tilde \psi = (\tilde Q, \tilde S)$, $\psi = ( Q, S)$, where we have decomposed them in the irreducible representations of $SO(4)$. In the holographic gauge, the fermion Lagrangian is\footnote{We thank Michele Redi and Andrea Tesi for having pointed out that  in their model $\Delta_{t_L}\neq \Delta_{t_R}$ in general.}
\be \begin{split}
	\Lag^{ferm} &\;= \Lag^{elem} +  i \bar{\tilde Q} \slashed{D} \tilde{Q} + i \bar{\tilde S} \slashed{D} \tilde S  +  i \bar{ Q} \slashed{D} Q  + i \bar{S} \slashed{D} S + \\
			&\; - m_T (\bar{Q} Q + \bar{S} S) - m_{\tilde{T}} (\bar{\tilde Q} \tilde Q + \bar{\tilde S} \tilde S) +\\
			&\; - (m_{Y_T} + Y_T) \bar{S}_L \tilde{S}_R - m_{Y_T} \bar{Q}_L \tilde{Q}_R + h.c. + \\
			&\; + \Delta_{t_L} \bar{\xi}_L U (Q_R + S_R) + \Delta_{t_R} \bar{\xi}_R U (\tilde{Q}_L + \tilde{S}_L) + h.c. \:.
\end{split} \ee
To compare this Lagrangian with our framework, we need to diagonalize the composite mass terms via biunitary transformations:
\be \begin{split}
	M_Q =&\; \left( \begin{array}{cc} m_T	& m_{Y_T} \\
						      0		& m_{\tilde{T}} \end{array} \right)
		= V_{Q_L} (\theta_{Q_L}) M_Q^d V_{Q_R} (\theta_{Q_R})^\dagger, \\
	M_S =&\; \left( \begin{array}{cc} m_T	& m_{Y_T} + Y_T \\
						      0		& m_{\tilde{T}} \end{array} \right)
		= V_{S_L} (\theta_{S_L}) M_S^d V_{S_R} (\theta_{S_R})^\dagger,
\end{split} \ee
where $M_Q^d = \text{diag}(m_{1Q}, m_{2Q})$, $M_S^d = \text{diag}(m_{1S}, m_{2S})$,
\be \begin{split}
	m_{1,2\:Q} =&\; \frac{1}{\sqrt{2}} \sqrt{m_T^2 + m_{\tilde T}^2 + m_{Y_T}^2 \mp \sqrt{ ( m_T^2 + m_{\tilde T}^2 + m_{Y_T}^2)^2 - 4 m_T^2 m_{\tilde T}^2 } }, \\
	\tan \theta_{Q_L} =&\; \frac{m_{\tilde T}^2 - m_T^2 - m_{Y_T}^2 - \sqrt{ (m_{\tilde T}^2 - m_T^2)^2 + m_{Y_T}^2 (m_{Y_T}^2 + 2m_{\tilde T} + 2 m_T^2 ) }  }{2 m_{\tilde T} m_{Y_T} },\\
	\tan \theta_{Q_R} =&\; \frac{m_{\tilde T}^2 - m_T^2 + m_{Y_T}^2 - \sqrt{ (m_{\tilde T}^2 - m_T^2)^2 + m_{Y_T}^2 (m_{Y_T}^2 + 2m_{\tilde T} + 2 m_T^2 ) }  }{2 m_T m_{Y_T} },
\end{split} \ee
and $m_{1,2\:S}$, $\tan \theta_{S_L}$ and $\tan \theta_{S_R}$ are the same as above with the substitution $m_{Y_T} \rightarrow Y_T + m_{Y_T}$.
Writing the Lagrangian in terms of the mass eigenstates (before EWSB),
\be
	\left\{ \begin{split}
		Q_L =&\; \cos \theta_{Q_L} Q_{1L} - \sin \theta_{Q_L} Q_{2L} \\
		\tilde{Q}_L =&\; \sin \theta_{Q_L} Q_{1L} + \cos \theta_{Q_L} Q_{2L}
	\end{split} \right. ,
\ee
and analogously for the other cases, we obtain the Lagrangian \eqref{eq:LGeneralFermions} for $N_Q = N_S = 2$ and
\be \left\{ \begin{split}
		\epsilon^1_{qQ} =&\; \Delta_{t_L} c_{\theta_{Q_R}}, \qquad \epsilon^2_{qQ} = - \Delta_{t_L} s_{\theta_{Q_R}}, \\
		\epsilon^1_{qS} =&\; \Delta_{t_L} c_{\theta_{S_R}}, \qquad \epsilon^2_{qS} = - \Delta_{t_L} s_{\theta_{S_R}}, \\
		\epsilon^1_{tQ} =&\; \sqrt{2} \Delta_{t_R} s_{\theta_{Q_L}}, \qquad \epsilon^2_{tQ} = \sqrt{2} \Delta_{t_R} c_{\theta_{Q_L}}, \\
		\epsilon^1_{tS} =&\; \sqrt{2} \Delta_{t_R} s_{\theta_{S_L}}, \qquad \epsilon^2_{tS} = \sqrt{2} \Delta_{t_R} c_{\theta_{S_L}}.
\end{split} \right. \ee
One can check that all the sum rules are satisfied by this model and therefore the Higgs potential is finite at 1-loop level. One can also check that the leading term in $\gamma_f$, quadratic in the mixing $\Delta_{t_{L,R}}$, is proportional to $Y_T (\Delta_{t_L}^2 m_T^2 - 2 \Delta_{t_R}^2  m_{\tilde T}^2) (2 m_{Y_T} + Y_{T})$.


\section{Results for Other Simple Models}
\label{App:results}

In all the models studied, and presented schematically below, EWSB is always due to a tuning between the fermionic and gauge contributions to $\gamma$. 
In the parameter scans we performed,  we have set $m_{top}({\rm TeV}) \simeq 150$ GeV and $\xi = 0.1$, solving these constraints for two of the input parameters. We have then imposed a cut for a light Higgs, $m_H \in [100,150]$ GeV. 
\begin{figure}[!t]
\vspace{-1cm}
\begin{center}
\fbox{\footnotesize $N_\rho = N_a = N_Q = N_S = 1$ - type 1 - $\xi = 0.1$} \\[0.05cm]
\hspace*{-0.65cm} 
\begin{minipage}{0.5\linewidth}
\begin{center}
	\includegraphics[width=70mm]{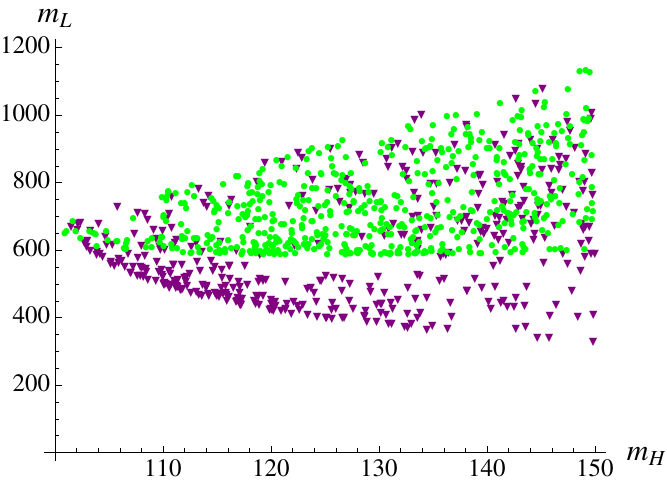}\\
	\mbox{\footnotesize (a)} \\
\end{center}
\end{minipage}
\hspace{0.25cm}
\begin{minipage}{0.5\linewidth}
\begin{center}
	\includegraphics[width=70mm]{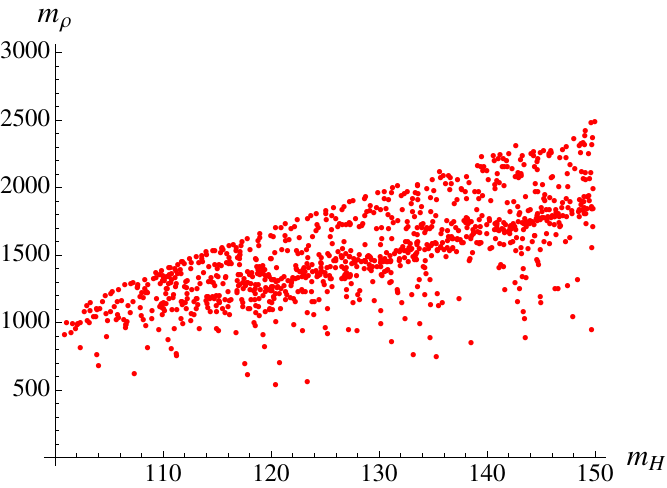}\\
	\mbox{\footnotesize (b)} \\
\end{center}
\end{minipage}
\\[0.65cm]
\fbox{\footnotesize $N_\rho = N_a = N_Q = N_S = 1$ - type 2 - $\xi = 0.1$} \\[0.05cm]
\hspace*{-0.65cm} 
\begin{minipage}{0.5\linewidth}
\begin{center}
	\includegraphics[width=70mm]{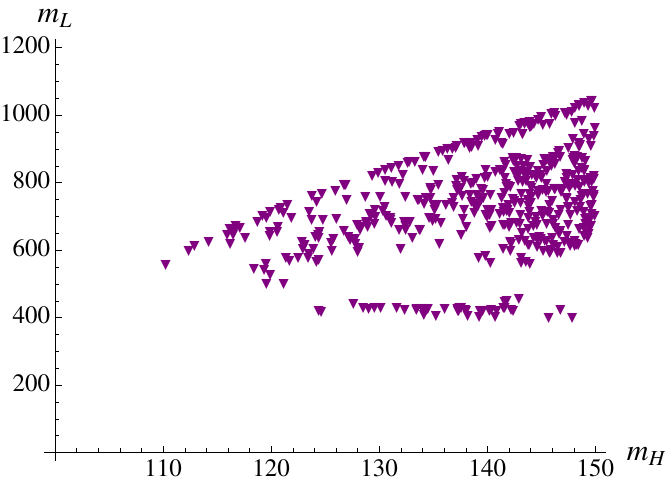}\\
	\mbox{\footnotesize (c)} \\
\end{center}
\end{minipage}
\hspace{0.25cm}
\begin{minipage}{0.5\linewidth}
\begin{center}
	\includegraphics[width=70mm]{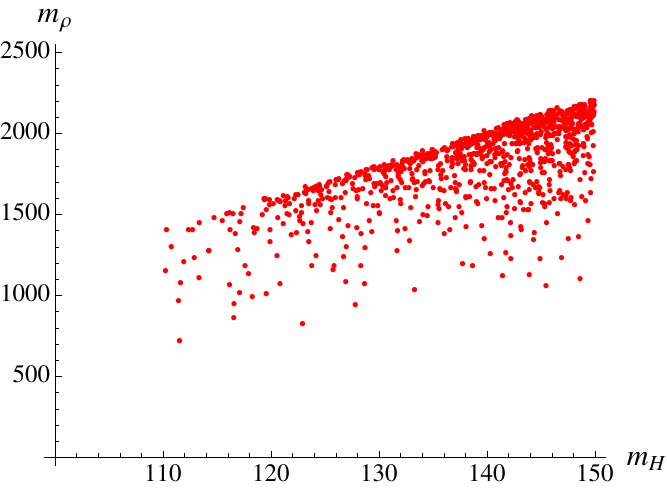}\\
	\mbox{\footnotesize (d)} \\
\end{center}
\end{minipage}
\end{center}
\vspace*{-0.2cm}
\caption{\label{fig:SpectrumMinModel} 
\small
(a,c) Mass of the LFR, before EWSB, as a function of the Higgs mass. The green circles represent the (lightest) singlet while the purple triangles represent the (lightest) exotic doublet with $Y=7/6$. (b,d) Mass of the $\rho_\mu$ vector as a function of the Higgs mass.  One can see that for $m_H \lesssim 130$ GeV, $m_\rho \lesssim 1.8$ TeV, which is too low for the model to pass the EWPT. In (a,b) we took the masses $m_Q, m_S \in [0, 5f]$, $a_\rho \in [1/\sqrt{2}, 2]$ while $\epsilon$ and $m_\rho$ have been obtained by fixing $m_{top}$ and $\xi$. In (c,d) the same range has been taken for the parameters $m, \epsilon_q$ and $a_\rho$, while $\epsilon_t$ and $m_\rho$ have been obtained by fixing $m_{top}$ and $\xi$. The direct search bound (\ref{DSbound}) has not been imposed.}
\end{figure}

\subsubsection*{Minimal model: $N_Q = 1$, $N_S = 1$, $N_\rho = 1$, $N_a = 1$}

For illustration, we consider here two versions of the minimal model, differing on how the Weinberg sum rules (\ref{SRIV}) are satisfied. We denote by ``type 1" the model where 
$\epsilon_{tS} = \epsilon_{tQ} = \epsilon_{qS} = - \epsilon_{qQ} = \epsilon $, $m_S\neq m_Q$   (as in eq.(\ref{Sol1})), and by ``type 2" the model where $\epsilon_{tS} = \epsilon_{tQ} \equiv \epsilon_t$,
$ \epsilon_{qS} = - \epsilon_{qQ} \equiv \epsilon_Q$, $m_Q=m_S\equiv m$. In the first model the LFR is either $t^\prime$ or $\chi$, while in the second one the LFR is necessarily $\chi$. In both cases the vector resonance's mass is bounded from above by $m_\rho \lesssim 2$ TeV, which implies that the S parameter is too big ($\Delta S \gtrsim 0.3$) and both models don't pass the EWPT, see fig.~\ref{fig:SpectrumMinModel}(b,d).

It is not difficult to see in more detail the tension present in this model. Let us for definiteness consider the type 1 model. The numerical scan show that EWSB mostly occurs in the region $\omega_L\ll1$, $\omega_R\simeq 1$. Taking $\omega_R=1$ and expanding at leading order in $\omega_L$, one finds
\be
\frac{m_\rho^2}{m_H^2} \simeq \frac{8\pi^2}{9\log 2}\frac{f(\omega_L)}{g^2\xi}\leq  \frac{8\pi^2}{9\log 2}\frac{1}{g^2\xi}
\label{mrhomh1model}
\ee
where 
\be
f(\omega_L) = \frac{8(1+\log \omega_L^2)}{1+8 \log \omega_L^2-\log 4/\xi}
\ee
is a smooth function $f(x)\leq 1$, for any  $x$. Using eq.(\ref{mrhomh1model}) for $m_H\simeq 125$ GeV, we immediately find an upper bound for $m_\rho$ (for $\xi=1/10$):
\be
m_\rho \lesssim 1.8 \, {\rm TeV}\,.
\label{mrhoBound}
\ee
Demanding $\Delta S\lesssim 0.2$ in eq.(\ref{SIplusII}), with $f_\rho\simeq f$,  gives $m_\rho\gtrsim 2.5$ TeV, in tension with the bound (\ref{mrhoBound}).
On the other hand, no problems from $\Delta S$ arise for $m_H\simeq 320$ GeV. A numerical scan shows indeed that this model, for $m_H\simeq 320$ GeV,
is able to pass the EWPT. The vector resonance mass is above 3 TeV and the LFR is the $t^\prime$ with $m_L\simeq 1.4$ TeV.

\begin{figure}[t]
\vspace{-1cm}
\begin{center}
\hspace*{0.25cm} 
\begin{minipage}{0.5\linewidth}
\begin{center}
	\fbox{\footnotesize $N_\rho = 2,\: N_a = N_Q =  N_S = 1$ - $\xi = 0.1$} \\[0.05cm]
	\includegraphics[width=70mm]{mh_mlightest_Nr2Na1Ns1Nq1_xi01.pdf}\\
	\mbox{\footnotesize (a)} \\
\end{center}
\end{minipage}
\hspace{-0.65cm}
\begin{minipage}{0.5\linewidth}
\begin{center}
	\fbox{\footnotesize $N_\rho = 1,\: N_a = 2,\: N_Q = N_S = 1$ - $\xi = 0.1$} \\[0.05cm]
	\includegraphics[width=70mm]{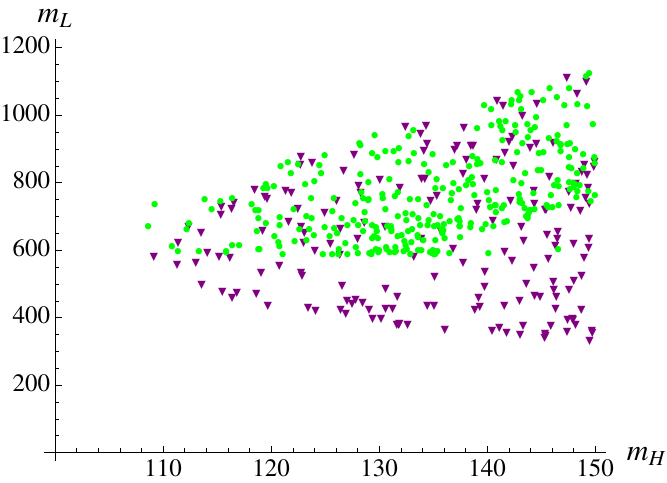}\\
	\mbox{\footnotesize (b)} \\
\end{center}
\end{minipage}
\end{center}
\vspace*{-0.2cm}
\caption{\label{fig:mLAllModels_1} 
\small
Mass of the LFR, before EWSB, as a function of the Higgs mass. The green circles represent the (lightest) singlet while the purple triangles represent the (lightest) exotic doublet with $Y=7/6$. EWPT and the bound (\ref{DSbound}) have not been imposed. In the model (a) the range in which we scanned the parameters is  the same as in fig.\ref{fig:mLNr2Na1Ns1Nq1}. For the model (b), instead, we took the fermionic masses in $[0, 5f]$, $a_\rho \in[1/\sqrt{2},2]$, $a_{1a} \in[0, \sqrt{a_\rho^2 - 1/2}]$ and $m_{a1}/m_\rho$ in a region $[0.2, 2]$ times the value for which $\Delta S$ vanishes. As usual, $m_\rho$ and $\epsilon$ have been obtained by fixing $\xi$ and $m_{top}$.} 
\end{figure}

\subsubsection*{Two vectors: $N_Q = 1$, $N_S = 1$, $N_\rho = 2$, $N_a = 1$}

We choose the type 1 finiteness condition for the fermionic sector. The numerical scan shows that the vector mass eigenstates and the axial vector can be arbitrarily heavy and therefore having a small $\Delta S$ is no longer a problem.
The LFR is either $\chi$, with $m_{7/6} \sim 500$ GeV, or $t^\prime$, with $m_0 \sim 600 - 1000$ GeV, see fig.\ref{fig:mLAllModels_1}(a). The EWPT selects points which are evenly distributed among the two regions, but the bound (\ref{DSbound}) rules out almost the whole region with a light $\chi$.
Among the points passing the EWPT, there are also ones with the lightest vector mass as light as $1.5$ TeV, while the axial is always heavier than $\sim 2.2$ TeV.

\subsubsection*{Two axials: $N_Q = 1$, $N_S = 1$, $N_\rho = 1$, $N_a = 2$}

We choose the type 1 finiteness condition for the fermionic sector. The results in this sector are completely analogue to the minimal model with the same type of finiteness condition. 
In particular, the vector resonance is always light: $m_\rho \lesssim 2$ TeV, see fig.\ref{fig:twoAxial}(a). The tree level $S$ parameter of this model can be written as  
\be
\Delta S = 8 \pi s_h^2 \frac{f^2 (m_{a1}^2 + m_{a2}^2) m_\rho^2 + 2 f_\rho^2 (m_{a1}^2 - m_\rho^2 ) (m_{a2}^2 - m_\rho^2 ) }{2 m_{a1}^2 m_{a2}^2 m_\rho^2 } \ ,
\label{Stwoaxials}
\ee   
after having solved the two Weinberg sum rules in terms of the two axial decay constants. We can see that $\Delta S$ can be made small or even negative by choosing the two masses of the axial resonances such that $m_{a1} < m_\rho < m_{a2}$. A closer inspection shows that the EWPT favour the region in parameter space
where the lighter axial resonances has sub-TeV masses. This is indeed reflected by the numerical scan, where we find that the lightest axial resonance has a mass $m_{a1} \sim 300 - 900$ GeV, see fig.\ref{fig:twoAxial}(b). This model has therefore a potentially interesting phenomenology, but it is fair to say that a model with light axial resonances 
and negative $S$ parameter looks quite ``exotic" and might not admit a consistent UV completion.

\begin{figure}[t]
\vspace{-1cm}
\begin{center}
\fbox{\footnotesize $N_a = 2,\: N_\rho=N_Q = N_S = 1$ - $\xi = 0.1$} \\[0.05cm]
\hspace*{0.25cm} 
\begin{minipage}{0.5\linewidth}
\begin{center}
	\includegraphics[width=80mm]{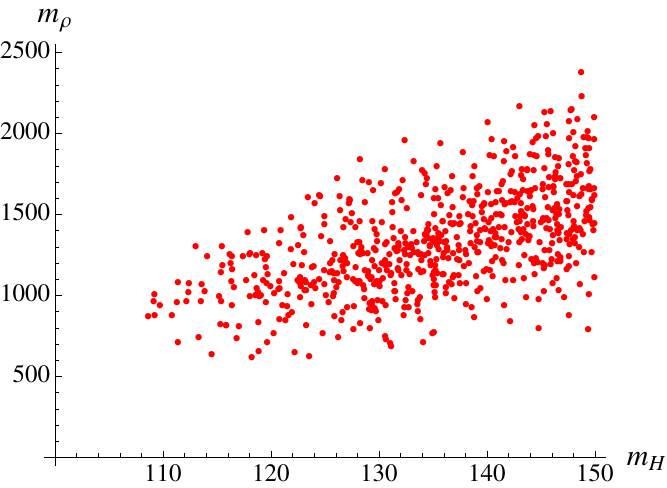}\\
	\mbox{\footnotesize (a)} \\
\end{center}
\end{minipage}
\hspace*{-0.65cm} 
\begin{minipage}{0.5\linewidth}
\begin{center}
	\includegraphics[width=80mm]{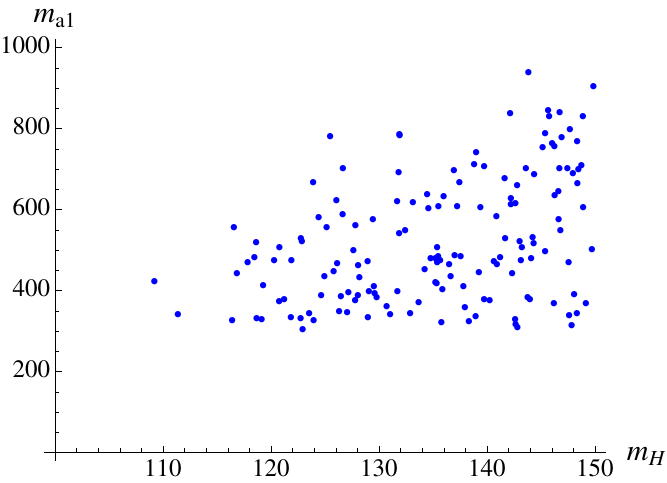}\\
	\mbox{\footnotesize (b)} \\
\end{center}
\end{minipage}
\end{center}
\vspace*{-0.2cm}
\caption{\label{fig:twoAxial} 
\small
(a) Mass of the vector resonance $\rho_\mu$  and (b) of the lightest axial vector, as a function of the Higgs mass. The points for the axial vector are the ones which pass the EWPT. The range of the parameters is the same as in fig.\ref{fig:mLAllModels_1}(b).
}
\end{figure}

\subsubsection*{Two singlets: $N_Q = 1$, $N_S = 2$, $N_\rho = 1$, $N_a = 1$}
See section \ref{Sec:results} for a more complete description of this model. In this case, the LFR is the singlet, with $m_0 \simeq 300 - 800$ GeV, see fig.\ref{fig:mLAllModels_2}(a), the second singlet being always much heavier. The vector resonance can be as heavy as 5-6 TeV, due to the fact that now $\gamma_f$ can be bigger than the minimal case. The points which pass the EWPT have $m_\rho > 2$ TeV and $t^\prime$ as the LFR, with $m_0 \simeq 500$ GeV, the other resonances being heavier than 1 TeV.

\begin{figure}[!t]
\vspace{-1cm}
\begin{center}
\hspace{0.25cm}
\begin{minipage}{0.5\linewidth}
\begin{center}
	\fbox{\footnotesize $N_\rho = 1,\: N_a = 1,\: N_Q = 1,\: N_S = 2$} \\[0.05cm]
	\includegraphics[width=70mm]{mh_mlightest_Nr1Na1Ns2Nq1_xi01.pdf}\\
	\mbox{\footnotesize (a)} \\
\end{center}
\end{minipage}
\hspace*{-0.65cm} 
\begin{minipage}{0.5\linewidth}
\begin{center}
	\fbox{\footnotesize $N_\rho = 1,\: N_a = 1,\: N_Q = 2,\: N_S = 1$} \\[0.05cm]
	\includegraphics[width=70mm]{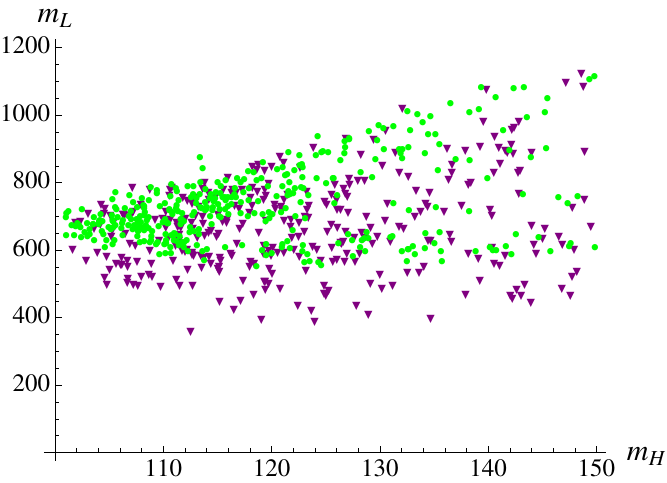}\\
	\mbox{\footnotesize (b)} \\
\end{center}
\end{minipage}
\\[0.65cm]
\hspace{0.25cm}
\begin{minipage}{0.5\linewidth}
\begin{center}
	\fbox{\footnotesize $N_\rho = 1,\: N_a = 1,\: N_Q = 2,\: N_S = 2$} \\[0.05cm]
	\includegraphics[width=70mm]{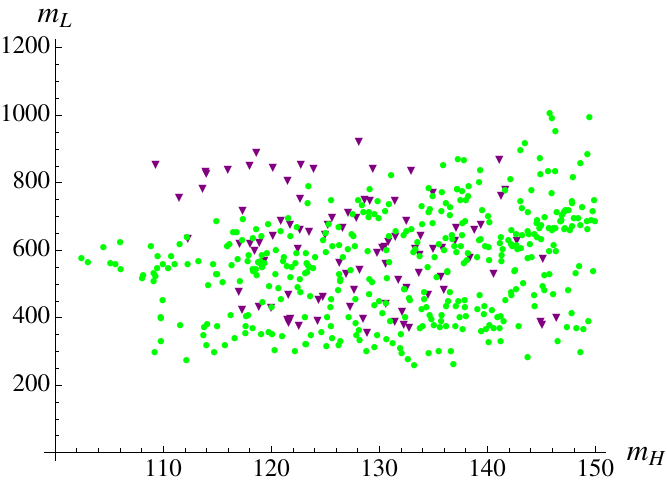}\\
	\mbox{\footnotesize (c)} \\
\end{center}
\end{minipage}
\hspace*{-0.65cm} 
\begin{minipage}{0.5\linewidth}
\begin{center}
	\fbox{\footnotesize $N_\rho = 1,\: N_a = 0,\: N_Q = 1,\: N_S = 1$ - $\log.$ div.} \\[0.05cm]
	\includegraphics[width=70mm]{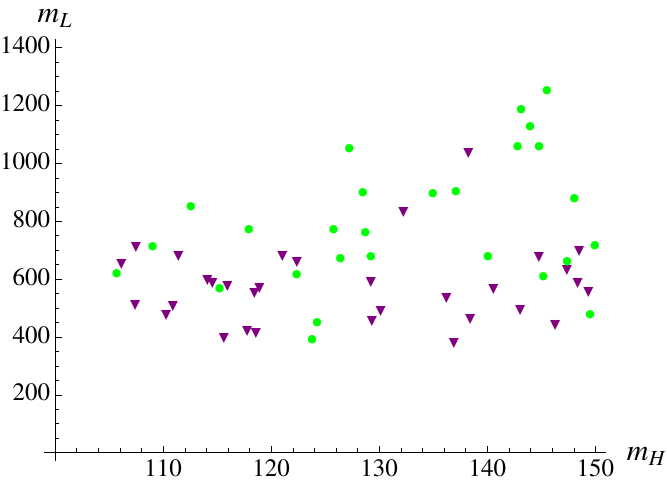}\\
	\mbox{\footnotesize (d)} \\
\end{center}
\end{minipage}
\end{center}
\vspace*{-0.2cm}
\caption{\label{fig:mLAllModels_2} 
\small
Mass of the LFR, before EWSB, as a function of the Higgs mass. The green circles represent the (lightest) singlet while the purple triangles represent the (lightest) exotic doublet with $Y=7/6$. The range of the parameters in the model (a) is the same as in fig.\ref{fig:mLNr1Na1Ns2Nq1}. For the  models (b,c) we took all the fermion masses $m_{iQ},m_{iS} \in [0, 8f]$ and $a_\rho \in [1/\sqrt{2}, 2]$, while $\epsilon_t$ and $m_{\rho}$ have been obtained by fixing respectively $m_{top}$ and $\xi$. In the log. divergent case, (d), the range is $m_Q, m_S, \epsilon_t\in [0, 8f]$ while $\epsilon_q$ has been obtained by fixing $m_{top}$. EWPT and the bound (\ref{DSbound}) have not been imposed.}
\end{figure}
\subsubsection*{Two bidoublets: $N_Q = 2$, $N_S = 1$, $N_\rho = 1$, $N_a = 1$}

In this case the LFR can be either the singlet or the lightest $Y=7/6$ doublet, their masses being always below $\sim 1$ TeV, see fig.\ref{fig:mLAllModels_2}(b). Analogously to the previous case, the vector resonance can be heavy and $\Delta S$ small. The EWPT select the points with the singlet as lightest state, $m_0 \simeq 500$ GeV, and with $m_\rho > 2$ TeV.

\subsubsection*{Two singlets and bidoublets: $N_Q = 2$, $N_S = 2$, $N_\rho = 1$, $N_a = 1$}

The most general solution for eq.\eqref{SRIII} is given in terms of four angles and two mixings:
\be \begin{split}
	\vec \epsilon_{qQ} = (\epsilon_q \cos \theta_{qQ},\; \epsilon_q\sin \theta_{qQ}), & \qquad	\vec \epsilon_{qS} = (\epsilon_q \cos \theta_{qS},\; \epsilon_q\sin \theta_{qS}), \\
	\vec \epsilon_{tQ} = (\epsilon_t \cos \theta_{tQ},\;\epsilon_t \sin \theta_{tQ}), & \qquad	\vec \epsilon_{tS} = (\epsilon_t \cos \theta_{tS},\; \epsilon_t \sin \theta_{tS}).
	\label{eq:WeinSolNs2Nq2}
\end{split}\ee
Now one can solve eq.\eqref{SRIV} for one of the remaining parameters, in the parameter scans we choose to solve it for $\epsilon_q$, as this allows us to go in the light singlet region.
The scan shows that the LFR tends to be the first singlet, see fig.\ref{fig:mLAllModels_2}(c). As in the previous two cases, the points which pass the EWPT  and the direct bound (\ref{DSbound}) have $m_\rho > 2$ TeV, $t^\prime$ as the LFR with $m_0 \simeq 400-1000$ GeV, the other resonances being generally heavier than 1 TeV.

\subsubsection*{Minimal Model with Logarithmic Divergence}

As we have seen above, the minimal model with $N_Q = N_S = N_\rho = N_a = 1$ is not viable because of a too light vector resonance, which implies a too big $S$ parameter.
This problem can be circumvented by relaxing the second Weinberg sum rules, so that the Higgs potential keeps a logarithmic divergence. This obviously implies that the MHP  hypothesis is no longer defendable, since local operators have to arise in order to renormalize the logarithmic divergence. In other words, the coefficients $\gamma_g^{(NDA)}$ and $\gamma_f^{(NDA)}$ introduced in eq.(\ref{NDAVH})  run and can be assumed to be vanishing only at a given energy scale. 
One could however hope that their impact is somehow small, so that it is still possible to make good estimates for the parameter $\xi$ integrating the form factors only up to the cutoff $\Lambda\sim 4 \pi f$.
To satisfy the first Weinberg sum rule in the fermion sector we can assume that 
\be
	\epsilon_{qS} = - \epsilon_{qQ} = \epsilon_q, \qquad   \epsilon_{tS} = \epsilon_{tQ} = \epsilon_t.
\ee
The logarithmically divergent term in $\gamma_f$ is proportional to the square of the mixing parameters, $\gamma_f \propto (\epsilon_t^2 - \epsilon_q^2) \log \Lambda / m$ where $m$ is a generic fermion mass. This is the same effect seen when adding more fermions which would allow higher values of $\gamma_f$ and, therefore, heavier vector masses.
Doing a numerical scan of such model we indeed obtain these results but, on the other side, we notice that the physics (that is, the value of $\xi$ and $m_H$) is too sensitive to the value of $\Lambda$: changing it by a factor of 2 has an $\mathcal O (1)$ effect on these observables, making the model unpredictable.

We can adopt another approach to deal with the logarithmic divergence, which is accepting that $\gamma$, and therefore $\xi$, is uncalculable. Assuming a given value of $\xi$ and using
eq.(\ref{eq:HiggsMass}) we can still compute the Higgs mass, being $\beta$ finite.
The relation $\gamma_f\simeq -\gamma_g$, connecting the fermion and the gauge sector in a crucial way, is now lost.
Given that $\beta_g \ll \beta_f$, as far as the Higgs potential is concerned, the gauge sector is completely negligible and thus unconstrained (see footnote \ref{foottR}). 
This allows the model to pass the EWPT, although in a somewhat trivial way. Neglecting the gauge sector  and performing a parameter scan for the minimal model presented above, we still obtain that a light Higgs implies light fermionic resonances, as can be seen from fig.\ref{fig:mLAllModels_2}(d).

Similar considerations would of course apply to the non-minimal models. 
As far as the Higgs sector is concerned, the price to be paid is high since EWSB is no longer under control. Moreover, as we have seen, non-minimal models are viable without the need of relaxing the second Weinberg sum rules.  For these reasons, we have decided to not explore any further models where a logarithmic divergence in the Higgs potential is kept.


\end{document}